\documentclass[aps,pra,twocolumn,english,preprintnumbers,amsmath,amssymb,nofootinbib,superscriptaddress]{revtex4-1}

\pdfoutput=1

\usepackage[latin1]{inputenc}
\usepackage{graphicx}

\usepackage{bm}
\usepackage{bbm}
\usepackage{amsmath}
\usepackage{slashed}
\usepackage{slashed}

\usepackage{dsfont}

\def\0#1#2{\frac{#1}{#2}}

\def\s0#1#2{\mbox{\small{$ \frac{#1}{#2} $}}}

\renewcommand*{\det}{\mbox{det}}

\newcommand{\be}{\begin{eqnarray}}
\newcommand{\ee}{\end{eqnarray}}

\newcommand{\nn}{\nonumber }

\newcommand{\FermiDotBox}{{\mathchoice{\raisebox{-14pt}{\includegraphics[height = 8ex]{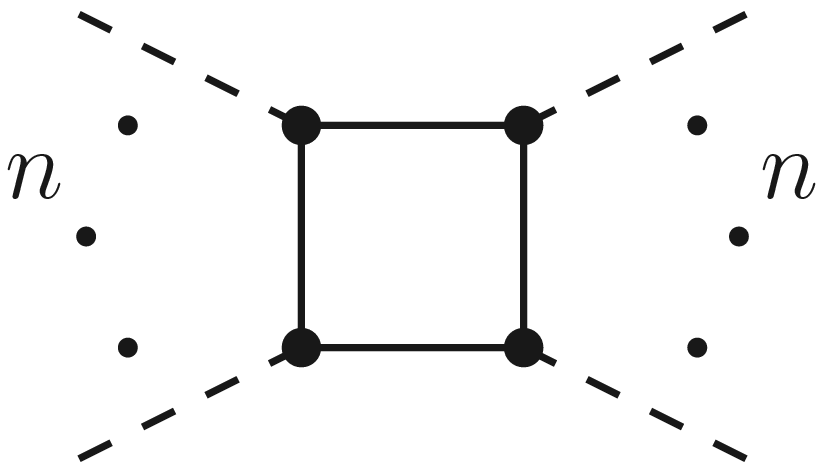}}}
{\includegraphics[height = 2ex]{./FermiDotBox.png}}
{\includegraphics[height = 1ex]{./FermiDotBox.png}}
{\includegraphics[height = 0.5ex]{./FermiDotBox.png}}}}

\newcommand{\dif}[1]{\ensuremath{\medspace \mbox{d} #1}}
\newcommand{\pdif}[2]{\ensuremath{\frac{\partial #1}{\partial #2}}}

\newcommand{\lfdif}[2]{\ensuremath{\frac{\overset{\rightarrow}{\delta} #1}{\delta #2}}}
\newcommand{\rfdif}[2]{\ensuremath{\frac{\overset{\leftarrow}{\delta} #1}{\delta #2}}}

\usepackage{babel}
\makeatother
\begin{document}

\title{Phase structure of mass- and spin-imbalanced unitary Fermi gases}

\author{Dietrich Roscher}
\affiliation{Institut f\"ur Kernphysik, Technische Universit\"at Darmstadt, 
64289 Darmstadt, Germany}
\author{Jens Braun} 
\affiliation{Institut f\"ur Kernphysik, Technische Universit\"at Darmstadt, 
64289 Darmstadt, Germany}
\affiliation{ExtreMe Matter Institute EMMI, GSI, Helmholtzzentrum f\"ur Schwerionenforschung GmbH,
64291 Darmstadt, Germany}
\author{Joaqu{\'\i}n E. Drut} 
\affiliation{Department of Physics and Astronomy, University of North Carolina, Chapel Hill, North Carolina, 27599, USA}
\begin{abstract}
We study the phase diagram of mass- and spin-imbalanced unitary Fermi gases, in search for the emergence of
spatially inhomogeneous phases. To account for fluctuation effects beyond the mean-field approximation, we employ renormalization group techniques.
We thus obtain estimates for critical values of the temperature, mass and spin imbalance, above which the system is in the normal phase. In the 
unpolarized, equal-mass limit, our result for the critical temperature is in accordance with state-of-the-art Monte Carlo calculations.
In addition, we estimate the location of regions in the phase diagram where inhomogeneous phases are likely to exist.
We show that an intriguing relation exists between the general structure of the many-body phase diagram and the binding energies of the 
underlying two-body bound-state problem, which further supports our findings. 
Our results suggest that inhomogeneous condensates form for 
mass imbalances ${m_\downarrow}/{m_\uparrow}\gtrsim 3$. The extent of the inhomogeneous phase in parameter space increases 
with increasing mass imbalance.
\end{abstract}

\maketitle

%
\section{Introduction}\label{sec:intro}
The past 15 years have witnessed tremendous advances in the experimental control and exploration of 
ultracold atomic Fermi gases. Since the first realization of a Bose-Einstein-condensate of paired fermions \cite{RegalGreinerJin_FermBEC04, *Jochim_etal_FermBEC03}, 
experimental techniques have been further developed and now allow for detailed studies of many-body phenomena,
controlled variations in temperature and polarization~\cite{Zwierlein27012006, *Partridge27012006,*2006Natur.442...54Z, *PhysRevLett.97.030401, *PhysRevLett.97.190407, *Schunck11052007, *2008Natur.451..689S}, 
studies of Bose-Fermi mixtures~\cite{Greiner2002,*Jordens2008,*Schneider05122008}, optical lattices~\cite{Greiner2008,*2011RvMP...83.1405C}, 
as well as precise determinations of the equation of state~\cite{PhysRevLett.106.215303,*2012NatPh...8..366V}, see 
Refs.~\cite{ketterle-review,Bloch:2008zzb,Giorgini:2008zz} for reviews. 
It has thus become possible to test theoretical descriptions of long-known effects such as Bardeen-Cooper-Schrieffer (BCS) 
superfluidity or Bose-Einstein-condensation (BEC) with high precision. Moreover,
many new experimental studies of many-body phenomena 
with mixtures of a variety of different fermion species (such as ${}^6$Li, ${}^{40}$K, ${}^{161}$Dy, ${}^{163}$Dy, and ${}^{167}$Er) are within reach 
in the near future (see e.g. Ref.~\cite{GrimmPC,2012PhRvL.108u5301L,2013PhRvA..88c2508F}), giving us an unprecedented opportunity to 
better our understanding of mass imbalances in strongly coupled Fermi gases and push our understanding of more exotic phenomena
such as the emergence of inhomogeneous phases~\cite{FuldeFerrell64,LarkinOvchinnikov64} to a whole new level.

In experiments, the particle density $n$ and $s$-wave scattering length $a_s$ are control parameters. In a sufficiently dilute gas, 
they represent the only scales of the systems, since the effective range $r_e$ of the interaction can safely be neglected. 
Experimentally, $a_s$ can be tuned at will by means of so-called magnetic Feshbach resonances. 
This opens up the possibility to explore many-body phenomena over a wide range of interaction strengths. 
Of particular interest is the strongly coupled ``unitary'' limit, where $a_s \rightarrow \infty$. In this regime, a non-perturbative treatment is 
inevitable due to the absence of a small expansion parameter \cite{Zwerger-book}. 

Systems with equal populations and particle masses for the different fermion species, i.e. spin- and mass-\emph{balanced} unitary Fermi gases, are
well under control by now from both an experimental and theoretical point of view. For example, lattice Monte Carlo studies 
of the equation of state and the critical temperature have reached a quantitative level~\cite{Bulgac:2005pj} 
and show good agreement with experimental data~\cite{Drut:2011tf}. 
For spin- and mass-imbalanced Fermi gases, which
are {at the heart of this work}, {less is known beyond}
the mean-field approximation, although great efforts have been made in recent years to study  
mass-imbalanced {(see, e.g., Refs.~\cite{CRLC,PhysRevLett.98.160402,Gezerlis:2009xp,Gandolfi:2010,2010PhRvA..82a3624B,PhysRevLett.106.166404,BaarsmaStoof12}) as well as
spin-imbalanced (see, e.g. Refs.~\cite{Chevy:2006,Lobo:2006,Bulgac:2007,ProkSvist07,Chevy,KBS,Schmidt:2011zu,hImb3DFRG})} unitary Fermi gases (see, e.g.,
Refs.~\cite{ChevyMora,StoofGubbels} for reviews). The difficulties encountered in studies of imbalanced systems beyond the mean-field limit are many.
For example, {\it ab initio} Monte Carlo studies are severely hampered by a so-called sign problem if spin and/or mass imbalances are 
{introduced \cite{GoulkoWingate_hImbMC10,Drut:2012md}.}
Techniques to surmount this problem have been {developed~\cite{ImhMC13,*ImMbarMC14}, but have so far 
focused on the zero-temperature equation of state of mass-imbalanced unitary Fermi gases~\cite{ImRidgeT0MC14}, and} their use is very recent.

The reasons for the interest in imbalanced systems are manifold as well. 
For example, the Fermi surfaces of the different species are mismatched in this case, possibly giving rise to exotic phenomena 
such as inhomogeneous phases of the Fulde-Ferrell-Larkin-Ovchinnikov (FFLO) type~\cite{FuldeFerrell64,LarkinOvchinnikov64} or Sarma~\cite{Sarma63} phases. 
For one-dimensional spin-imbalanced Fermi gases, where inhomogeneous pairing is expected to exist for a wide range in parameter space~\cite{Inho1DMF}, observation of an FFLO phase has indeed been claimed~\cite{Liao_etal_1DFFLO10}. 
In three dimensions, however, the inhomogeneous phase is expected to occupy only a thin layer of parameter space 
between the homogeneous superfluid and the normal fluid in parameter space, 
if at all~\cite{SheehyRadzihovskyPRL06,HuLiuMFh06,Bulgac:2008tm}. This renders the experimental detection of such phases quite challenging. The utilization of mass-imbalanced mixtures is expected to alleviate this situation somewhat 
due to the larger parameter space for inhomogeneous pairing in this case~\cite{Wang_etal_mIFFLO14}.

Most of the studies, especially of FFLO phases, have so far relied on the mean-field approximation. However, even in three dimensions, these studies
yield at best qualitative insights into the phase structure of the system. In fact, even in the balanced case it
is known that the critical temperature (measured in units of the chemical potential) is 
overestimated by a factor $\sim 1.6$ in mean-field studies. This can be traced back to the omission of order-parameter fluctuations. 
However, the fate of FFLO-type phases upon inclusion of such fluctuation effects remains largely an open question.

In this work, we analyze the phase structure of a unitary Fermi gas with spin and mass imbalance at finite temperature. 
For this purpose, we employ functional renormalization group (fRG) techniques, which allow us to include order-parameter fluctuations. 
Results from previous studies for the superfluid critical temperature of the mass- and spin-imbalanced 
case are in good agreement with experimental data and {\it ab initio} MC studies~\cite{hImb3DFRG}. 
Here, we extend the theoretical framework developed in earlier RG works~\cite{Diehl:2007th,*Diehl:2007ri,Diehl:2009ma,Scherer:2010sv,Boettcher:2012cm,hImb3DFRG} in order to investigate mass- and spin-imbalanced systems. 
Although our setup does not yet allow us to explicitly resolve inhomogeneous phases, strong hints of their existence 
beyond the mean-field approximation can already be detected and we discuss them here.

The rest of the paper is organized as follows. In Sec.~\ref{sec:model} we introduce the microscopic model and define the scales and
dimensionless parameters.
Since the identification of inhomogeneous phases is challenging, we begin our discussion of the phase structure of mass- and spin-imbalanced Fermi gases
by revisiting the two-body bound-state problem in Sec.~\ref{subsec:2}.
In a study of one-dimensional gases~\cite{Inho1DMF}, it was found that the {two-body problem in the presence of Fermi spheres provides} useful information 
about the many-body problem. In fact, it was then found that the phase structure is in approximate quantitative agreement with the associated many-body study 
in the mean-field approximation. In Sec.~\ref{sec:MF} we present the mean-field phase diagram. Corrections beyond the mean-field approximation are discussed in 
Sect.~\ref{sec:bmf} and the phase diagram resulting from those corrections is shown.

\section{Microscopic model}\label{sec:model}
Microscopically, the two-component spin- and mass-imbalanced Fermi gas in the vicinity of a broad Feshbach resonance is described by the following action:
\begin{equation}\begin{aligned}
S^{}_{\rm{F}}[\{\psi_{\sigma}\}] =& \int_{\tau,\vec{x}}\left[\sum_{\sigma = \uparrow,\downarrow} \psi^*_\sigma \left(\partial_\tau - \frac{\nabla^2}{2m_\sigma} - \mu_\sigma\right)\psi_\sigma \right.\\
&+ g\left(\psi^*_\uparrow\psi_\uparrow\psi^*_\downarrow\psi_\downarrow\right)\bigg].
\end{aligned}\label{SBase}
\end{equation}
Here,~$\int_{\tau,\vec{x}}=\int d\tau\int d^3x$.
The two fermion species are represented by the Grassmann-valued fields $\psi_{\sigma}$ that depend on spatial coordinates $\vec{x}$ and compact 
Euclidean time $\tau$. A contact interaction of strength $g$ couples both species. Its strength can be tuned through its dependence on the $s$-wave scattering length $a_s$:
\begin{equation}
g = \frac{4\pi \Lambda}{a_s^{-1}-c_{\rm{reg}}\Lambda},
\label{FermCoup_as}
\end{equation}
where $\Lambda$ is the ultraviolet cutoff and $c_{\rm{reg}}$ is a constant that depends on the regularization scheme. 

It is convenient to trade in the two fermion masses for an imbalance parameter~$\bar{m}$ using the following {definitions:
\begin{equation}
m_+ \equiv \frac{4m_\uparrow m_\downarrow}{m_\uparrow + m_\downarrow}\,,
\quad m_- \equiv \frac{4m_\uparrow m_\downarrow}{m_\downarrow - m_\uparrow }\,,
\quad \bar{m} \equiv \frac{m_-}{m_+}\,.\label{mbarDef}
\end{equation}
In} this work we set~$m_+=1$ which corresponds to the choice~$m_{\uparrow,\downarrow}=1/2$ in the mass-balanced case. 
Note that the mass imbalance parameter~$\bar{m}$ is thus normalized such that $0\leq\bar{m}<1$. 
The chemical potentials of the fermion species can be expressed in terms of an average chemical 
potential $\mu$ and the (normalized) spin imbalance parameter or so-called Zeeman field $\bar{h}$ via
\begin{equation}
\begin{aligned}
\mu = \frac{\mu_\uparrow + \mu_\downarrow}{2}, \quad h = \frac{\mu_\uparrow - \mu_\downarrow}{2}, \quad \bar{h} \equiv \frac{h}{\mu}.
\label{hDef}
\end{aligned}
\end{equation}
It is also convenient to define a dimensionless temperature parameter
\begin{equation}
\bar{T} \equiv \frac{T}{\mu}.
\end{equation}
Finally, we choose units such that $\hbar = k_{\rm{B}} = 1$.

The action $S^{}_{\rm{F}}$ defined above 
features a global $U(1)$ symmetry associated with particle number conservation. This symmetry is spontaneously broken
by a non-vanishing field expectation value $\langle \psi_\uparrow \psi_\downarrow\rangle  \neq 0$ in the superfluid phase. 
The main goal of the present work is to identify the region of parameter space where $\langle  \psi_\uparrow \psi_\downarrow\rangle$, as a 
function of $(\bar{h},\bar{m},\bar{T})$, becomes nonzero and possibly position-dependent, indicating the breakdown of translational invariance.

%
\section{Two-body Bound States}\label{subsec:2}
It is particularly challenging to identify those regimes in the phase diagram where the $U(1)$ symmetry and translational invariance are broken
simultaneously. For example, it may be reasonable to assume that the order parameter $\langle\psi_\uparrow\psi_\downarrow\rangle(\vec{x})$ is a periodic function.
However, neither its precise functional form nor the length scale associated with its period (i.e. the characteristic momentum $\vec{Q}$ associated with
the inhomogeneity in Fourier space) are known {\it a priori}. Therefore, it is worthwhile to perform preparatory analyses to help identify domains 
in parameter space where $U(1)$ symmetry breaking may appear in the full many-body problem and, in particular, where such breaking 
is accompanied by a spontaneous breakdown of translation invariance.

The physical interpretation of a finite order parameter $\langle\psi_\uparrow\psi_\downarrow\rangle \neq 0$ is a condensate of paired fermions. 
Naturally, those (bosonic) pairs have to be formed in the first place in order to condense. It thus seems reasonable to assume
that the characteristics of the energetically most 
favorable pairing state (for a given parameter set $(\bar{h},\bar{m},\bar{T})$) will strongly influence the properties of a (potential) condensate in the
many-body problem. A quantum mechanical 
bound-state calculation in Ref.~\cite{Inho1DMF} has indeed shown good qualitative {and even semi-quantitative} agreement with the many-body mean-field analysis 
in one dimension. Here, we perform a similar calculation for the three-dimensional case. 

Following Ref.~\cite{Inho1DMF},
the Schr\"odinger equation for the wave function $\Psi$ of two distinct fermions in the presence of their respective Fermi surfaces is given by
\begin{equation}
\left[\sum_{\sigma = \uparrow,\downarrow}\epsilon_\sigma(\partial_{x_\sigma}) - g\delta(x_\uparrow - x_\downarrow) + E_B \right]\Psi(x_\uparrow,x_\downarrow) = 0\,,
\label{Schroegl}
\end{equation}
where~$E_{B}=\epsilon_{{\rm F},\uparrow}+ \epsilon_{{\rm F},\downarrow}-E$ and
the delta-shaped potential is the two-body equivalent of the contact interaction in Eq.~\eqref{SBase}. Thus, the relation between the coupling strength $g$ 
and the $s$-wave scattering length is given by Eq.~\eqref{FermCoup_as}. The kinetic energy is measured with respect to the Fermi surfaces: 
$\epsilon_\sigma(\partial_{x_\sigma}) = |-(2m_\sigma)^{-1}\partial^2_{x_\sigma} - \epsilon_{{\rm F},\sigma}|$, where $\epsilon_{{\rm F},\sigma}$ 
corresponds to the Fermi energies of the two species, respectively.
Note that the general setup is well known and has been previously used to determine the properties of a single Cooper pair in the context of 
balanced BCS theory, see e.g. Refs.~\cite{Cooper56, PitaevskiiBook}.

In momentum space, Eq.~\eqref{Schroegl} can be recast into a (renormalized) integral equation for the binding energy~$E_B$:
\begin{equation}
\int_{\vec{p}}\left[\frac{1}{\epsilon_\uparrow\left(\frac{\vec{P}}{2}+ \vec{p}\right) +\epsilon_\downarrow\left(\frac{\vec{P}}{2}-\vec{p}\right)+E_B(|\vec{P}|)} - \frac{1}{2}\right]=0\,,
\label{IntSchroegl}
\end{equation}
where $\int_{\vec{p}}=\int d^3p/(2\pi)^3$.
The vectors $\vec{p}$ and $\vec{P}$ denote the relative and center-of-mass momenta, respectively. 
Since the energy of the two-particle state $\Psi$ depends on the magnitude of the center-of-mass momentum~$\vec{P}$, 
the ground-state solution of Eq.~\eqref{IntSchroegl} determines whether a bound state exists for a specific set of parameters
and, crucially for this work, provides information about the energetically favored center-of-mass momentum. 

\begin{figure}[t]
\includegraphics[width=.48\textwidth]{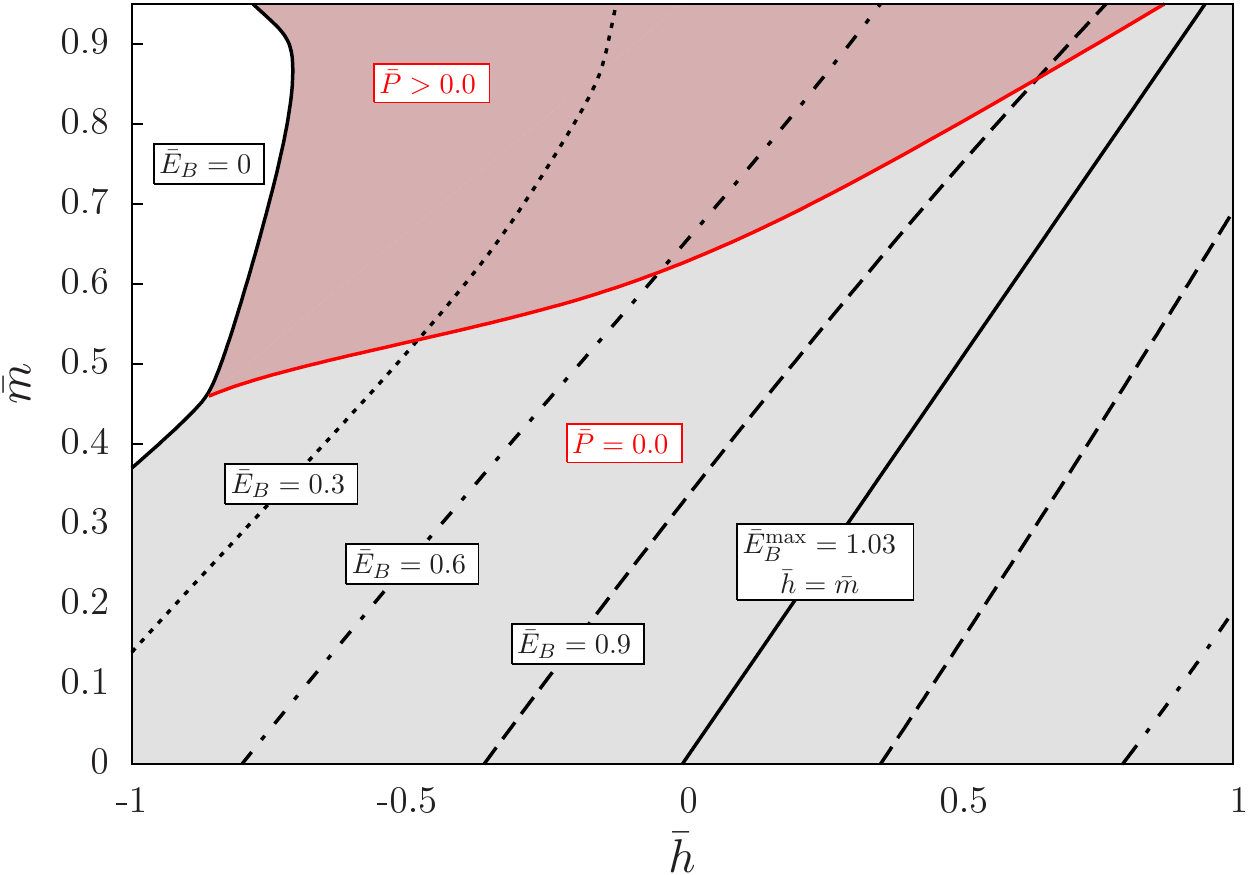}
\caption{Dimensionless two-particle binding energies $\bar{E}_B = E_B/\mu$ as obtained from Eq.~\eqref{IntSchroegl}. Domains in parameter space 
where bound states are found ($\bar{E}_B > 0$) are (gray) shaded. For red/dark shading, bound-state formation with a finite center-of-mass momentum $\bar{P} =|\vec{P}|/\sqrt{\mu}$ is favored. 
The line~$\bar{h} = \bar{m}$ corresponds to the case of equal Fermi momenta for the two species. {The two-body states found along this line are found 
to be the overall most deeply bound states.}}
\label{SchroeRes}
\end{figure}

In Fig.~\ref{SchroeRes} we show the results for the ground-state binding energies.
We find that for small $\bar{m}$, {and around the} line of equal Fermi momenta~$k_{{\rm F},\uparrow}=k_{{\rm F},\downarrow}$ at $\bar{h} =\bar{m}$,
the bound-state formation with zero center-of-mass momentum is {favored (light-gray shaded area).}
Due to the increasing mismatch of Fermi surfaces away from
the $\bar{h} = \bar{m}$ line, the (dimensionless) binding energy $\bar{E}_B = E_B/\mu$ decreases monotonically until it reaches zero for high mass imbalances and negative spin 
imbalances. If, on the other hand, $\bar{m} \gtrsim 0.46$, pairing with finite $\bar{P} = P/\sqrt{\mu}$ is favored for sufficiently small spin imbalances (dark/red-shaded area). 
Note that this seems to stabilize the binding energy away from equal Fermi momenta for very large mass imbalances, 
manifested by a slight back-bending of the $\bar{E}_B$ isolines. Similar behavior was observed in the one-dimensional case~\cite{Inho1DMF},
where it was found to have strong influence on the structure of the many-body phase diagram.

Loosely speaking, a condensate of pairs with finite center-of-mass momentum would break translational invariance. Our observation of the existence of a region 
of parameter space associated {with two-body bound states with a finite center-of-mass momentum suggest that}
a region governed by an inhomogeneous ground state may exist in the many-body phase diagram.

We close this section with a word of caution as to the relevance of our two-body analysis for the actual
many-body problem: The existence of bound states in the two-body problem in the presence of Fermi surfaces 
does not necessarily entail spontaneous symmetry breaking in the associated many-body problem. The latter 
requires, additionally, Bose-Einstein condensation of said bound states. 
Furthermore, the consideration of inert Fermi surfaces in our two-body study is questionable in the strongly coupled unitary regime.\footnote{Strictly speaking,
the assumption of inert Fermi surfaces is only justified if the Fermi momenta/energies of the non-interacting system entering our two-body study are of the order of the
Fermi momenta/energies attributed to the fully interacting many-body problem.} 
For example, the so-called 
Fermi-polaron {problem~\cite{Chevy:2006, Chevy, CRLC, Lobo:2006, Bulgac:2007, ProkSvist07, KBS,Schmidt:2011zu}, which} constitutes the limiting case of a single spin-up impurity 
in a sea of indistinguishable spin-down fermions, is known to have a negative chemical potential. This implies 
that energy is gained when a spin-up impurity is added to a system of non-interacting spin-down fermions. 
As discussed in Ref.~\cite{Chevy:2006}, this energy gain determines the value of the Zeeman field~$\bar{h}$ above which a mixed-phase may emerge in experimental studies. Moreover,
this value of~$\bar{h}$ can be viewed as a strict lower bound for the critical value of~$\bar{h}$ above which the ground state of the system is superfluid.
Note that the interaction of the impurity with all the spin-down fermions is taken into account in these studies. 
Our study of the two-body problem in the presence of Fermi spheres cannot reproduce the results of the above-mentioned Fermi-polaron problem,
as we only allow for an interaction between one spin-up and one spin-down fermion.\footnote{Note also that the non-interacting Fermi spheres enter our computation and that the associated
Fermi momentum~$k_{{\rm F},\uparrow}$ ($k_{{\rm F},\downarrow}$) becomes {complex for $\bar{h}< -1$ ($\bar{h}>1$).}}
Our approach should therefore be viewed as complementary to the Fermi-polaron problem. In fact, our
analysis gives direct access to the center-of-mass momentum of the bound state and therefore enables us to estimate the regions in the many-body phase diagram where inhomogeneous phases are likely to exist.
In any case, all results so far are by 
construction valid only at strictly zero temperature. Therefore, a proper many-body treatment is mandatory in order to obtain the actual phase diagram. However, as 
we will see below, 
our predictions resulting from Eq.~\eqref{Schroegl} for the position of inhomogeneous phases 
turn out to be astonishingly good, which emphasizes the importance of few-body physics for our 
understanding of complex many-body phenomena.

%
\section{Mean-field analysis}\label{sec:MF}
%
\subsection{Formalism and Fulde-Ferrell Ansatz}\label{sec:MFForm}
For this work, the properties of the order parameter~$\langle \psi_\uparrow\psi_\downarrow\rangle$ are essential. We therefore 
formulate the problem of spontaneous symmetry breaking in terms of the associated order-parameter field. 
This can be achieved by means of a judiciously chosen Hubbard-Stratonovich transformation~\cite{Hubbard59,*Stratonovich57},
upon which we obtain the following microscopic action
\be
S^{}_{\rm{B}}[\{\psi_{\sigma}\},\bar{\varphi}] &=& \int_{\tau,\vec{x}}\Big[\sum_{\sigma = \uparrow,\downarrow} \psi^*_\sigma \Big(\partial_\tau - \frac{\nabla^2}{2m_\sigma} - \mu_\sigma\Big)\psi_\sigma\nn \\
&&\!\!\!\!\!\! + \bar{m}_\varphi^2 \bar{\varphi}^*\bar{\varphi} - \bar{h}_\varphi\left(\bar{\varphi}^*\psi_\uparrow \psi_\downarrow - \bar{\varphi} \psi^*_\uparrow\psi^*_\downarrow\right) \Big]\,.
\label{S_HST_Base}
\ee
{This action is equivalent to the purely fermionic action $S^{}_{\rm{F}}$ in Eq.~\eqref{SBase} and is the basis for all the calculations in this work.}
The boson field $\bar{\varphi} \sim \psi_\uparrow\psi_\downarrow$ can be viewed as mediating the fermionic self-interaction 
$g (\psi^*_\uparrow\psi_\uparrow\psi^*_\downarrow\psi_\downarrow)
\sim \bar{h}_\varphi\bar{\varphi}^*\psi_\uparrow \psi_\downarrow$. Note that no kinetic term for the order-parameter field appears in the classical 
action~$S^{}_{\rm B}$. Such a term is generated dynamically when fermion fluctuations are integrated out, due to the presence of the Yukawa-type interaction,
as we will see in more detail below.

From a field theoretical point of view, the field $\bar{\varphi}$ is nothing but an auxiliary field, introduced by the Hubbard-Stratonovich transformation to facilitate the computations by removing the 
quartic fermion term in favor of a Yukawa-type interaction. From a phenomenological point of view, however, $\bar{\varphi}$ may be interpreted as a collective state of two fermions, 
corresponding to the closed channel of the Feshbach resonance, see, e.g., Refs.~\cite{Diehl:2005ae,Diehl:2007th}.

In our partially bosonized formulation, 
the order parameter for $U(1)$ symmetry breaking can now be identified {with $\bar{\varphi}_0 \sim \langle\psi_\uparrow \psi_\downarrow\rangle$.}
In the ground state, these field expectation values can be obtained by minimizing the quantum effective action $\Gamma \sim -\ln \mathcal{Z}$ with respect {to $\bar{\varphi}\sim \psi_\uparrow\psi_\downarrow$, where}
\begin{equation}
\mathcal{Z} = \int\mathcal{D}\psi_\uparrow\mathcal{D}\psi_\downarrow \mathcal{D}\bar{\varphi} e^{-S^{}_{\rm{B}}[\psi_\sigma,\bar{\varphi}]} = \int\mathcal{D}\bar{\varphi}\, \det_\psi[\bar{\varphi}]
\label{Partfunc}
\end{equation}
is the partition function in the path-integral representation. Note that the determinant on the right-hand side can in principle be used to define a purely bosonic action~$S_{\rm PB}[\bar{\varphi}]=-\ln \det_\psi[\bar{\varphi}]$,
which is the starting point for
lattice MC {calculations~\cite{Chen:2003vy,Bulgac:2005pj,GoulkoWingate_hImbMC10} (see Ref.~\cite{Drut:2012md} for a review).}
While the effective action~$\Gamma$ shares the symmetries of the microscopic action $S^{}_{\rm{B}}$ 
by construction,\footnote{The path integral measure respects the symmetries of the theory.} it 
includes the effects of all thermal and quantum fluctuations. Thus, it is composed of renormalized fields and couplings which are in general momentum dependent. 
Since the fermion determinant $\det_\psi$ involves a generally complicated dependence on $\bar{\varphi}$, an exact computation of $\Gamma$ is highly non-trivial. 
Therefore, systematic approximation schemes are needed to gain insight into the physical content of the theory. 

In this sense, the widely used mean-field approximation can be considered as a lowest-order approximation to the effective action. 
It is obtained by shifting the field~$\bar{\varphi}\to \bar{\varphi}+\delta\bar{\varphi}$, where $\delta\bar{\varphi}$ {now represents} the fluctuation field, 
and performing a saddle-point approximation of the path integral about~$\bar{\varphi}$.
This renders the analytic computation of the fermion determinant more feasible.
However, it is still non-trivial to carry out if one allows for a general space-dependent ``background" field~$\bar{\varphi}(\vec{x})$, 
which is needed to enable the detection of inhomogeneous phases. 
In the following, we employ an analytically feasible Fulde-Ferrell {ansatz (FF)~\cite{FuldeFerrell64},}
\begin{equation}
\bar{\varphi}(\vec{x}) =\bar{\varphi} e^{i2\vec{Q}\cdot\vec{x}}\,.
\label{FFAnsatz}
\end{equation}
Since~$\bar{\varphi}$ is chosen to be a constant amplitude, we are left with the standard mean-field ansatz in the limit of vanishing 
momentum~$\vec{Q}$.
Note that the ansatz~\eqref{FFAnsatz} may be regarded as the first term of a Fourier expansion of a more general $\bar{\varphi}(\vec{x})$. {It is hence considered to} 
be a particularly good approximation to the full solution in the vicinity of a second order phase transition to the normal phase, where higher order contributions {are expected to be}
small (see e.g. Refs.~\cite{ThiesInhoRev06,BasarDunnePRD08,Inho1DMF,GNDTrick}).

Using the ansatz~\eqref{FFAnsatz}, the order-parameter potential~$U$ becomes
\begin{equation}
\begin{aligned}
U&(\bar{\Delta},Q) = {\int_{\vec{q}}\left\{E_{\bar{\Delta}=0} - E_{\bar{\Delta}} + \frac{\bar{\Delta}^2}{2q^2}\right.}\\
& {-\left.\sum_{\sigma = \pm 1}T\log\left(1+e^{-\frac{1}{T}E_{\bar{\Delta}}-\frac{\sigma}{T}\left[\left(q^2+Q^2\right)\bar{m} - 2\vec{q}\cdot\vec{Q} - h \right]}\right)\right\}}
\end{aligned}
\label{MFPot}
\end{equation}
where 
\begin{equation}
E_{\bar{\Delta}} = \sqrt{(q^2+Q^2-2\bar{m}\vec{q}\cdot\vec{Q} - \mu)^2 + \bar{\Delta}^2}\,, 
\end{equation}
and $Q =|\vec{Q}|$ as well as $q =|\vec{q}|$. The quantity $\bar{\Delta} = \bar{h}_\varphi^2 \bar{\varphi}^*\bar{\varphi}$ evaluated at the (global) minimum~$\bar{\varphi}_0$ 
of~$U$ is nothing but the fermion gap. Note that the minimum of~$U$ in $\vec{Q}$-direction is not necessarily an extremum. A global (numerical)
minimization of the potential is therefore required to find the {ground state of the theory.}

\subsection{Results}\label{sec:MFRes}
\begin{figure}[t]
\includegraphics[width=.48\textwidth]{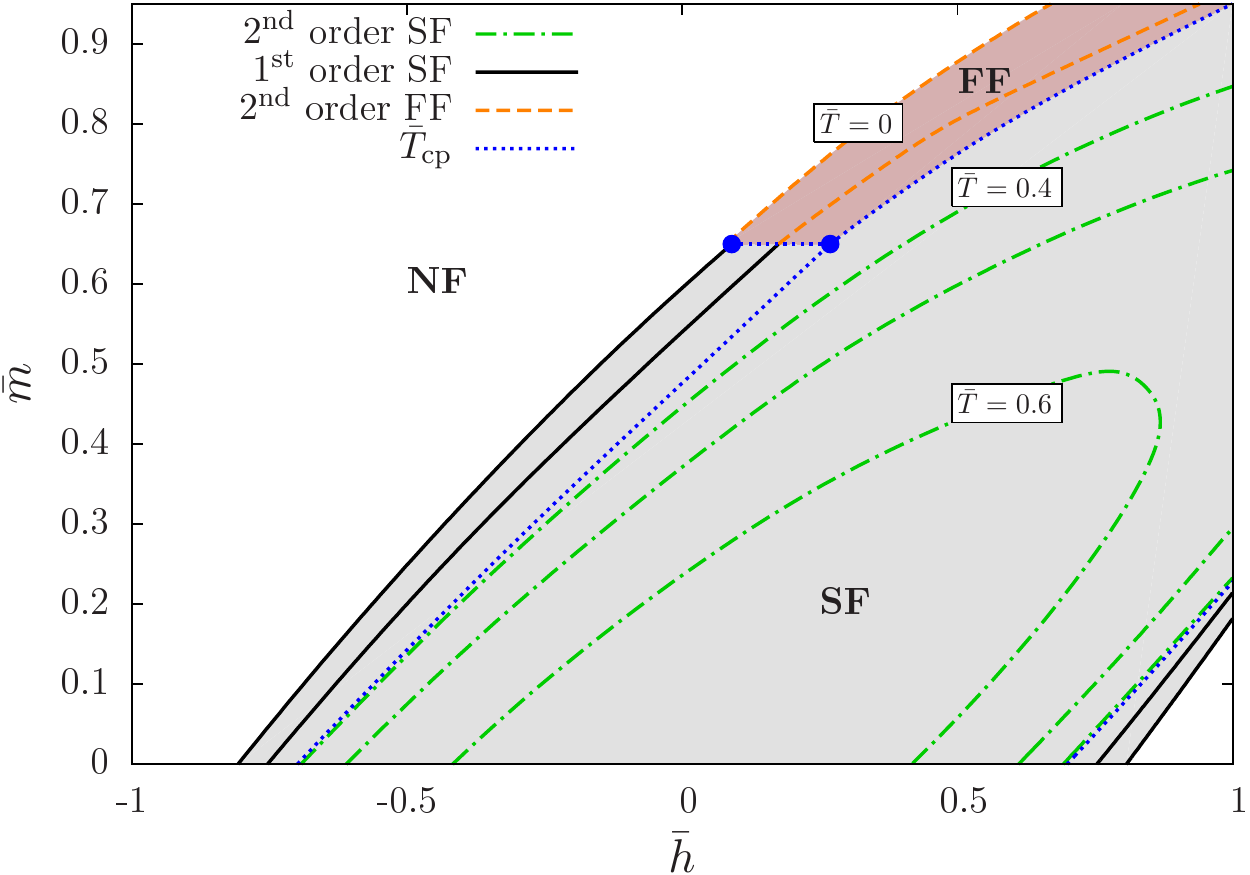}
\caption{{Mean-field phase diagram} in the plane spanned by $\bar{h}$ and $\bar{m}$ with $\bar{T}$-isolines. The (light) gray-shaded region corresponds to a homogeneous condensate, i.e. $\Delta_0 >0$ and $Q_0 = 0$ (SF);
the dark/red-shaded area depicts a Fulde-Ferrell regime, i.e. $\Delta_0> 0$ and $Q_0 > 0$ (FF). The (blue) dotted lines correspond to lines of multicritical points~$\bar{T}^{}_{\rm cp}$, where either a surface associated with first-order transitions 
from the homogeneous to the inhomogeneous phase meets a surface associated with second-order transitions from the homogeneous to the normal fluid (NF) phase as well as a surface from the
inhomogeneous to the normal fluid phase (NF), or where a surface associated with second-order transitions from the homogeneous to the NF phase meets a surface associated with  
first-order transitions from the homogeneous to the NF phase.}
\label{MFDiag}
\end{figure}

In Fig.~\ref{MFDiag} we show the phase diagram as obtained from a direct minimization of $U(\Delta,Q)$ in
Eq.~\eqref{MFPot}. Here, we mainly discuss aspects related to the emergence of inhomogeneous phases.
For detailed discussions on the extent and the properties of the homogeneous phases at zero and finite temperature, 
we refer the reader to earlier work, {e.g. Refs.~\cite{PhysRevLett.98.160402,WuPaoYip06,2010PhRvA..82a3624B,2010PhRvA..82a3624B,Braun:2014ewa}.} 

We begin by briefly discussing some characteristic features of the phase diagram. 
The region of homogeneous symmetry breaking is roughly centered around the line of equal Fermi momenta, $\bar{h}=\bar{m}$, 
as suggested by the analysis of the two-body problem of Sec.~\ref{subsec:2}. At $\bar{T}=0$, the transition from the homogeneous superfluid 
to the normal phase is always of first order for the parameter space considered in the present work. At sufficiently high temperatures, on the other hand, 
the transition from the superfluid to 
the normal phase changes into a second-order transition. The surfaces in parameter space associated with these two types of transitions meet at a 
critical line denoted by $\bar{T}^{}_{\rm{cp}}$. Note that we limit ourselves to spin imbalances $\bar{h} \in [-1,1]$.
Contrary to the case of the two-body analysis in Sec.~\ref{subsec:2}, this constraint is not imposed by the method itself. 
In fact, the domain of high mass imbalance, $\bar{m} \gtrsim 0.8$ for $\bar{h} > 1$ is 
of interest for an investigation of the physics of \emph{Sarma} phases~\cite{Sarma63}. In particular, it was found in mean-field calculations that 
a second order transition occurs in this regime {even at $T=0$~\cite{PhysRevLett.98.160402}.}\footnote{By employing the functional RG scheme 
discussed in Sec.~\ref{sec:FRGForm}, we do indeed find strong hints that a zero-temperature Sarma phase in the regime of large $\bar{m}$ and $\bar{h}>1$ 
exists even beyond the mean-field approximation.
For a discussion of the existence of Sarma phases in the phase diagram of spin-imbalanced but mass-balanced Fermi gases beyond the mean-field limit
see Ref.~\cite{BoettcherSarma}.} 
We leave a detailed discussion of Sarma phases for future work. 

We now turn to a discussion of the inhomogeneous phase depicted by the dark/red-shaded area in Fig.~\ref{MFDiag}. 
Studies similar to ours have been performed for
specific values for~$\bar{m}$ {(e.g., Li-K mixture~\cite{BaarsmaStoof12})} as well as arbitrary values of~$\bar{m}$ employing a $T$-matrix 
approach~\cite{Wang_etal_mIFFLO14}. While our mean-field results agree well with those of {Ref.~\cite{BaarsmaStoof12}, we} do not find an 
inhomogeneous phase for mass imbalances down to $\bar{m} = 0$ as 
in Ref.~\cite{Wang_etal_mIFFLO14}. There is, in fact, disagreement in the literature on this point (see e.g., Refs.~\cite{SheehyRadzihovskyPRL06,HuLiuMFh06}), which suggests that the appearance of an inhomogeneous phase is very 
sensitive to the details of the approximation scheme as well as the specific ansatz for the inhomogeneity. This strengthens our motivation to consider alternative sources of information on pairing behavior such as our few-body analysis of
Sec.~\ref{subsec:2} and RG methods to account fluctuation effects (Sec.~\ref{sec:bmf} below).
\begin{figure}[t]
\includegraphics[width=.48\textwidth]{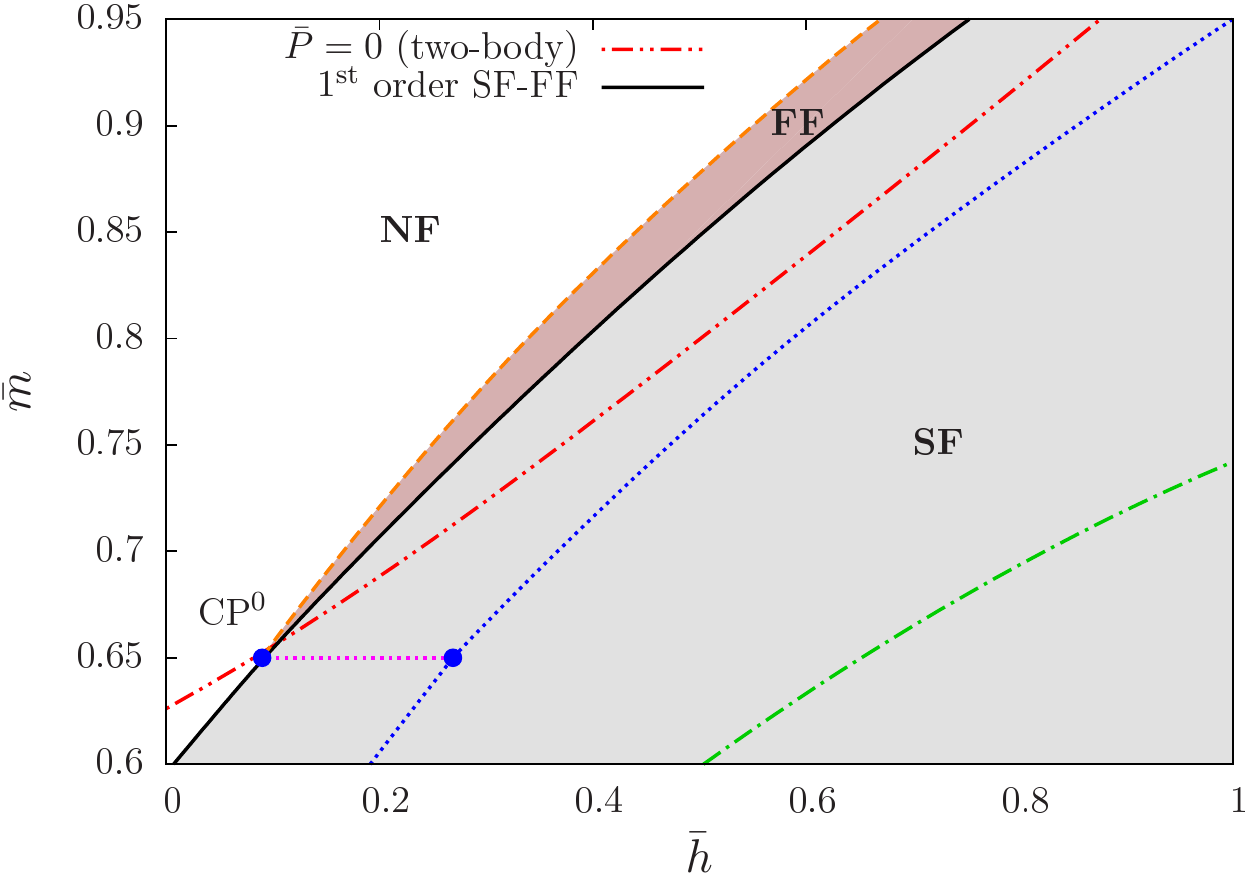}
\caption{Zoom of Fig.~\ref{MFDiag} showing the regime in parameter space {where inhomogeneous phases (FF) are found in our mean-field study.} 
The multicritical (dotted) and the second-order (dashed and dot-dashed) lines are defined as in Fig.~\ref{MFDiag} and included here to guide the eye. 
The black/solid line 
represents the transition line between the {homogeneous (SF) and inhomogeneous or normal phases (NF) at $\bar{T}=0$,}
respectively. Finally, the (red) dot-dot-dashed line is the lower bound for pairing with $\bar{P} > 0$ 
as obtained from our two-body analysis, see Fig.~\ref{SchroeRes}.}
\label{MFInhoZoom}
\end{figure}

Fig.~\ref{MFInhoZoom} displays an enlarged view of the region with inhomogeneous superfluidity, focusing on the domain of parameter space governed by inhomogeneous pairing at $\bar{T}=0$.
As evident from the figure, the inhomogeneous phase occupies only a thin shell in parameter space close the phase with a homogeneous condensate. It is separated from the normal phase 
by a second-order transition and from the homogeneous phase by a first-order transition.\footnote{Both the phase with a homogeneous condensate
and the one with an inhomogeneous condensate are associated with spontaneous $U(1)$ symmetry breaking. The boundary between these two phases is solely associated with
translation symmetry breaking where the value of~$Q=|\vec{Q}|$ of the condensate acts as an order parameter. On the other hand, $U(1)$ symmetry is restored in
the normal phase. The boundary between the normal phase and the phase with a (in)homogeneous condensate is therefore associated with $U(1)$ symmetry breaking where~$\bar{\varphi}_0$ 
plays the role of the order parameter.}
Note that, for finite negative scattering lengths $a_s$, a similar behavior has been found previously also for the mass-balanced case~\cite{HuLiuMFh06}.

Within numerical errors, the position of the first-order transition line from the homogeneous to the inhomogeneous phase coincides almost perfectly with the transition line from the 
homogeneous superfluid to the normal phase, which is found if one does not
allow for a space-dependent order parameter.\footnote{For fixed~$\bar{m}$, we find that the extent of the homogeneous phase which has been superseded by an inhomogeneous phase is as narrow as $\delta\bar{h} \approx 0.006$ for the largest 
considered $\bar{m}=0.95$ and decreases rapidly towards the zero-temperature endpoint of the critical line at CP$^0\approx (0.09,0.65)$.}
The observed first-order nature of the homogeneous-inhomogeneous transition may be an artifact of the FF-ansatz in our study. In fact, 
studies of analytically solvable relativistic models in 1D found that the corresponding transition line is of second order rather than first order~\cite{ThiesInhoRev06,BasarDunnePRD08}.
The position of the associated transition line was found to be reasonably well described when the space dependence
of the condensate is approximated by the first term of the Fourier expansion~\cite{GNDTrick}, as done by the FF ansatz.
However, the order of the transition is spuriously predicted to be of first oder.\footnote{Note that, in addition
to a first-order transition measured by the parameter~$Q$, we observe a discontinuity in the $U(1)$ order-parameter~$\bar{\varphi}_0$ as well.}

The very small extent of the inhomogeneous phase in $\bar{h}$-direction for fixed~$\bar{m}$ close to the zero-temperature critical endpoint CP$^0$ renders the precise determination
of its coordinates very difficult from a numerical point of view. Following our line of arguments from Sec.~\ref{subsec:2}, the quick disappearance of the 
inhomogeneous phase close to this point is not unexpected: 
The two-body binding energy away from its maximum at $\bar{h} = \bar{m}$ is ``stabilized" for finite pair-momentum only at very large mass imbalances, as indicated by the characteristic back-bending of the lines of constant binding energy in Fig.~\ref{SchroeRes}.
Since deeply bound pairs should be less sensitive to (quantum) fluctuations that tend to destroy pairing correlations, the formation of a condensate in the many-body problem may be expected to be favored in this regime as well. Thus, 
the inhomogeneous phase is expected to widen towards smaller $\bar{h}$ only at large~$\bar{m}$, whereas it is expected to narrow down when~$\bar{m}$ is decreased. This is indeed what we find in Fig.~\ref{MFInhoZoom}.

In the same figure, we also show the lower bound (dot-dot-dashed line) for pair formation with a finite center-of-mass momentum 
$\bar{P} > 0$, see also Fig.~\ref{SchroeRes}. We find that there is an obvious discrepancy between this line (which we argued above is 
important for the occurrence of inhomogeneous condensates) and the actual many-body phase boundary
associated with a transition from a homogeneous {to an inhomogeneous phase.} 

The critical endpoint CP$^0$ of the inhomogeneous phase appears to coincide quite well with the intersection point of the lower bound for 
{finite-momentum pair formation (dot-dot-dashed line)}
and the phase boundary of the phase with a homogeneous condensate (black solid line). A high precision determination of the position of CP$^0$ is beyond our reach at the 
moment due to the numerical complications mentioned above; we therefore refrain from drawing rigorous conclusions about this stunning coincidence.
For larger values of $\bar{m}$ and $\bar{h}$ in the domain between the (black) solid and (red) dot-dot-dashed line, we find that a homogeneous condensate is energetically favored
over an inhomogeneous condensate (see Fig.~\ref{MFInhoZoom}), even though bound-state formation with finite center-of-mass momentum should still be preferred. 
However, as discussed above, the simple FF ansatz~\eqref{FFAnsatz} for $\bar{\varphi}(x)$ is not expected to yield precise results away from the transition to the normal phase. 
While our two-body calculation is not hampered by assumptions of this kind, it {\it is} limited by the assumption of inert Fermi surfaces. In contrast, 
our mean-field study of the full many-body problem takes into account interactions between all spin-up {and spin-down fermions.
We shall come back to this issue below when we discuss the results of our RG study.}

In summary, the need to investigate the dynamics of the many-body problem beyond the mean-field limit seems inevitable. Furthermore,
the above analysis indicates that one must include more sophisticated ans\"atze for the condensate $\bar{\varphi}(x)$ to gain reliable 
insights into the problem of inhomogeneous superfluidity. Our fRG analysis, as detailed next, aims to provide a systematic way to approach this goal.

\section{Beyond the Mean-field Approximation}\label{sec:bmf}
%
\subsection{Functional Renormalization Group: Formalism}\label{sec:FRGForm}
To understand spontaneous symmetry breaking in quantum systems, it is essential to account for quantum fluctuations of the order-parameter.
In order to accomplish this in a systematic way, we employ a fRG approach. More specifically, the Wetterich equation~\cite{Wetterich93}
\begin{equation}
\partial_k \Gamma_k = \frac{1}{2} \mbox{STr} \left[\frac{\partial_k R_k}{\Gamma_k^{(2)} + R_k}\right]
\label{WetterichEq}
\end{equation}
will be central to our analysis. Here, $k$ is the RG scale introduced by the regulator function $R_k$. 
The latter specifies the details of the Wilsonian momentum-shell integration. In particular, it provides for a suitable infrared (IR) and ultraviolet (UV) regularization,
see App.~\ref{app:FEq} for details. 
The Wetterich equation determines the change of the effective average action~$\Gamma_k$ under a change of the scale~$k$ and therefore allows to 
interpolate between the microscopic action $S$ at a given UV 
cutoff scale~$\Lambda$ and the full quantum effective action~$\Gamma\equiv\Gamma_{k\to 0}$ in the long-range (i.e. IR) limit:
\begin{equation}
S = \lim_{k\rightarrow\Lambda\to\infty} \Gamma_k,\quad \Gamma = \lim_{k\rightarrow 0} \Gamma_k\,.
\end{equation}
{For reviews and introductions to this approach with respect to an application to ultracold Fermi gases, see 
Refs.~\cite{Diehl:2009ma,Scherer:2010sv,Braun:2011pp,Boettcher:2012cm}.}

While Eq.~\eqref{WetterichEq} is exact, in general one must consider an ansatz for $\Gamma_k$ (i.e. a truncation of the full~$\Gamma_k$) in order to 
arrive at a solution. Here, we choose
\begin{equation}\begin{aligned}
\Gamma_k[\{\psi_\sigma\},\varphi] =& \int_{\tau,\vec{x}}\Big[\sum_{\sigma = \uparrow,\downarrow} \psi^*_\sigma \Big(\partial_\tau - \frac{\nabla^2}{2m_\sigma} - \mu_\sigma \Big) \psi_\sigma\\
&+ \varphi^*\Big(\frac{Z_{\varphi,k}}{A_{\varphi,k}}\partial_\tau -\frac{1-\bar{m}^2}{2}\nabla^2 \Big)\varphi\\ 
&+ U_k(\rho) - h_\varphi\Big(\varphi^*\psi_\uparrow \psi_\downarrow - \varphi \psi^*_\uparrow\psi^*_\downarrow\Big) \Big]
\end{aligned}\label{GammaAnsatz}
\end{equation}
with renormalized quantities
\begin{equation}
\varphi = A_{\varphi,k}^{\frac{1}{2}}\, \bar{\varphi}\,,\quad h_\varphi = A_{\varphi,k}^{-\frac{1}{2}}\,\bar{h}_\varphi\,.
\label{ExpRenorm}
\end{equation}
This ansatz is an extension of a class of ans\"atze which has been successfully applied to spin-balanced 
(see, e.g., Refs.~\cite{Diehl:2007th,*Diehl:2007ri,Diehl:2009ma,Bartosch:2009zr,Scherer:2010sv,Boettcher:2012cm,PhysRevA.89.053630}) 
and spin-imbalanced (see, e.g., Refs.~\cite{Schmidt:2011zu,Krippa:2014kra,hImb3DFRG,BoettcherSarma}) systems across the whole BEC-BCS crossover. 

In addition to an ansatz for~$\Gamma_k$, the solution of the RG equation~\eqref{WetterichEq} requires an initial condition~$\Gamma_{k=\Lambda}$. 
In our case this implies choosing initial conditions for the so-called wavefunction renormalizations~$Z_{\varphi,k}$ and~$A_{\varphi,k}$,
the Yukawa coupling~$h_\varphi$, and the effective order-parameter potential~$U_k$. The latter depends only on the 
$U(1)$ invariant $\rho = \varphi^*\varphi$. At the UV scale~$\Lambda$, it is fixed by the two-body problem in the vacuum and is therefore
simply given by~\cite{Diehl:2007th,*Diehl:2007ri}
\be
{U_\Lambda(\rho) = m^2_{\varphi,\Lambda} \rho\,,}
\ee
where
\begin{equation}
m^2_{\varphi,\Lambda} = \nu_\Lambda + \frac{h_{\varphi,\Lambda}^2}{6\pi^2}\Lambda\,.
\label{InitialUh}
\end{equation}
and
\be
h_{\varphi,\Lambda} = \sqrt{6\pi^2\Lambda}\,.
\ee
Here, the initial value of the Yukawa coupling $h_\varphi$ has been chosen such that the system is set to be close to a broad Feshbach resonance~\cite{Diehl:2007th,*Diehl:2007ri}.
The parameter $\nu_\Lambda \sim (B-B_0)$ measures the detuning of the system with respect to the resonance.\footnote{Here, $B$ denotes the external magnetic field.} 
Since we are only interested in the unitary limit 
(i.e. the system {\it at} resonance), we set $\nu_\Lambda \equiv 0$ for the remainder of this work. 
Higher order terms~$\sim\rho^n$ in the potential~$U_k$ are generated during the RG flow
due to quantum {fluctuations and are taken into account in our analysis.}

At this point we are left with the determination of the initial conditions for the wavefunction renormalizations. 
The momentum and frequency dependence of the boson-field propagator {effectively allows to resolve} at least part of the momentum and frequency dependence 
of the fermionic self-interactions (see, e.g., Ref.~\cite{Braun:2011pp} for a more general introduction).
Per our ansatz~\eqref{GammaAnsatz}, the inverse boson propagator for the bare/unrenormalized field~$\bar{\varphi}$ 
in momentum space is\footnote{For convenience, we do not show contributions to the propagator stemming from the insertion of the regulator function into the path integral.}
{\be
&& \!\!\!\!\!\! \bar{P}_\varphi (q_0,\vec{q}^{\,2}) \nn\\
&& \; = \Big(iZ_{\varphi,k} q_0 + \frac{1\!-\!\bar{m}^2}{2} A_{\varphi,k} \vec{q}^{\,2} 
+ \frac{\partial^2 U}{\partial \bar{\varphi}\partial \bar{\varphi}^{\ast}}\Big|_{\bar{\varphi}_0}\Big)
\,,
\label{BoseProp}
\ee}
where~$\bar{\varphi}_0$ denotes 
the $k$-dependent value of the position of the ground-state 
of the potential~$U_k$.
The wavefunction renormalizations $Z_{\varphi,k}$ and~$A_{\varphi,k}$ are assumed to be independent {of~$q_0$ and~$\vec{q}$. 
Here, we only take into account the scale dependence of the wavefunction renormalizations, which is minimally required to
detect the emergence of an inhomogeneous ground state within our present setup (see our discussion below).}
The structure of this propagator can be understood as arising from a derivative expansion of the effective action that has been truncated at the lowest nontrivial order. 

The RG flow of the wavefunction renormalizations~$Z_{\varphi,k}$ and~$A_{\varphi,k}$ is conveniently parameterized with the aid of the associated anomalous dimensions~$\eta_{Z,k}$
and~$\eta_{A,k}$, respectively. For example,
\begin{equation}
\eta_{A,k} = -k\partial_k \ln A_{\varphi,k}\,.
\label{etaADef}
\end{equation}
In the present work, we keep the ratio $Z_{\varphi,k}/A_{\varphi,k}=1$ fixed in the RG flow and compute 
only the flow {of~$A_{\varphi,k}$, see also Ref.~\cite{hImb3DFRG}.}
The initial condition for~$A_{\varphi,k}$ is given by
\be
\lim_{k\to\Lambda\to\infty}A_{\varphi,k}=0\,.\label{eq:aphiic}
\ee
From a phenomenological point of view, this implies that the boson field is not dynamical at the UV scale~$\Lambda$.\footnote{In our numerical studies, 
we have chosen a finite value for~$\Lambda=1000\sqrt{\bar{\mu}}$ and therefore a finite value for the initial condition~$A_{\varphi,k=\Lambda}=1$ such 
that it is consistent with Eq.~\eqref{eq:aphiic}.}
Its dynamics as ``measured" by
the wavefunction renormalization~$A_{\varphi,k}$ {is solely generated by 
quantum fluctuations.} Our choice for the initial conditions ensures that our ansatz for the effective average action~$\Gamma_k$
is identical to the well-known partially bosonized action~\eqref{S_HST_Base} at the UV scale~$\Lambda$, as it should be.

By applying the Wetterich equation~\eqref{WetterichEq} to the ansatz~\eqref{GammaAnsatz}, the RG flow equations for the various couplings 
can now be derived. Details as well as our choices for the regulator functions can be found in App.~\ref{sec:regfct} and~\ref{app:FEq}. In the following, we only discuss the general properties
of these equations.

The general form of the flow equation for the Yukawa coupling reads
\begin{equation}
k\partial_k h_\varphi = \frac{1}{2}\eta_{A,k} h_\varphi\,.
\label{YukawaFlow}
\end{equation}
Note that we have dropped terms on the right-hand side which are only non-zero in the regime with broken $U(1)$ symmetry but vanish identically otherwise.
Since we are primarily aiming at a determination of the phase boundaries, but not at a quantitative prediction of a particular observable 
(e.g. the {\it Bertsch} parameter) for which these terms may become important, we expect this approximation to be justified. The flow of the 
Yukawa coupling is then purely driven by the anomalous dimension of the boson field, which be can be split into two distinct contributions:
\begin{equation}
\eta_{A,k} = \eta_{A,k}^\psi + \eta_{A,k}^\varphi\,.
\label{etaAStruc}
\end{equation}
The contribution $\eta_{A,k}^\psi$ is built up by one-particle irreducible (1PI) diagrams with no internal boson lines but only fermion lines. 
On the other hand, $\eta_{A,k}^\varphi$ receives contributions from 1PI diagrams with at least one internal boson line and, in our present truncation, 
it is found to be non-zero only in the regime with broken $U(1)$ symmetry. 

Finally, the flow of the renormalized effective order-parameter potential can be conveniently split up into three terms:
\begin{equation}
k\partial_k U_k = {\eta_{A,k}}\rho U'_k + \left[k\partial_k U_k\right]^\psi + \left[k\partial_k U_k\right]^\varphi\,.
\label{dtUStruc}
\end{equation}
The first term accounts for the renormalization of the field~$\bar{\varphi}$, the second term is simply a pure fermion loop, and the third
term is nothing but a pure boson loop. The last two are both dressed with suitable regulator insertions (see also Appendix~~\ref{app:FEq}).

Using the initial conditions detailed above and integrating the coupled system of flow equations given in Eqs.~\eqref{YukawaFlow}-\eqref{dtUStruc}, 
we can extract values of physical observables such as the {\it Bertsch} parameter, the fermion gap, or the
critical temperature. As in the mean-field case, a nontrivial global minimum at $U_{k=0}(\rho_0)$ signals spontaneous $U(1)$-symmetry breaking associated 
with a finite order {parameter $\rho_0=\varphi^{\ast}_0\varphi_0$ or, equivalently, a} finite fermion gap $\Delta_0^2 = h_\varphi^2\rho_0$.

Due to the highly nonlinear, coupled structure of the system of flow equations, the solutions need to be found numerically. As discussed 
in Sec.~\ref{sec:MFRes} above, a mean-field analysis suggests the existence of first-order phase transitions. Therefore, we choose to discretize 
the $k$-dependent effective potential on a grid in field space (see also Ref.~\cite{hImb3DFRG} for details on our numerical implementation).

Before discussing the numerical results from our RG flow equations, we briefly discuss more precisely in what sense 
Eqs.~\eqref{YukawaFlow}-\eqref{dtUStruc} represent an extension of the conventional 
mean-field approximation of Sec.~\ref{sec:MF}. Diagrammatically, Eq.~\eqref{WetterichEq} 
has a simple one-loop structure, where the internal lines 
of Feynman diagrams are given by the so-called \emph{full} propagators $\sim (\Gamma^{(2)}_k + R_k)^{-1}$. 
The mean-field approximation, on the other hand, takes into account Feynman diagrams only with internal fermion lines, but with no internal boson lines. 
In our RG approach, this implies that the flow equation for the wavefunction renormalization and for the effective potential simplify considerably:
\be
\eta_{A,k} = \eta_{A,k}^\psi\,,
\ee
and
\be
k\partial_k U_k = {\eta_{A,k}}\rho U'_k + \left[k\partial_k U_k\right]^\psi\,,
\ee
where, diagrammatically,
\begin{equation}
[\partial_k U_k]^\psi \sim {\sum_{n=0}^\infty} \FermiDotBox \cdot\left(\varphi^*\varphi\right)^n.
\label{dtUMFPart}
\end{equation}
Moreover, we observe that the flow equation for the effective potential reduces to
\be
k\partial_k U_k(\bar{\varphi}) = \left[k\partial_k U_k(\bar{\varphi})\right]^\psi\,,\label{eq:mfpotrg}
\ee
if we rewrite it in terms of the unrenormalized anologue of~$\rho$, 
namely~$\bar{\rho}=\rho/A_{\varphi,k}$. For the unrenormalized Yukawa coupling, we 
find $\partial_k \bar{h}_k=0$. Thus, in this limit the RG flow of the effective order-parameter potential~$U_k$ and wavefunction 
renormalization~$A_{\varphi,k}$ are no longer coupled, and so the flow of the Yukawa coupling is trivial. This implies that the flow equations of~$U_k$ and~$A_{\varphi,k}$
can be solved independently. Diagrammatically, we find that a fermion loop with two external bosonic lines attached to it
contributes not only to the flow of~$U_k$ (see Eq.~\eqref{dtUMFPart}), but also to the flow of~$A_{\varphi,k}$, i.e. to~$\eta_{A,k}$.\footnote{Note that~$A_{\varphi,k}$
directly contributes to the inverse boson propagator being the second functional derivative of the effective action with respect to the boson fields.}
The consequences of this observation will be discussed in detail below. At this point, we note that 
we do indeed recover the well-known mean-field solution for the effective potential at $Q=0$ from integrating Eq.~\eqref{eq:mfpotrg} with respect to the RG scale~$k$, as one should.
Here, the restriction on the case $Q=0$ comes from the need to project onto constant $\varphi$ in order to obtain explicit expressions for the flow 
equations~\eqref{etaAStruc} and~\eqref{dtUStruc}. 
Thus, at the level of our present approximations, our RG approach is unable to resolve inhomogeneous phases explicitly. However, 
our present setup is not ``blind" to the fact that inhomogeneous condensates may emerge and govern the ground-state dynamics, as we discuss next.
\begin{figure}[t]
\includegraphics[width = .48\textwidth]{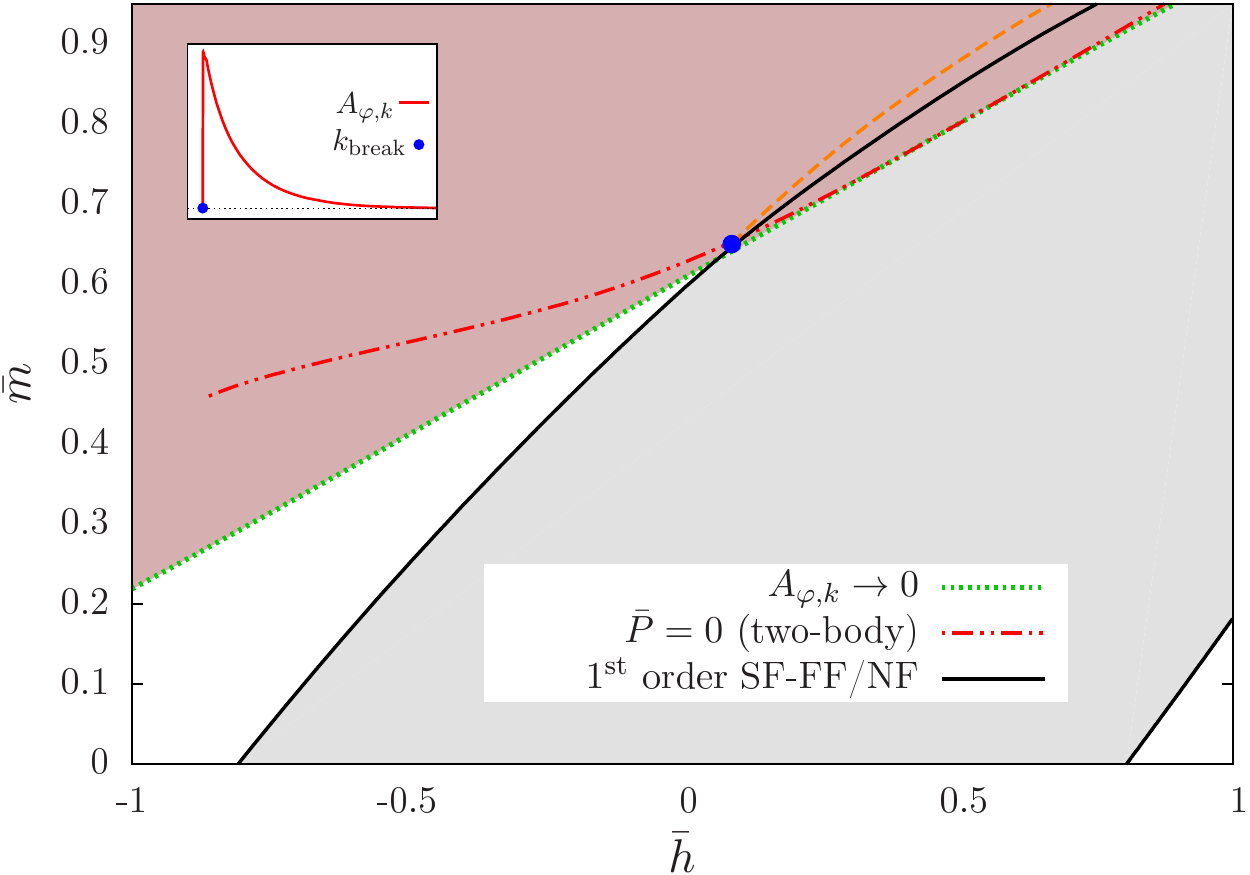}
\caption{Mean-field phase diagram at $T=0$ showing the region of vanishing $A_{\varphi,k_{\rm{break}}}$ (red/dark-shaded area) bounded by the green/dotted line.
For large~$\bar{m}\gtrsim 0.6$, this line associated with inhomogeneous pairing (see main text) agrees very well with the (red) dot-dot-dashed line 
depicting the lower bound for {bound-state formation} with $\bar{P} > 0$ 
as obtained from our two-body analysis, see Sec.~\ref{subsec:2} and Fig.~\ref{SchroeRes}.
{The orange/dashed line depicts the transition line from the inhomogeneous phase to the normal phase as obtained with a FF ansatz,
see also Figs.~\ref{MFDiag} and~\ref{MFInhoZoom}.}} 
\label{MFDiagBreak}
\end{figure} 
%

%
\subsection{Results}\label{sec:FRGRes}
%
%
\subsubsection{Flow of the fermion loop: mean-field and beyond}\label{subsec:FRGMF}
As already mentioned above, integrating Eq.~\eqref{dtUMFPart} (or rather Eq.~\eqref{dkUkFermi}) and 
minimizing $U_{k=0}(\Delta^2)$ yields the well-known mean-field phase diagram of mass- and spin-imbalanced Fermi gases 
for the case where the possibility of inhomogeneous phases was not taken into account explicitly
{(see e.g. Refs.~\cite{PhysRevLett.98.160402,WuPaoYip06,2010PhRvA..82a3624B,Braun:2014ewa}).}
However, as mentioned above, a fermion loop with two external bosonic legs contributes not only to the flow of $U_k$, but also to $\eta_{A,k}$, see Eq.~\eqref{etaAStruc}. 
Recall that the flow equations for~$U_k$ and~$\eta_{A,k}$ in the mean-field approximation are decoupled,
implying that the fermion gap $\Delta_0^2 = h_\varphi^2\rho_0 = \bar{h}_\varphi^2\bar{\rho_0}$ is independent of $A_{\varphi,k}$. 
Although $\eta_{A,k}$ has therefore no direct impact on the value of $\Delta_0^2$, the flow of~$\eta_{A,k}$ still contains important information about ground-state properties. 

Indeed, solving the flow equation for $A_{\varphi,k}$ alongside with the one for $U_k$, it turns out that the long-range limit ($k\to 0$) cannot always be reached 
due to a peculiar behavior of the RG flow. This is seen in Fig.~\ref{MFDiagBreak}, where we show a red/dark-shaded region bounded by a green/dotted line. Within that region, $A_{\varphi,k}$ becomes zero for a finite value~$k=k_{\rm{break}}$,
(see inset of Fig.~\ref{MFDiagBreak} for illustration). Strictly speaking, the RG flow breaks down at this point. Had we only considered the flow of~$U_k$, we would have
recovered the well-known mean-field phase diagram and easily overlooked this instability, which occurs at the same level of truncation in terms of Feynman diagrams.
On the other hand, this instability can be expected to be an artifact of our {ansatz~\eqref{GammaAnsatz}, which} yields Eq.~\eqref{BoseProp} for the boson propagator. 
The latter requires $A_{\varphi,k} > 0$ in order to be physically meaningful. If the {boson propagator 
evaluated at~$\varphi_0$ (or, equivalently, $\rho_0$) turns out to be not positive (semi-)definite, then}
the configuration~$\varphi_0$ extracted from the computation of~$U_k$ cannot be the true ground-state of the theory. Of course, this instability
does not really come as a surprise. In Sec.~\ref{sec:MF}, we have already shown that, for a simple FF-ansatz, a regime governed by an inhomogeneous condensate exists 
in the phase diagram. Within this regime, a homogeneous condensate does not represent the true ground state, but rather an ``excited" state in the spectrum of the theory. 
\begin{figure}[t]\
\includegraphics[width = .40\textwidth]{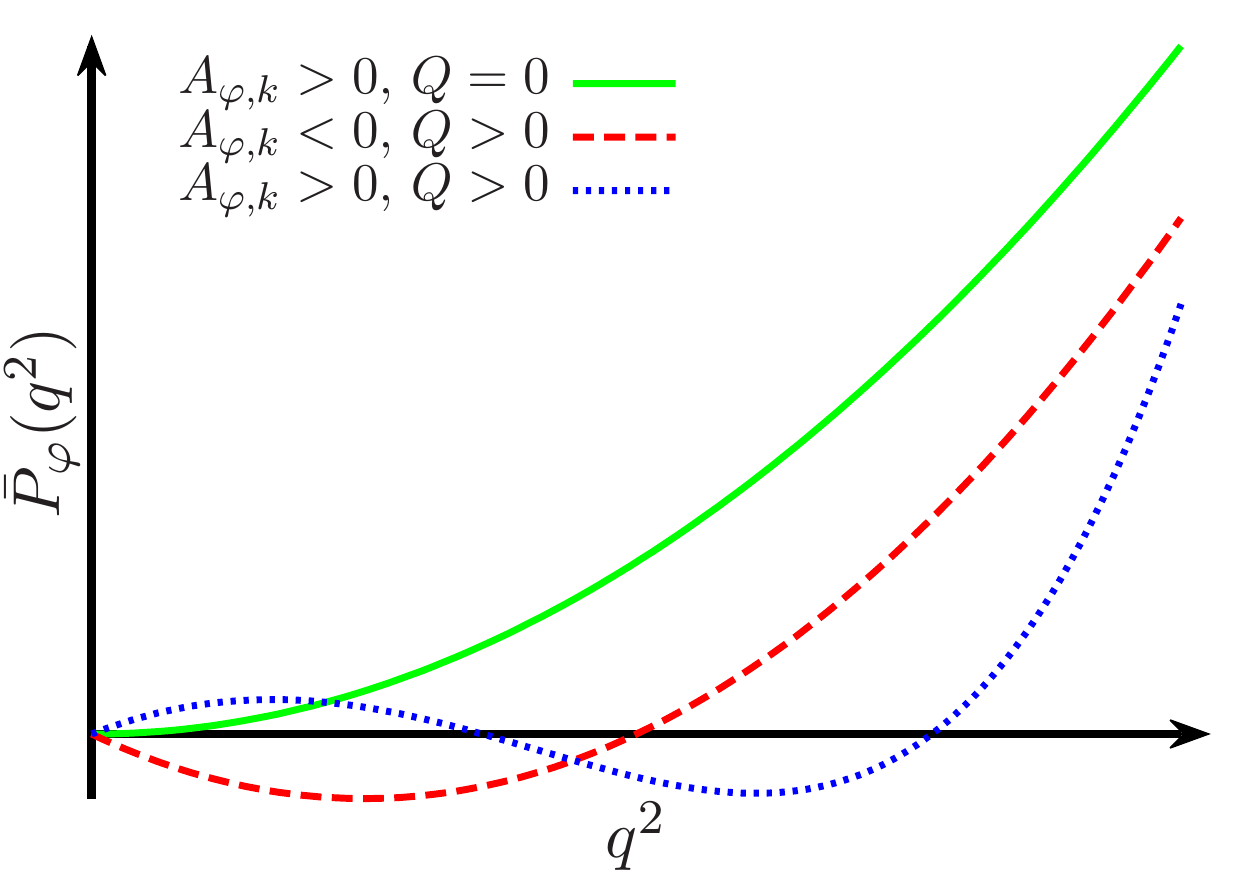}
\caption{Sketches of possible functional forms of the inverse bosonic propagator in terms of spatial momentum. The green/solid 
line corresponds to a homogeneous phase, whereas the red/dashed and blue/dotted lines sketch two cases with
a nontrivial minimum at $q_{\rm{min}} \sim {\mathcal O}(Q) > 0$ favoring the dominance of modes associated with bound states with a finite center-of-mass
momentum. In case of the red/dashed curve, this is linked to a negative $A_{\varphi,k}$.}
\label{PropGuess}
\end{figure}

In our present RG study, we do not employ a specific ansatz to parametrize the spatial dependence of the condensate, such as an FF ansatz. Rather, we study the stability
of the RG flow under the assumption that the condensate does not exhibit spatial dependence.
The momentum structure in Eq.~\eqref{BoseProp} can be seen as the first nontrivial term in a derivative expansion of the exact propagator {about zero momentum}.
The fundamental assumption for this ansatz to be valid is that bosonic modes with spatial momentum $|\vec{q}| \approx 0$ contribute predominantly to the flow. 
For an inhomogeneous phase this is obviously not the case, as modes with momenta around $|\vec{Q}|$ are rather expected to dominate the dynamics. 
The inverse boson propagator close to $k = k_{\rm{break}}$ is therefore expected to be shaped as the (red) dashed line in Fig.~\ref{PropGuess} instead of the 
green (solid) curve which corresponds to Eq.~\eqref{BoseProp} with $A_{\varphi,k} >0$ (see also Ref.~\cite{KrahlFriedWett_Hubbard09}). 
It might therefore be tempting to associate the existence of a finite scale $k_{\rm{break}}$ with the formation of an inhomogeneous condensate directly. 
For the following reasons, however, a sign change of~$A_{\varphi,k}$ should only be considered as a ``hint'' for the emergence of an inhomogeneous condensate:
\begin{itemize}
\item Only for $k\rightarrow 0$ (long-range limit), strictly speaking, results extracted from the RG flow are physically observable. Here, 
we find~$A_{\varphi,k} \rightarrow 0$ at $k_{\rm{break}} \gtrsim \sqrt{\mu}$. Now it may be the case that~$A_{\varphi,k} \rightarrow 0$ is negative only
for a finite $k$-range below~$k_{\rm{break}}$ and then becomes positive again in the limit~$k\to 0$. This 
behavior of~$A_{\varphi,k}$ may be considered as indicating the existence {of an ``inhomogeneous pre-condensation" phenomenon, in} analogy to the well-known
precondensation effect at finite temperature~\cite{PhysRevLett.71.3202,Randeria19981754} and finite spin imbalance~\cite{hImb3DFRG}.
As discussed in Ref.~\cite{KrahlFriedWett_Hubbard09}, it is necessary to employ a modified ansatz for the momentum dependence of the inverse boson propagator for 
scales $k\le k_{\rm{break}}$ in order to reach the long-range limit, $k\to 0$. Otherwise, it is {\it a priori} unclear whether we indeed have $A_{\varphi,k} < 0$ for all $k<k_{\rm{break}}$, 
i.e. if the dominance of modes favoring an inhomogeneous ground state persists in the deep IR. An investigation of this issue is beyond the scope of the present work.
\item The agreement between the boundaries derived from our two-body study on the one hand, and the criterion associated with the appearance of a zero in the flow 
of~$A_{\varphi,k}$ on the other, is stunningly good for $\bar{m} \gtrsim 0.6$ (see Fig.~\ref{MFDiagBreak}). It can already be deduced from Fig.~\ref{SchroeRes} 
that inhomogeneous pairing is particularly preferred for such highly mass-imbalanced systems. In fact, our conventional 
mean-field study with a FF ansatz already suggested the existence of an inhomogeneous phase in this regime (see the domain 
enclosed by the orange/dashed line and the black line in Fig.~\ref{MFDiagBreak}).
However, we also observe that
bound states with finite center-of-mass
momenta as well as the regime with $k_{\rm{break}}>0$ are found to exist well beyond this FF-type phase.
This discrepancy may be traced back to the fact that a simple FF ansatz may be insufficient in some parts of the phase diagram. However,
this can also be viewed as a manifestation of the existence of bosonic bound states (or the dominance of finite-momentum bosonic modes), which does not 
necessarily imply condensate formation, even at $T=0$. 
\item For $\bar{h} \lesssim 0$, the {domain bounded by} the criterion of~$A_{\varphi,k} \rightarrow 0$ extends down to $\bar{m} \approx 0.22$ for $\bar{h} = -1$, whereas no 
bound states with finite center-of-mass momentum are found below $\bar{m} = 0.46$ and $\bar{h} = -0.86$, see Figs.~\ref{SchroeRes} and~\ref{MFDiagBreak}. The discrepancy 
between these two lines should not be too surprising for small~$\bar{h}$. In fact, the behavior of $A_{\varphi,k}$ is only indirectly linked to the appearance 
of bound states. It is reasonable to expect that the momentum dependence, i.e. the functional form, of the propagator is dominated by
the existence of deeply bound two-body states as they are present for~$\bar{m}\gtrsim 0.6$, where the two lines are in very good agreement. In this regime, condensation
of these states may indeed be energetically favored and the emergence of an instability in the RG flow associated with the 
observation~$A_{\varphi,k} \rightarrow 0$ may then be viewed as a strong indication that the formation of an inhomogeneous condensate is most favorable.
{However, once the binding energy becomes smaller, the functional form} of the propagator becomes progressively more 
dominated by modes with momenta
significantly different from the center-of-mass momentum of the lowest lying two-body bound state for given values of~$\bar{h}$ and~$\bar{m}$. 
Even if there is no bound state at all in our analysis of the two-body problem, {there is no reason why bosonic fluctuations in general, and} those with finite momentum in particular, 
should become irrelevant. Consequently, in domains with only shallowly bound two-body states, as it is the case for decreasing~$\bar{h}$, 
we may expect at best qualitative agreement between our predictions from the two-body problem and the study of the RG flow of~$A_{\varphi,k}$.
\end{itemize}
In summary, $A_{\varphi,k} \rightarrow 0$ signals the dominance of finite-momentum bosonic fluctuations at least for a certain range of values of the RG scale $k\leq k_{\rm{break}}$. 
Strictly speaking, however, even positive {values of~$A_{\varphi,k\to 0}$ do not imply 
that a low-order derivative expansion} of the effective action captures the correct dynamics with respect to 
inhomogeneous phases. In fact, the true inverse propagator may, e.g., be shaped as depicted by the blue/dotted line in Fig.~\ref{PropGuess}. In any case, it is also clear that 
it is not possible to discriminate irrevocably between normal, homogeneous and inhomogeneous phases only
with the aid of the criterion~$A_{\varphi,k}\rightarrow 0$. If, on the other hand, an existing homogeneous condensate (as found in a study of the RG flow of the potential~$U_k$) 
overlaps with a domain where $A_{\varphi,k} \rightarrow 0$ is also observed in the flow, this is a strong indication of an inhomogeneous phase in that same domain. 
This point will be detailed further in the next section, where we present results from the RG flow including bosonic fluctuations.

%
\subsubsection{Full Flow Equations: Impact of Bosonic Fluctuations}\label{subsec:FullFRG}
Let us now study the effect of the bosonic contributions $\eta_{A,k}^\varphi$ and $[\partial_k U_k]^\varphi$ in our set of flow equations.
Due to the derivatives of $U_k$ with respect to~$\rho$ in the flow equations for~$A_{\varphi,k}$ and~$U_k$, and the nontrivial dependence of
these equations on $h_\varphi$ (see Eqs.~\eqref{dkUkBose} and~\eqref{etaABos}), the flow equations~\eqref{YukawaFlow}-\eqref{dtUStruc} are now coupled in a highly 
nontrivial way. As discussed in Sec.~\ref{sec:FRGForm}, we solve this set of flow equations numerically. However, it does not appear feasible to either start the integration 
at $k=\Lambda\rightarrow\infty$, nor to reach the IR limit $k \rightarrow 0$ exactly. 
In the IR regime, we have stopped the RG flow at~$k_{\rm IR} \approx 10^{-3}\sqrt{\mu}$ in those domains of parameter space that are close to a second-order phase transition. 
The remaining uncertainty in the critical temperature, as measured by the fermion gap~$\Delta_k$, turned out to be well below the percent level. In domains close to a first-order transition, 
the introduction of a finite IR cutoff~$k_{\rm IR}$ is even less severe as the order parameter~$\Delta_k$ is found to jump typically on scales $k\sim\mathcal{O}(\sqrt{\bar{\mu}}/10)$
(see also Ref.~\cite{hImb3DFRG} for a detailed discussion of the numerical implementation in the mass-balanced limit).
\begin{figure}[t]
\includegraphics[width = .48\textwidth]{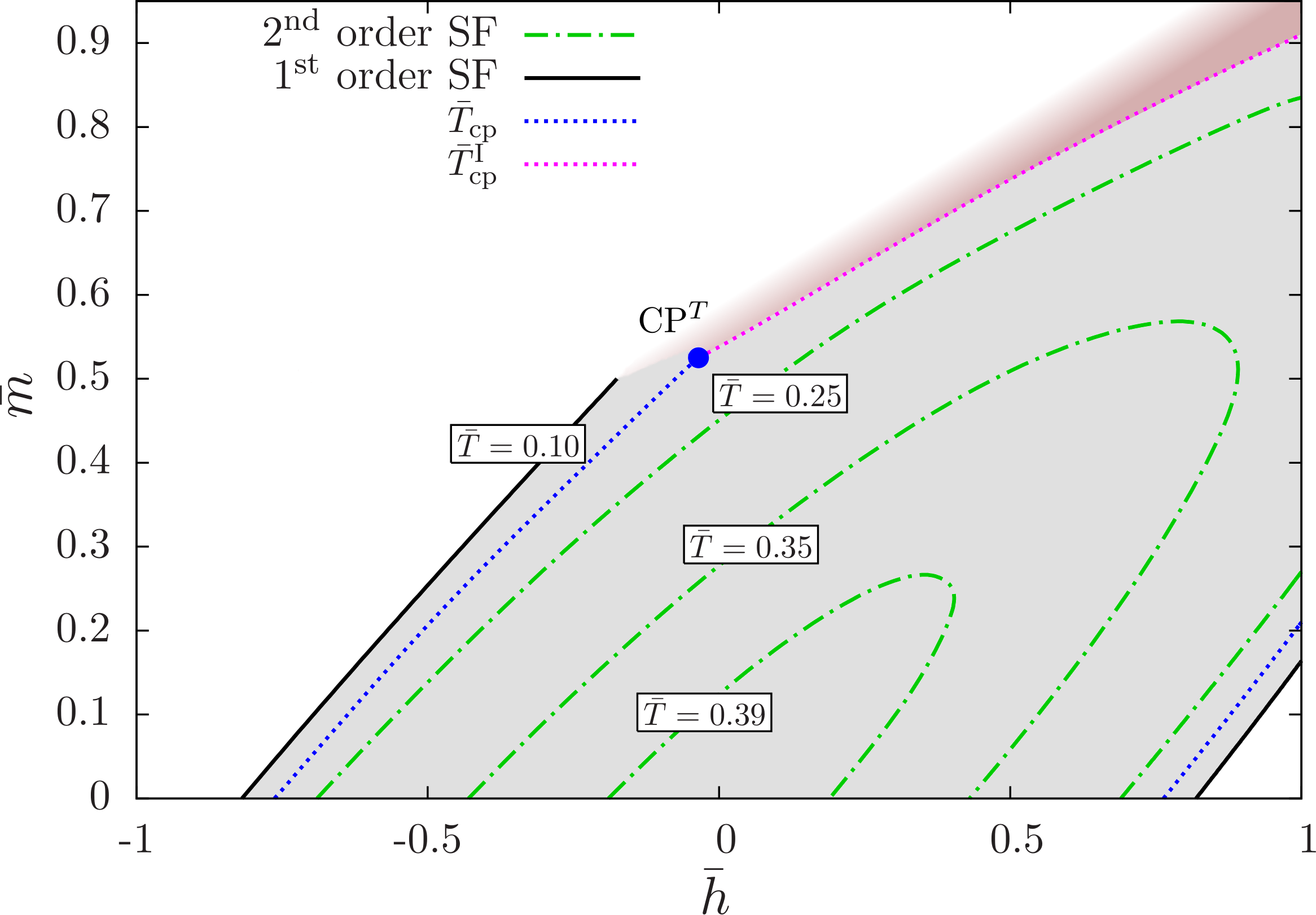}
\caption{Phase diagram obtained from the full set of flow equations~\eqref{YukawaFlow}-\eqref{dtUStruc} {including bosonic fluctuations}.
The red/dark-shaded area marks the domain where an inhomogeneous phase is likely to exist, see main text {for a detailed discussion.}}
\label{FRGDiag}
\end{figure}

The phase diagram from our RG study including order-parameter fluctuations is shown in Fig.~\ref{FRGDiag}.
We first note that the qualitative features of the phase diagram are similar to the mean-field phase diagram discussed above. However, quantitative corrections are found to be sizable. 
For~$\bar{m}=\bar{h}=0$, for example, we find that the critical temperature~$\bar{T}$ is significantly changed from~$\bar{T}_{\rm c}\approx 0.66$ in the mean-field approximation 
to~$\bar{T}_{\rm c}\approx 0.40$, in good agreement with recent experiments~\cite{Ku_etal12}, Monte Carlo studies~\cite{Bulgac:2005pj}, 
as well as with recent RG studies~\cite{PhysRevA.89.053630,hImb3DFRG}. 

Lowering~$\bar{h}$, starting from~$\bar{h}=\bar{m}$ for fixed~$\bar{m} \lesssim 0.53$,
the order of the phase transition changes from second to first along a critical line $\bar{T}_{\rm{c}}(\bar{h},\bar{m})$. 
In this $\bar{m}$-regime, we find $T_{\rm{cp}}(\bar{h},\bar{m}) \approx 0.19 ... 0.20$ for
the critical point where the nature of the transition changes from second to first order.
For temperatures below $\bar{T}^{}_{\rm{cp}}(\bar{h},\bar{m})$ and given $\bar{m}$, the superfluid phase is enlarged in the $\bar{h}$-direction compared to the mean-field result, as 
also found in the mass-balanced case~\cite{hImb3DFRG}. For $\bar{T} \ll \bar{T}_{\rm{c}}$ the integration of Eqs.~\eqref{YukawaFlow}-\eqref{dtUStruc} becomes numerically more and more involved, 
as the right-hand sides become discontinuous at $\bar{T} = 0$ (see also Eqs.~\eqref{dkUkFermi} and~\eqref{NFT0Limit}). Contrary to the mean-field case, 
{we therefore do not present} 
results for the strict zero-temperature limit here but restrict ourselves to temperatures $\bar{T} \geq 0.1$. However, we expect the position of the phase transition lines in the zero-temperature limit to still be in reasonable agreement with the ones obtained for~$\bar{T}=0.1$.

Lowering~$\bar{h}$, starting from~$\bar{h}=\bar{m}$ for fixed $\bar{m} \gtrsim 0.53$ and $\bar{T} < T_{\rm{cp}}^{\rm I}(\bar{m})$ 
with $T_{\rm{cp}}^{\rm I}(\bar{m}) \approx 0.19 ... 0.21$, 
we find that a critical value $\bar{h}_{\rm{break}}(\bar{m},\bar{T})$ exists at which~$A_{\varphi,k}$
tends to zero, potentially indicating a transition from a homogeneous superfluid phase to an inhomogeneous phase. 
In fact, we find that~$A_{\varphi,k}$ tends to zero at a {\it finite} RG scale~$k_{\text{break}}$ for all $\bar{h} \in [-1,\bar{h}_{\rm{break}}(\bar{m},\bar{T})]$.
As discussed in Sec.~\ref{subsec:FRGMF}, this does not necessarily 
imply that there exists an inhomogeneous phase for the whole domain~$\bar{h} \in [-1,\bar{h}_{\rm{break}}(\bar{m},\bar{T})]$ for a given~$\bar{m}$ and~$\bar{T}$. 
Therefore the red/dark-shaded area in Fig.~\ref{FRGDiag} only represents a domain where an inhomogeneous phase is most likely to be found. 
Our mean-field study supports this interpretation: In Fig.~\ref{MFDiagBreak}, we find reasonable agreement for large~$\bar{m}$ between the 
onset of inhomogeneous condensation and the line
defined by the largest value of~$\bar{h}$ for which~$A_{\varphi,k}$ still tends to zero at a finite scale~$k_{\text{break}}$, for a given value of~$\bar{m}$.
For this reason, we consider our $A_{\varphi,k}$-criterion to provide a reasonable estimate for the boundary separating the phase with a homogeneous condensate 
from the inhomogeneous phase, even beyond the mean-field limit.
However, within our present setting, it is not possible to detect the transition from the 
inhomogeneous phase to the normal phase. The associated phase-transition line is therefore not shown in Fig.~\ref{FRGDiag}. 
Instead, we indicate the presumed existence of this phase-transition line by a fading out of the red/dark-shaded 
domain.\footnote{Of course, {\it a priori}, it is unclear whether a transition from an inhomogeneous to a normal phase 
exists after all. However, our mean-field study in Sec.~\ref{sec:MF}, our analysis of the two-body problem
in Sec.~\ref{subsec:2}, and studies of the {Fermi polaron~\cite{Chevy:2006, Chevy, CRLC, Lobo:2006, Bulgac:2007, ProkSvist07, KBS,Schmidt:2011zu}}
suggest that such a transition indeed exists.} 

\begin{figure}[t]
\includegraphics[width = .48\textwidth]{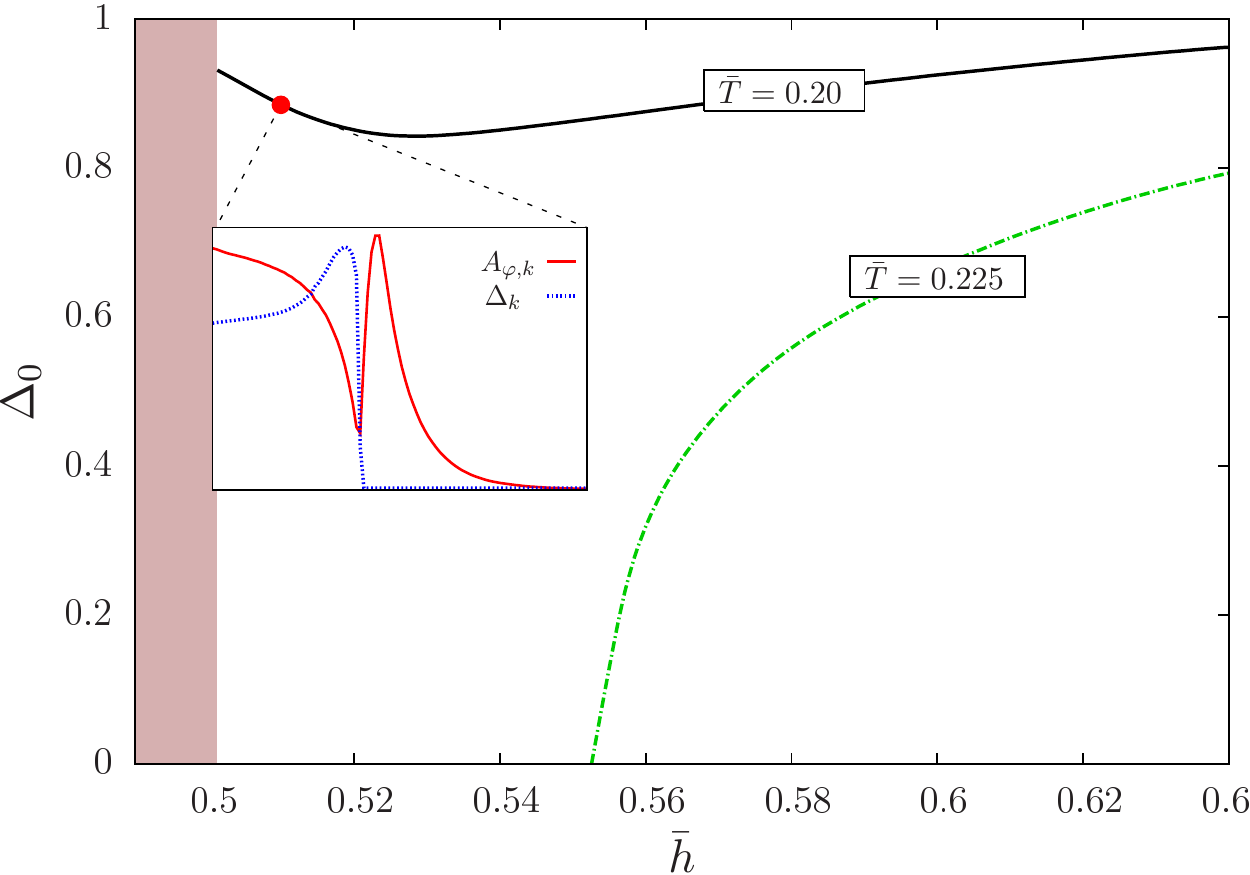}
\caption{Fermion gap for two different values of~$\bar{T}$ and fixed $\bar{m} = 0.74$ (corresponding to a Li-K mixture) 
as a function of $\bar{h}$. The red/dark-shaded area indicates the regime where $A_{\varphi,k}\rightarrow 0$ for $\bar{T}= 0.20$. 
The inset shows the $k$-evolution of $A_{\varphi,k}$ and $\Delta_k$ for $(\bar{T},\bar{h}) = (0.20,0.51)$.}	
\label{DeltaBend}
\end{figure}

In our mean-field studies, conventional and RG, we have considered the zero-temperature limit explicitly. This allowed us to compare directly the
lower bound for pairing with finite center-of-mass momentum and the bound from our $A_{\varphi,k}$-criterion.
Since the strict zero-temperature limit is difficult to reach from a numerical point of view when bosonic fluctuations are taken into account, we have restricted ourselves 
to temperatures~$\bar{T}\geq 0.1$. However, this renders a direct comparison with the results from our two-body analysis impossible.
Moreover, taking into account bosonic fluctuations, our $A_{\varphi,k}$-criterion used to detect the onset of inhomogeneous phases has to be taken {with some care:
In} Fig.~\ref{DeltaBend}, the fermion gap $\Delta_0$ is depicted as a function of $\bar{h}$ for two different temperatures and fixed $\bar{m} = 0.74$ corresponding 
to a mixture of ${}^6$Li and ${}^{40}$K.
For $\bar{T} = 0.225$, the green/dot-dashed line exhibits the typical behavior expected for a second order transition to the normal phase. 
Only slightly below this temperature, at $\bar{T} = 0.20$, 
the behavior of the condensate is now found to be quite different. Already before reaching the regime defined by $\bar{h}\leq \bar{h}_{\rm{break}}$, 
where~$A_{\varphi,k}$ tends to zero in the RG flow (indicated by the red/dark-shaded area), 
the gap $\Delta_0$ is {found to increase rather than decrease.} Such a behavior is not seen in the mean-field approximation. In fact, for any finite $\bar{T}$ and fixed $\bar{m}$, 
we find that $\Delta_0$ is a strictly concave function.
One may speculate that the observed increase in the gap is due, e.g., to numerical artifacts. Indeed, the value of~$\Delta_0$ is more sensitive to numerical inaccuracies 
in this region than anywhere else in the phase diagram and should therefore be taken with some care. {However, we have checked that 
the general observation of an increasing gap cannot be traced back to numerical instabilities.}

The inset of Fig.~\ref{DeltaBend} hints at the true reason for the unexpected behavior of the gap as a function of~$\bar{h}$ for given~$\bar{m}$ at low temperatures:
It shows the $k$-evolution of $A_{\varphi,k}$ (red/solid line) and $\Delta_k$ (blue/dotted line) for a point in a region of the phase diagram where the gap increases again 
but $A_{\varphi,k}$ is still positive on all scales~$k$. In this region (as exemplified by the inset), we observe that $A_{\varphi,k}$ undergoes a sharp decrease. From 
the dependence of the RG flow of the Yukawa coupling on~$A_{\varphi,k}$ via the anomalous dimension~$\eta_{A,k}$ (see Eq.~\eqref{YukawaFlow}),
it is clear that $h_\varphi$ experiences a sharp increase when~$A_{\varphi,k}$ decreases. {The sharp increase of the Yukawa coupling also
increases the gap up to values larger than those in which} no (strong) decrease of~$A_{\varphi,k}$ is present at all (e.g. deep in the 
homogeneous superfluid phase). In the inset of~Fig.~\ref{DeltaBend}, we also observe that the increase of the gap induced by the decrease of~$A_{\varphi,k}$ can
potentially counterbalance the decrease of~$A_{\varphi,k}$. This effect is intimately related to the presence of bosonic fluctuations. In fact, the strong increase of the gap
leads to a suppression of Feynman diagrams with internal fermion lines. In particular, purely fermionic diagrams present in mean-field studies are parametrically suppressed
by a (strong) increase of the gap. Thus, once a gap is generated, the RG flows of the potential, the Yukawa coupling, and the wavefunction renormalization~$A_{\varphi,k}$
are mainly driven by purely bosonic diagrams (without any internal fermion lines). As (purely) bosonic and fermionic contributions come in general with opposite signs, bosonic 
fluctuations tend to counterbalance the decrease of~$A_{\varphi,k}$ as well as the increase of the gap, which is induced by purely fermionic diagrams
(see also Appendix~\ref{app:FEq} for the RG flow equations of the various quantities).
Since the class of purely fermionic diagrams is the only one present in the mean-field approximation, 
this explains why such a counterbalancing effect is not observed in our mean-field study.

The observed decrease in~$A_{\varphi,k}$ can {at least partially} 
be traced back to the fact that a change of the spin 
imbalance parameter~$\bar{h}$ induces a mismatch in the Fermi momenta and Fermi energies
associated with the spin-up and spin-down fermions. 
Within our RG framework this implies (loosely speaking) that the two spin components contribute to the RG flow over different
ranges of scales~$k$.
In any case, the decrease in~$A_{\varphi,k}$ actually becomes stronger when we decrease~$\bar{h}$ for a given~$\bar{m}$
starting from the line of equal Fermi momenta~$\bar{h}=\bar{m}$. Eventually, we observe that the mismatch between the Fermi momenta becomes so large that the 
associated decrease of~$A_{\varphi,k}$ can no longer be compensated by the presence of bosonic fluctuations, resulting in $A_{\varphi,k} \to 0$ 
for~$\bar{h}\leq \bar{h}_{\rm{break}}(\bar{m},\bar{T})$ at a given finite RG scale~$k_{\rm{break}}$, at least within our present truncation. 

We close this section with a word of caution regarding the increase of the gap towards smaller values of~$\bar{h}$ 
in some domains of the phase diagram (see Fig.~\ref{DeltaBend}).
At present, it is unclear whether this behavior is a physical effect and traces of it 
should be expected in future experimental studies, or whether it is ``cured" when
considering a suitable extension of our present truncation. For example, a resolution of the full momentum dependence of the inverse boson propagator~$P_{\varphi}$ might be
required to resolve this issue since, in the relevant domain, $P_{\varphi}$ may assume a form as exemplified by the blue/dotted line in Fig.~\ref{PropGuess}. However, a
detailed analysis of this issue is left to future work. Here, we restrict ourselves to conclude that our RG study already suggests that 
bosonic fluctuation effects are of great importance in studies of mass- and spin-imbalanced unitary Fermi gases. We have found that the extent of the 
various phases in parameter space (normal, homogeneous superfluid, inhomogeneous superfluid) can change significantly relative to mean-field studies. In particular,
our RG analysis indicates that the size of the inhomogeneous phase is extended to smaller values of~$\bar{m}$. More specifically, our mean-field studies suggest 
that~$\bar{m}\gtrsim 0.65$ ($m_{\downarrow}/m_{\uparrow} \gtrsim 4.7$) is required to form an inhomogeneous condensate. 
Recall that~$\bar{m}\approx 0.74$ ($m_{\downarrow}/m_{\uparrow} \gtrsim 6.7$) for a Li-K mixture. Taking bosonic fluctuations into account, we find that
inhomogeneous condensates may already appear for mass imbalances~$\bar{m}\gtrsim 0.53$ ($m_{\downarrow}/m_{\uparrow} \gtrsim 3.2$).

%
\section{Summary}\label{sec:sum}
We have studied the phase diagram of mass- and spin-imbalanced unitary Fermi gases in three 
dimensions, with an emphasis on the detection of inhomogeneous condensates. To this end, we have employed various approaches. 
Since the formation of two-body bound states can be viewed as a necessary condition for condensation, 
we have analyzed the two-body problem in the presence of (inert) Fermi spheres associated with
the two spin components, and computed the binding energy as well as the center-of-mass momentum of the lowest lying bound state. This
helped us guide and analyze our studies of the many-body phase diagram. Indeed, we have found that our two-body bound-state problem
allows us to understand many features of the many-body phase diagram, {such as the emergence of a domain} in parameter space governed
by a homogeneous condensate. Moreover, our study of two-body bound states suggests the existence of a region in the
many-body phase diagram where an inhomogeneous condensate is formed, see Fig.~\ref{SchroeRes}.

Our explicit studies of the many-body problem indeed confirm the qualitative picture of the phase diagram suggested by the
two-body problem. In particular, we have only found indications of inhomogeneous phases in those regimes of the many-body
phase diagram in which also the center-of-mass momentum of the lowest-lying two-body bound state is finite. The 
agreement of the phase boundary of the homogeneous superfluid and inhomogeneous phases with the lower bound for pairing with
finite center-of-mass momentum is only qualitative when we consider a Fulde-Ferrell ansatz for the inhomogeneity in our mean-field study. On the other hand,
the agreement of this lower bound with the instability line\footnote{Recall that, for a fixed temperature, this line is
defined by the largest value of~$\bar{h}$ for which~$A_{\varphi,k}$ still tends to zero at a finite scale~$k_{\text{break}}$ for a given value of~$\bar{m}$.}
associated with the homogeneous-inhomogeneous transition is stunning for large~$\bar{m}\gtrsim 0.5$ 
in our RG mean-field study, which does not rely on an explicit ansatz for the inhomogeneity. 
Taking bosonic fluctuations into account in our RG study, the latter two lines are still in accordance.\footnote{Note that a direct 
comparison of both lines is not possible at this stage since we restricted
ourselves to temperature~$\bar{T}\geq 0.1$ in our RG study including bosonic fluctuations. As discussed above, a study of even lower temperatures is 
numerically challenging and costly and is therefore left to future work.} For mass imbalances~$\bar{m}\lesssim 0.5$, the analysis of the two-body problem
still suggests the existence of a regime with bound states with finite center-of-mass momentum. Our many-body studies, however, suggest that 
inhomogeneous condensates are unlikely to be found in this regime. This can be traced back to the fact that only shallowly bound two-body states
exist there. {Moreover, it also indicates that macroscopic} condensation of such bound states is not energetically favored.

For large~$\bar{m}\gtrsim 0.5$, on the other hand, a regime with deeply bound two-body states with finite center-of-mass momentum can be identified, suggesting
that the formation of an inhomogeneous condensate is energetically favored. Indeed, we find that this regime overlaps with the regime
where homogeneous condensation is found in our many-body analysis whenever inhomogeneous pairing is not taken into account. Therefore,
the instability line found in our RG study including bosonic fluctuations may indeed be identified with the transition line from a superfluid phase with
a homogeneous condensate to one with an inhomogeneous condensate, at least for~$\bar{m}\gtrsim 0.5$.
Finally, for small mass imbalances~$\bar{m}\lesssim 0.25$ we 
do not find any indication of inhomogeneous phases, neither in our analysis of the two-body problem nor in our many-body studies.

In summary, our RG study beyond the mean-field limit suggests significant modifications to the mean-field extent of the various phases in parameter space.
For example, for the mass- and spin-balanced Fermi gas, we find that the critical temperature~$\bar{T}_c$ agrees well with state-of-the-art Monte Carlo studies,
as already found in recent RG studies~\cite{PhysRevA.89.053630,hImb3DFRG}. For small mass imbalances, moreover, we find that the phase with a homogeneous condensate
is extended to larger spin imbalances~$|\bar{h}|$ in agreement with a previous RG study of the mass-balanced case~\cite{hImb3DFRG}. For large mass imbalances, however,
we observe that the extent of the phase with a homogeneous condensate shrinks in the~$\bar{h}$ direction. Moreover, we find strong indications for the
emergence of an inhomogeneous phase in this regime. The extent of the latter phase in the~$\bar{h}$-direction actually increases for increasing mass imbalance.
{Our estimate for a lower bound on $\bar{m}$ for the occurrence of an inhomogeneous condensate is~$\bar{m}\approx 0.53$. 
This value} of~$\bar{m}$ corresponds to a mass ratio~$m_{\downarrow}/m_{\uparrow} \approx 3.2$ which is considerably lower than the mass ratio associated
with a Li-K mixture ($m_{\downarrow}/m_{\uparrow} \approx 6.7$). Since three-body effects are expected to be significant for mass ratios associated with the Li-K mixture and beyond 
(see, e.g., Refs.~\cite{Efimov1,Efimov2,NRBH}), future experimental searches for, e.g., signals of inhomogeneous 
condensation in the regime~$m_{\downarrow}/m_{\uparrow} \gtrsim 6.7$ are highly challenging. 
In this respect, our estimate for the lower $\bar{m}$-bound for the occurrence of inhomogeneous phases may help to guide future experiments. Our
results suggest the existence of a much larger regime in parameter space where inhomogeneous condensation may be detected with a new variety of mixtures
of fermion species other than Li-K, but without suffering from three-body effects.


\acknowledgments
The authors thank I.~Boettcher, H.-W. Hammer, T.~K.~Herbst, and P. Niemann for helpful discussions.
J.B. and D.R. acknowledge support by the DFG under Grant BR 4005/2-1. Moreover, the authors acknowledge support
by HIC for FAIR within the LOEWE program of the State of Hesse. 
This work was supported in part by the U.S. National Science Foundation under Grant No. PHY1306520.

%
\appendix
%

%
\section{Regulator Functions}\label{sec:regfct}
Apart from the very defining properties of a regulator function~$R_k$ in the fRG formalism (see Ref.~\cite{Wetterich93}), there is considerable freedom in the choice concerning the precise shape of $R_k$ which
even allows to optimize the RG flows, see, {e.g., Refs.~\cite{Litim:2000ci,*Litim:2001fd,*Litim:2001up,Pawlowski20072831}. Guided} by these powerful principles of optimization, 
we employ the following set of functions for the fermion and boson fields, respectively:
\begin{equation}
R_k^\psi(z) = k^2\left(\mbox{sign}(z)-z\right) \theta\left(1-|z|\right),\,\, z = \frac{\vec{q}^{\,2} - \mu}{k^2}\,,
\label{FermiReg}
\end{equation}
{and
\begin{equation}
R_k^{\varphi}(y) = A_{\varphi,k}k^2(1-y) \theta(1-y),\,\, y = \frac{\vec{q}^{\,2}}{2k^2}\left(1-\bar{m}^2\right)\,.
\label{BoseReg}
\end{equation}
Note} that in both cases only spatial momenta are regularized which allows us to perform the Matusbara sums in the loop integrals analytically.

The bosonic regulator function~$R_k^{\varphi}$ is the same as in previous studies~\cite{Diehl:2007th,*Diehl:2007ri}. 
Only the argument has been modified by a factor~$(1-\bar{m}^2)$ to accommodate for the generalized dispersion relation
in the mass-imbalanced case.

Intuitively, the two fermion species should be regulated separately, since their dispersion relations differ 
by construction in the imbalanced case. This appears to be particularly important since the presence of Fermi seas in general requires fluctuations below and 
above the respective Fermi levels to be treated differently. However, the analytic computation of the frequency sum reveals that in the critical cases, only an ``averaged" Fermion dispersion appears in the RG flow equations. 
Thus, the ``averaged" regulator~\eqref{FermiReg} is sufficient to render the integration over spatial momenta finite. The structure of the flow equations is thereby greatly simplified. In contrast to the 
mass-balanced but spin-imbalanced case~\cite{hImb3DFRG}, to our knowledge, an analytical computation of the momentum integrals is not possible. 
We add that IR observables from a RG flow controlled by the regulator 
function~\eqref{FermiReg} on the one hand and by separately regulated fermion propagators on the other hand has found to agree extremely well in the mass-balanced case, 
see Ref.~\cite{hImb3DFRG}. We therefore expect that our choice~\eqref{FermiReg} can safely be employed in the mass-imbalanced case as well.

%
\section{Flow Equations}\label{app:FEq}
In this appendix we provide explicit expressions for the flow equations of the effective potential and the boson anomalous dimension $\eta_{A,k}$. 
Since they can be derived largely along the lines of the balanced or spin-imbalanced cases, we focus mainly on the peculiarities of the mass-imbalanced case. 
For further details, we refer the reader to earlier work, see, e.g., Refs.~\cite{Diehl:2007th,*Diehl:2007ri,Scherer:2010sv,Boettcher:2012cm,Diehl:2009ma,hImb3DFRG}.

\subsection{Flow of the effective potential}
The flow equation of the order-parameter potential $U_k(\rho)$ is obtained from
\be
\partial_k U_k\!=\!\frac{1}{2}{\text{STr}} \big[({\partial_k R_k})\!\cdot\!({\Gamma_k^{(2)} \!+\! R_k})^{-1}\big]_{\psi^* = \psi = \varphi_2 = 0,\varphi_1 =\sqrt{2\rho}}\,.\nn
\label{UkProj}
\ee
Note that, for convenience, we have written the complex boson field $\varphi$ as a {sum of its real and imaginary parts,}
$\varphi(\tau,\vec{x}) = \frac{1}{\sqrt{2}}\left[\varphi_1(\tau,\vec{x})+ i\varphi_2(\tau,\vec{x})\right]$. Moreover, we add that,
due to the the fact that $U_k$ only depends on~$\rho = \varphi^*\varphi$, the projection on $\varphi_1 =2\sqrt{\rho}$ and $\varphi_2 = 0$ can be applied without loss of generality. 
Eventually, this yields
\be
\partial_k U_k = \frac{1}{k}\eta_{A,k}\rho U'_k+ [\partial_k U_k]^\psi + [\partial_k U_k]^\varphi\,
\ee
with
\begin{widetext}
\begin{equation}
\left[\partial_k U_k\right]^\psi = \frac{k^4}{2\pi^2}\int_{\max[-\tilde{\mu},-1]}^1\dif{\tilde{z}}\frac{\sqrt{\tilde{z} +\tilde{\mu}}}{\sqrt{1+w_3}}\sum_{\sigma=\pm 1}\sigma N_{\rm{F}}\left(\bar{m}(\tilde{z}+\tilde{\mu}) - \tilde{h} + \sigma\sqrt{1+w_3}\right) ,
\label{dkUkFermi}
\end{equation}
\begin{equation}
\left[\partial_k U_k\right]^\varphi = {\frac{\sqrt{2}k^4}{3\pi^2}}\frac{1}{\left(1-\bar{m}^2\right)^{\frac{3}{2}}}\left(1-\frac{\eta_A}{5}\right)\left[\sqrt{\frac{1+w_1}{1+w_2}} +\sqrt{\frac{1+w_2}{1+w_1}}\right]\left[\frac{1}{2} + N_{\rm{B}}\left(\sqrt{1+w_1}\sqrt{1+w_2}\right)\right].
\label{dkUkBose}
\end{equation}
\end{widetext}
Here, $N_{\rm{F}}$ and $N_{\rm{B}}$ correspond to the Fermi and Bose-Einstein distribution functions, respectively:
\begin{equation}
N_{\rm{F}}(x) = \frac{1}{e^{x/\tilde{T}}+1},\quad N_{\rm{B}} = \frac{1}{e^{x/\tilde{T}} - 1}\,.
\label{Thermfunc}
\end{equation}
Furthermore, the following quantities have been introduced:
\begin{equation}
w_1 = \frac{U'_k}{k^2},\quad w_2 = \frac{U'_k + 2\rho U''_k}{k^2},\quad w_3 = \frac{h_\varphi^2\rho}{k^4}\,.
\label{wDef}
\end{equation}
Quantities~$\mathcal{Q}$ divided by a factor~$k^2$ are written as~$\tilde{\mathcal{Q}}$, e.g. $\tilde{\mu} =\mu/k^2$.

Since $\Delta_k = h_\varphi^2\rho = k^4 w_3$, Eq.~\eqref{dkUkFermi} does not depend solely on $\rho$. This allows us to identify 
the solution from standard mean field theory with the solution of the purely fermionic flow of $U_k$ in the limit~$k\to 0$, 
as discussed at the end of sect.~\ref{sec:FRGForm}. Furthermore, we note that
\begin{equation}
\lim_{T\rightarrow 0}N_{\rm{F}}(x) = \theta(-x)\,.
\label{NFT0Limit}
\end{equation}
Thus, the right-hand side of Eq.~\eqref{dkUkFermi} becomes non-analytic in the zero-temperature limit. 
While the flow of $U_k$ in principle exists for arbitrarily small but finite $T$, 
the ``steepness" of the Fermi function makes the numerical treatment increasingly challenging as $T$ is lowered.

\subsection{Boson anomalous dimension}
In order to extract the running of the wavefunction renormalization parameter $A_{\varphi,k}$ via the boson anomalous 
dimension $\eta_{A,k} = -k\partial_k \ln A_{\varphi,k}$, the following projection rule is employed:
\begin{equation}
\begin{aligned}
\eta_{A,k} &= -\frac{k}{A_{\varphi,k}}\frac{2}{1-\bar{m}^2}\pdif{}{\vec{q}^{\,2}}\partial_k\left[\left(\bar{P}_{\varphi}\right)_{12}(0,\vec{q})\right]_{\vec{q}=0}\\
&\equiv \eta_{A,k}^{\psi,1} + \eta_{A,k}^{\psi,2} + \eta_{A,k}^{\varphi}\,,
\end{aligned}
\label{etaAProj}
\end{equation}
where
\begin{equation}
\begin{aligned}
&\left(\bar{P}_\varphi\right)_{12}(q_0,\vec{q})\delta^{(4)}(p-q)\\
= &\lfdif{}{\bar{\varphi}_1(-p)}\Gamma_k \rfdif{}{\bar{\varphi}_2(q)}\Bigg|_{\psi^*=\psi=\varphi_2 = 0,\varphi_1=\sqrt{2\rho_{0,k}}}\,.
\end{aligned}
\label{DAProp}
\end{equation}
Note that we do not consider a $\rho$-dependent $\eta_{A,k}$. Instead, $\eta_{A,k}$ is projected onto the $k$-dependent minimum $\rho_{0,k}$ of the effective potential. 
The explicit expressions for the different contributions to $\eta_{A,k}$ are given by:
\begin{widetext}
\begin{equation}
\begin{aligned}
\eta_{A,k}^{\psi,1} = \frac{1}{1-\bar{m}^2}&\frac{h_\varphi^2}{6\pi^2 k(1+w_3)^{\frac{3}{2}}} \sum_{\sigma,\kappa = \pm 1}\left(\tilde{\mu} + \kappa\right)^{\frac{3}{2}}\theta\left(\tilde{\mu} + \kappa\right)\\
&{\left[\sigma N_{\rm{F}}\left(\bar{m}(\tilde{\mu}+\kappa) - \tilde{h} -\sigma\sqrt{1+w_3}\right) + \sqrt{1+w_3}N'_{\rm{F}}\left(\bar{m}(\tilde{\mu}+\kappa) - \tilde{h} -\sigma\sqrt{1+w_3}\right)\right]\,,}
\end{aligned}
\label{etaAFerm1}
\end{equation}
\begin{equation}
\begin{aligned}
\eta_{A,k}^{\psi,2} =& -\frac{h_\varphi^2\bar{m}^2}{6\pi^2 k(1-\bar{m}^2)}\int_{\max[-\tilde{\mu},-1]}^1\dif{\tilde{z}}\frac{(\tilde{z} + \tilde{\mu})^{\frac{3}{2}}}{(1+w_3)^{\frac{5}{2}}}\sum_{\sigma =\pm 1}\left[3\sigma N_{\rm{F}}\left(\bar{m}(\tilde{z} + \tilde{\mu})-\tilde{h}+\sigma\sqrt{1+w_3}\right)\right.\\
&\left. - 3\sqrt{1+w_3}N'_{\rm{F}}\left(\bar{m}(\tilde{z} + \tilde{\mu})-\tilde{h}-\sigma\sqrt{1+w_3}\right) - \sigma(1+w_3)N''_{\rm{F}}\left(\bar{m}(\tilde{z} + \tilde{\mu})-\tilde{h} - \sigma\sqrt{1+w_3}\right) \right]\,,
\end{aligned}
\label{etaAFerm2}
\end{equation}
\begin{equation}
\eta_{A,k}^\varphi = \frac{\rho_0 {U''_k}^2}{\left(1-\bar{m}^2\right)^{\frac{3}{2}}}\frac{\sqrt{2}}{3\pi^2 k}\frac{1}{\left[(1+w_1)(1+w_2)\right]^{\frac{3}{2}}}\left[1+2N_{\rm{B}}\left(\sqrt{1+w_1}\sqrt{1+w_2}\right) -2N'_{\rm{B}}\left(\sqrt{1+w_1}\sqrt{1+w_2}\right)\right]\,.
\label{etaABos}
\end{equation}
\end{widetext}
Here, the primes denote derivatives of the thermal distribution functions with respect to their arguments, $N_{\rm F/B}^{\prime}(x)\equiv \partial N_{\rm F/B}(x)/\partial x$, 
and the $w_i$'s are understood to be evaluated at $\rho = \rho_{0,k}$. 

The contributions in Eqs.~\eqref{etaAFerm1} and~\eqref{etaABos} represent straightforward extensions of their mass-balanced analogues. 
The situation is different for $\eta_{A,k}^{\psi,2}$ since it is not even present in the mass-balanced case. It becomes important only at relatively large mass imbalances since it is proportional to 
$\bar{m}^2/(1-\bar{m}^2)$.

Taking a closer look at the bosonic contribution $\eta_{A,k}^\varphi$, it becomes clear why a finite $\rho_{0,k}$ may have such an enormous influence on the 
RG flow of the theory (see also Fig.~\ref{DeltaBend}): $\eta_{A,k}^\varphi$ vanishes identically in the $U(1)$-symmetric regime, i.e. for~$\rho_{0,k} = 0$.

It is not too surprising that bosonic fluctuations are generally amplified by mass imbalance. This becomes apparent by the fact that both $\eta_{A,k}^\varphi$ and $[\partial_k U_k]^\varphi$ are 
proportional to $(1-\bar{m}^2)^{-\frac{3}{2}}$. A detailed analysis how this affects the properties and the extent of inhomogeneous phases for large $\bar{m}$ is left to future work.

\bibliography{bib_source}

\begin{thebibliography}{89}%
\makeatletter
\providecommand \@ifxundefined [1]{%
 \@ifx{#1\undefined}
}%
\providecommand \@ifnum [1]{%
 \ifnum #1\expandafter \@firstoftwo
 \else \expandafter \@secondoftwo
 \fi
}%
\providecommand \@ifx [1]{%
 \ifx #1\expandafter \@firstoftwo
 \else \expandafter \@secondoftwo
 \fi
}%
\providecommand \natexlab [1]{#1}%
\providecommand \enquote  [1]{``#1''}%
\providecommand \bibnamefont  [1]{#1}%
\providecommand \bibfnamefont [1]{#1}%
\providecommand \citenamefont [1]{#1}%
\providecommand \href@noop [0]{\@secondoftwo}%
\providecommand \href [0]{\begingroup \@sanitize@url \@href}%
\providecommand \@href[1]{\@@startlink{#1}\@@href}%
\providecommand \@@href[1]{\endgroup#1\@@endlink}%
\providecommand \@sanitize@url [0]{\catcode `\\12\catcode `\$12\catcode
  `\&12\catcode `\#12\catcode `\^12\catcode `\_12\catcode `\%12\relax}%
\providecommand \@@startlink[1]{}%
\providecommand \@@endlink[0]{}%
\providecommand \url  [0]{\begingroup\@sanitize@url \@url }%
\providecommand \@url [1]{\endgroup\@href {#1}{\urlprefix }}%
\providecommand \urlprefix  [0]{URL }%
\providecommand \Eprint [0]{\href }%
\providecommand \doibase [0]{http://dx.doi.org/}%
\providecommand \selectlanguage [0]{\@gobble}%
\providecommand \bibinfo  [0]{\@secondoftwo}%
\providecommand \bibfield  [0]{\@secondoftwo}%
\providecommand \translation [1]{[#1]}%
\providecommand \BibitemOpen [0]{}%
\providecommand \bibitemStop [0]{}%
\providecommand \bibitemNoStop [0]{.\EOS\space}%
\providecommand \EOS [0]{\spacefactor3000\relax}%
\providecommand \BibitemShut  [1]{\csname bibitem#1\endcsname}%
\let\auto@bib@innerbib\@empty
\bibitem [{\citenamefont {Regal}\ \emph {et~al.}(2004)\citenamefont {Regal},
  \citenamefont {Greiner},\ and\ \citenamefont
  {Jin}}]{RegalGreinerJin_FermBEC04}%
  \BibitemOpen
  \bibfield  {author} {\bibinfo {author} {\bibfnamefont {C.~A.}\ \bibnamefont
  {Regal}}, \bibinfo {author} {\bibfnamefont {M.}~\bibnamefont {Greiner}}, \
  and\ \bibinfo {author} {\bibfnamefont {D.~S.}\ \bibnamefont {Jin}},\ }\href
  {\doibase 10.1103/PhysRevLett.92.040403} {\bibfield  {journal} {\bibinfo
  {journal} {Phys. Rev. Lett.}\ }\textbf {\bibinfo {volume} {92}},\ \bibinfo
  {pages} {040403} (\bibinfo {year} {2004})}\BibitemShut {NoStop}%
\bibitem [{\citenamefont {Jochim}\ \emph {et~al.}(2003)\citenamefont {Jochim},
  \citenamefont {Bartenstein}, \citenamefont {Altmeyer}, \citenamefont {Hendl},
  \citenamefont {Riedl}, \citenamefont {Chin}, \citenamefont
  {Hecker~Denschlag},\ and\ \citenamefont {Grimm}}]{Jochim_etal_FermBEC03}%
  \BibitemOpen
  \bibfield  {author} {\bibinfo {author} {\bibfnamefont {S.}~\bibnamefont
  {Jochim}}, \bibinfo {author} {\bibfnamefont {M.}~\bibnamefont {Bartenstein}},
  \bibinfo {author} {\bibfnamefont {A.}~\bibnamefont {Altmeyer}}, \bibinfo
  {author} {\bibfnamefont {G.}~\bibnamefont {Hendl}}, \bibinfo {author}
  {\bibfnamefont {S.}~\bibnamefont {Riedl}}, \bibinfo {author} {\bibfnamefont
  {C.}~\bibnamefont {Chin}}, \bibinfo {author} {\bibfnamefont {J.}~\bibnamefont
  {Hecker~Denschlag}}, \ and\ \bibinfo {author} {\bibfnamefont
  {R.}~\bibnamefont {Grimm}},\ }\href {\doibase 10.1126/science.1093280}
  {\bibfield  {journal} {\bibinfo  {journal} {Science}\ }\textbf {\bibinfo
  {volume} {302}},\ \bibinfo {pages} {2101} (\bibinfo {year}
  {2003})}\BibitemShut {NoStop}%
\bibitem [{\citenamefont {Zwierlein}\ \emph {et~al.}(2006)\citenamefont
  {Zwierlein}, \citenamefont {Schirotzek}, \citenamefont {Schunck},\ and\
  \citenamefont {Ketterle}}]{Zwierlein27012006}%
  \BibitemOpen
  \bibfield  {author} {\bibinfo {author} {\bibfnamefont {M.~W.}\ \bibnamefont
  {Zwierlein}}, \bibinfo {author} {\bibfnamefont {A.}~\bibnamefont
  {Schirotzek}}, \bibinfo {author} {\bibfnamefont {C.~H.}\ \bibnamefont
  {Schunck}}, \ and\ \bibinfo {author} {\bibfnamefont {W.}~\bibnamefont
  {Ketterle}},\ }\href {\doibase 10.1126/science.1122318} {\bibfield  {journal}
  {\bibinfo  {journal} {Science}\ }\textbf {\bibinfo {volume} {311}},\ \bibinfo
  {pages} {492} (\bibinfo {year} {2006})}\BibitemShut {NoStop}%
\bibitem [{\citenamefont {Partridge}\ \emph
  {et~al.}(2006{\natexlab{a}})\citenamefont {Partridge}, \citenamefont {Li},
  \citenamefont {Kamar}, \citenamefont {Liao},\ and\ \citenamefont
  {Hulet}}]{Partridge27012006}%
  \BibitemOpen
  \bibfield  {author} {\bibinfo {author} {\bibfnamefont {G.~B.}\ \bibnamefont
  {Partridge}}, \bibinfo {author} {\bibfnamefont {W.}~\bibnamefont {Li}},
  \bibinfo {author} {\bibfnamefont {R.~I.}\ \bibnamefont {Kamar}}, \bibinfo
  {author} {\bibfnamefont {Y.-a.}\ \bibnamefont {Liao}}, \ and\ \bibinfo
  {author} {\bibfnamefont {R.~G.}\ \bibnamefont {Hulet}},\ }\href {\doibase
  10.1126/science.1122876} {\bibfield  {journal} {\bibinfo  {journal}
  {Science}\ }\textbf {\bibinfo {volume} {311}},\ \bibinfo {pages} {503}
  (\bibinfo {year} {2006}{\natexlab{a}})}\BibitemShut {NoStop}%
\bibitem [{\citenamefont {{Zwierlein}}\ \emph {et~al.}(2006)\citenamefont
  {{Zwierlein}}, \citenamefont {{Schunck}}, \citenamefont {{Schirotzek}},\ and\
  \citenamefont {{Ketterle}}}]{2006Natur.442...54Z}%
  \BibitemOpen
  \bibfield  {author} {\bibinfo {author} {\bibfnamefont {M.~W.}\ \bibnamefont
  {{Zwierlein}}}, \bibinfo {author} {\bibfnamefont {C.~H.}\ \bibnamefont
  {{Schunck}}}, \bibinfo {author} {\bibfnamefont {A.}~\bibnamefont
  {{Schirotzek}}}, \ and\ \bibinfo {author} {\bibfnamefont {W.}~\bibnamefont
  {{Ketterle}}},\ }\href {\doibase 10.1038/nature04936} {\bibfield  {journal}
  {\bibinfo  {journal} {\nat}\ }\textbf {\bibinfo {volume} {442}},\ \bibinfo
  {pages} {54} (\bibinfo {year} {2006})},\ \Eprint
  {http://arxiv.org/abs/cond-mat/0605258} {cond-mat/0605258} \BibitemShut
  {NoStop}%
\bibitem [{\citenamefont {Shin}\ \emph {et~al.}(2006)\citenamefont {Shin},
  \citenamefont {Zwierlein}, \citenamefont {Schunck}, \citenamefont
  {Schirotzek},\ and\ \citenamefont {Ketterle}}]{PhysRevLett.97.030401}%
  \BibitemOpen
  \bibfield  {author} {\bibinfo {author} {\bibfnamefont {Y.}~\bibnamefont
  {Shin}}, \bibinfo {author} {\bibfnamefont {M.~W.}\ \bibnamefont {Zwierlein}},
  \bibinfo {author} {\bibfnamefont {C.~H.}\ \bibnamefont {Schunck}}, \bibinfo
  {author} {\bibfnamefont {A.}~\bibnamefont {Schirotzek}}, \ and\ \bibinfo
  {author} {\bibfnamefont {W.}~\bibnamefont {Ketterle}},\ }\href {\doibase
  10.1103/PhysRevLett.97.030401} {\bibfield  {journal} {\bibinfo  {journal}
  {Phys. Rev. Lett.}\ }\textbf {\bibinfo {volume} {97}},\ \bibinfo {pages}
  {030401} (\bibinfo {year} {2006})}\BibitemShut {NoStop}%
\bibitem [{\citenamefont {Partridge}\ \emph
  {et~al.}(2006{\natexlab{b}})\citenamefont {Partridge}, \citenamefont {Li},
  \citenamefont {Liao}, \citenamefont {Hulet}, \citenamefont {Haque},\ and\
  \citenamefont {Stoof}}]{PhysRevLett.97.190407}%
  \BibitemOpen
  \bibfield  {author} {\bibinfo {author} {\bibfnamefont {G.~B.}\ \bibnamefont
  {Partridge}}, \bibinfo {author} {\bibfnamefont {W.}~\bibnamefont {Li}},
  \bibinfo {author} {\bibfnamefont {Y.~A.}\ \bibnamefont {Liao}}, \bibinfo
  {author} {\bibfnamefont {R.~G.}\ \bibnamefont {Hulet}}, \bibinfo {author}
  {\bibfnamefont {M.}~\bibnamefont {Haque}}, \ and\ \bibinfo {author}
  {\bibfnamefont {H.~T.~C.}\ \bibnamefont {Stoof}},\ }\href {\doibase
  10.1103/PhysRevLett.97.190407} {\bibfield  {journal} {\bibinfo  {journal}
  {Phys. Rev. Lett.}\ }\textbf {\bibinfo {volume} {97}},\ \bibinfo {pages}
  {190407} (\bibinfo {year} {2006}{\natexlab{b}})}\BibitemShut {NoStop}%
\bibitem [{\citenamefont {Schunck}\ \emph {et~al.}(2007)\citenamefont
  {Schunck}, \citenamefont {Shin}, \citenamefont {Schirotzek}, \citenamefont
  {Zwierlein},\ and\ \citenamefont {Ketterle}}]{Schunck11052007}%
  \BibitemOpen
  \bibfield  {author} {\bibinfo {author} {\bibfnamefont {C.~H.}\ \bibnamefont
  {Schunck}}, \bibinfo {author} {\bibfnamefont {Y.}~\bibnamefont {Shin}},
  \bibinfo {author} {\bibfnamefont {A.}~\bibnamefont {Schirotzek}}, \bibinfo
  {author} {\bibfnamefont {M.~W.}\ \bibnamefont {Zwierlein}}, \ and\ \bibinfo
  {author} {\bibfnamefont {W.}~\bibnamefont {Ketterle}},\ }\href {\doibase
  10.1126/science.1140749} {\bibfield  {journal} {\bibinfo  {journal}
  {Science}\ }\textbf {\bibinfo {volume} {316}},\ \bibinfo {pages} {867}
  (\bibinfo {year} {2007})}\BibitemShut {NoStop}%
\bibitem [{\citenamefont {{Shin}}\ \emph {et~al.}(2008)\citenamefont {{Shin}},
  \citenamefont {{Schunck}}, \citenamefont {{Schirotzek}},\ and\ \citenamefont
  {{Ketterle}}}]{2008Natur.451..689S}%
  \BibitemOpen
  \bibfield  {author} {\bibinfo {author} {\bibfnamefont {Y.-I.}\ \bibnamefont
  {{Shin}}}, \bibinfo {author} {\bibfnamefont {C.~H.}\ \bibnamefont
  {{Schunck}}}, \bibinfo {author} {\bibfnamefont {A.}~\bibnamefont
  {{Schirotzek}}}, \ and\ \bibinfo {author} {\bibfnamefont {W.}~\bibnamefont
  {{Ketterle}}},\ }\href {\doibase 10.1038/nature06473} {\bibfield  {journal}
  {\bibinfo  {journal} {\nat}\ }\textbf {\bibinfo {volume} {451}},\ \bibinfo
  {pages} {689} (\bibinfo {year} {2008})},\ \Eprint
  {http://arxiv.org/abs/0709.3027} {arXiv:0709.3027 [cond-mat.soft]}
  \BibitemShut {NoStop}%
\bibitem [{\citenamefont {Greiner}\ \emph {et~al.}(2002)\citenamefont
  {Greiner}, \citenamefont {Mandel}, \citenamefont {Esslinger}, \citenamefont
  {Hansch},\ and\ \citenamefont {Bloch}}]{Greiner2002}%
  \BibitemOpen
  \bibfield  {author} {\bibinfo {author} {\bibfnamefont {M.}~\bibnamefont
  {Greiner}}, \bibinfo {author} {\bibfnamefont {O.}~\bibnamefont {Mandel}},
  \bibinfo {author} {\bibfnamefont {T.}~\bibnamefont {Esslinger}}, \bibinfo
  {author} {\bibfnamefont {T.~W.}\ \bibnamefont {Hansch}}, \ and\ \bibinfo
  {author} {\bibfnamefont {I.}~\bibnamefont {Bloch}},\ }\href
  {http://dx.doi.org/10.1038/415039a} {\bibfield  {journal} {\bibinfo
  {journal} {Nature}\ }\textbf {\bibinfo {volume} {415}},\ \bibinfo {pages}
  {39} (\bibinfo {year} {2002})}\BibitemShut {NoStop}%
\bibitem [{\citenamefont {Jordens}\ \emph {et~al.}(2008)\citenamefont
  {Jordens}, \citenamefont {Strohmaier}, \citenamefont {Gunter}, \citenamefont
  {Moritz},\ and\ \citenamefont {Esslinger}}]{Jordens2008}%
  \BibitemOpen
  \bibfield  {author} {\bibinfo {author} {\bibfnamefont {R.}~\bibnamefont
  {Jordens}}, \bibinfo {author} {\bibfnamefont {N.}~\bibnamefont {Strohmaier}},
  \bibinfo {author} {\bibfnamefont {K.}~\bibnamefont {Gunter}}, \bibinfo
  {author} {\bibfnamefont {H.}~\bibnamefont {Moritz}}, \ and\ \bibinfo {author}
  {\bibfnamefont {T.}~\bibnamefont {Esslinger}},\ }\href
  {http://dx.doi.org/10.1038/nature07244} {\bibfield  {journal} {\bibinfo
  {journal} {Nature}\ }\textbf {\bibinfo {volume} {455}},\ \bibinfo {pages}
  {204} (\bibinfo {year} {2008})}\BibitemShut {NoStop}%
\bibitem [{\citenamefont {Schneider}\ \emph {et~al.}(2008)\citenamefont
  {Schneider}, \citenamefont {Hackerm\"uller}, \citenamefont {Will},
  \citenamefont {Best}, \citenamefont {Bloch}, \citenamefont {Costi},
  \citenamefont {Helmes}, \citenamefont {Rasch},\ and\ \citenamefont
  {Rosch}}]{Schneider05122008}%
  \BibitemOpen
  \bibfield  {author} {\bibinfo {author} {\bibfnamefont {U.}~\bibnamefont
  {Schneider}}, \bibinfo {author} {\bibfnamefont {L.}~\bibnamefont
  {Hackerm\"uller}}, \bibinfo {author} {\bibfnamefont {S.}~\bibnamefont
  {Will}}, \bibinfo {author} {\bibfnamefont {T.}~\bibnamefont {Best}}, \bibinfo
  {author} {\bibfnamefont {I.}~\bibnamefont {Bloch}}, \bibinfo {author}
  {\bibfnamefont {T.~A.}\ \bibnamefont {Costi}}, \bibinfo {author}
  {\bibfnamefont {R.~W.}\ \bibnamefont {Helmes}}, \bibinfo {author}
  {\bibfnamefont {D.}~\bibnamefont {Rasch}}, \ and\ \bibinfo {author}
  {\bibfnamefont {A.}~\bibnamefont {Rosch}},\ }\href@noop {} {\bibfield
  {journal} {\bibinfo  {journal} {Science}\ }\textbf {\bibinfo {volume}
  {322}},\ \bibinfo {pages} {1520} (\bibinfo {year} {2008})}\BibitemShut
  {NoStop}%
\bibitem [{\citenamefont {Greiner}\ and\ \citenamefont
  {Folling}(2008)}]{Greiner2008}%
  \BibitemOpen
  \bibfield  {author} {\bibinfo {author} {\bibfnamefont {M.}~\bibnamefont
  {Greiner}}\ and\ \bibinfo {author} {\bibfnamefont {S.}~\bibnamefont
  {Folling}},\ }\href {http://dx.doi.org/10.1038/453736a} {\bibfield  {journal}
  {\bibinfo  {journal} {Nature}\ }\textbf {\bibinfo {volume} {453}},\ \bibinfo
  {pages} {736} (\bibinfo {year} {2008})}\BibitemShut {NoStop}%
\bibitem [{\citenamefont {{Cazalilla}}\ \emph {et~al.}(2011)\citenamefont
  {{Cazalilla}}, \citenamefont {{Citro}}, \citenamefont {{Giamarchi}},
  \citenamefont {{Orignac}},\ and\ \citenamefont
  {{Rigol}}}]{2011RvMP...83.1405C}%
  \BibitemOpen
  \bibfield  {author} {\bibinfo {author} {\bibfnamefont {M.~A.}\ \bibnamefont
  {{Cazalilla}}}, \bibinfo {author} {\bibfnamefont {R.}~\bibnamefont
  {{Citro}}}, \bibinfo {author} {\bibfnamefont {T.}~\bibnamefont
  {{Giamarchi}}}, \bibinfo {author} {\bibfnamefont {E.}~\bibnamefont
  {{Orignac}}}, \ and\ \bibinfo {author} {\bibfnamefont {M.}~\bibnamefont
  {{Rigol}}},\ }\href {\doibase 10.1103/RevModPhys.83.1405} {\bibfield
  {journal} {\bibinfo  {journal} {Reviews of Modern Physics}\ }\textbf
  {\bibinfo {volume} {83}},\ \bibinfo {pages} {1405} (\bibinfo {year}
  {2011})},\ \Eprint {http://arxiv.org/abs/1101.5337} {arXiv:1101.5337
  [cond-mat.str-el]} \BibitemShut {NoStop}%
\bibitem [{\citenamefont {Nascimb\`ene}\ \emph {et~al.}(2011)\citenamefont
  {Nascimb\`ene}, \citenamefont {Navon}, \citenamefont {Pilati}, \citenamefont
  {Chevy}, \citenamefont {Giorgini}, \citenamefont {Georges},\ and\
  \citenamefont {Salomon}}]{PhysRevLett.106.215303}%
  \BibitemOpen
  \bibfield  {author} {\bibinfo {author} {\bibfnamefont {S.}~\bibnamefont
  {Nascimb\`ene}}, \bibinfo {author} {\bibfnamefont {N.}~\bibnamefont {Navon}},
  \bibinfo {author} {\bibfnamefont {S.}~\bibnamefont {Pilati}}, \bibinfo
  {author} {\bibfnamefont {F.}~\bibnamefont {Chevy}}, \bibinfo {author}
  {\bibfnamefont {S.}~\bibnamefont {Giorgini}}, \bibinfo {author}
  {\bibfnamefont {A.}~\bibnamefont {Georges}}, \ and\ \bibinfo {author}
  {\bibfnamefont {C.}~\bibnamefont {Salomon}},\ }\href {\doibase
  10.1103/PhysRevLett.106.215303} {\bibfield  {journal} {\bibinfo  {journal}
  {Phys. Rev. Lett.}\ }\textbf {\bibinfo {volume} {106}},\ \bibinfo {pages}
  {215303} (\bibinfo {year} {2011})}\BibitemShut {NoStop}%
\bibitem [{\citenamefont {{van Houcke}}\ \emph {et~al.}(2012)\citenamefont
  {{van Houcke}}, \citenamefont {{Werner}}, \citenamefont {{Kozik}},
  \citenamefont {{Prokof'ev}}, \citenamefont {{Svistunov}}, \citenamefont
  {{Ku}}, \citenamefont {{Sommer}}, \citenamefont {{Cheuk}}, \citenamefont
  {{Schirotzek}},\ and\ \citenamefont {{Zwierlein}}}]{2012NatPh...8..366V}%
  \BibitemOpen
  \bibfield  {author} {\bibinfo {author} {\bibfnamefont {K.}~\bibnamefont {{van
  Houcke}}}, \bibinfo {author} {\bibfnamefont {F.}~\bibnamefont {{Werner}}},
  \bibinfo {author} {\bibfnamefont {E.}~\bibnamefont {{Kozik}}}, \bibinfo
  {author} {\bibfnamefont {N.}~\bibnamefont {{Prokof'ev}}}, \bibinfo {author}
  {\bibfnamefont {B.}~\bibnamefont {{Svistunov}}}, \bibinfo {author}
  {\bibfnamefont {M.~J.~H.}\ \bibnamefont {{Ku}}}, \bibinfo {author}
  {\bibfnamefont {A.~T.}\ \bibnamefont {{Sommer}}}, \bibinfo {author}
  {\bibfnamefont {L.~W.}\ \bibnamefont {{Cheuk}}}, \bibinfo {author}
  {\bibfnamefont {A.}~\bibnamefont {{Schirotzek}}}, \ and\ \bibinfo {author}
  {\bibfnamefont {M.~W.}\ \bibnamefont {{Zwierlein}}},\ }\href {\doibase
  10.1038/nphys2273} {\bibfield  {journal} {\bibinfo  {journal} {Nature
  Physics}\ }\textbf {\bibinfo {volume} {8}},\ \bibinfo {pages} {366} (\bibinfo
  {year} {2012})},\ \Eprint {http://arxiv.org/abs/1110.3747} {arXiv:1110.3747
  [cond-mat.quant-gas]} \BibitemShut {NoStop}%
\bibitem [{\citenamefont {Ketterle}\ \emph {et~al.}(1999)\citenamefont
  {Ketterle}, \citenamefont {Durfee},\ and\ \citenamefont
  {Stamper-Kurn}}]{ketterle-review}%
  \BibitemOpen
  \bibfield  {author} {\bibinfo {author} {\bibfnamefont {W.}~\bibnamefont
  {Ketterle}}, \bibinfo {author} {\bibfnamefont {D.}~\bibnamefont {Durfee}}, \
  and\ \bibinfo {author} {\bibfnamefont {D.}~\bibnamefont {Stamper-Kurn}},\
  }\href@noop {} {\bibfield  {journal} {\bibinfo  {journal} {Proceedings of the
  International School of Physics "Enrico Fermi", Course CXL, edited by M.
  Inguscio, S. Stringari and C.E. Wieman}\ ,\ \bibinfo {pages} {67}} (\bibinfo
  {year} {1999})}\BibitemShut {NoStop}%
\bibitem [{\citenamefont {Bloch}\ \emph {et~al.}(2008)\citenamefont {Bloch},
  \citenamefont {Dalibard},\ and\ \citenamefont {Zwerger}}]{Bloch:2008zzb}%
  \BibitemOpen
  \bibfield  {author} {\bibinfo {author} {\bibfnamefont {I.}~\bibnamefont
  {Bloch}}, \bibinfo {author} {\bibfnamefont {J.}~\bibnamefont {Dalibard}}, \
  and\ \bibinfo {author} {\bibfnamefont {W.}~\bibnamefont {Zwerger}},\ }\href
  {\doibase 10.1103/RevModPhys.80.885} {\bibfield  {journal} {\bibinfo
  {journal} {{Rev. Mod. Phys.}}\ }\textbf {\bibinfo {volume} {80}},\ \bibinfo
  {pages} {885} (\bibinfo {year} {2008})},\ \Eprint
  {http://arxiv.org/abs/0704.3011} {arXiv:0704.3011 [cond-mat.other]}
  \BibitemShut {NoStop}%
\bibitem [{\citenamefont {Giorgini}\ \emph {et~al.}(2008)\citenamefont
  {Giorgini}, \citenamefont {Pitaevskii},\ and\ \citenamefont
  {Stringari}}]{Giorgini:2008zz}%
  \BibitemOpen
  \bibfield  {author} {\bibinfo {author} {\bibfnamefont {S.}~\bibnamefont
  {Giorgini}}, \bibinfo {author} {\bibfnamefont {L.~P.}\ \bibnamefont
  {Pitaevskii}}, \ and\ \bibinfo {author} {\bibfnamefont {S.}~\bibnamefont
  {Stringari}},\ }\href {\doibase 10.1103/RevModPhys.80.1215} {\bibfield
  {journal} {\bibinfo  {journal} {{Rev. Mod. Phys.}}\ }\textbf {\bibinfo
  {volume} {80}},\ \bibinfo {pages} {1215} (\bibinfo {year} {2008})},\ \Eprint
  {http://arxiv.org/abs/0706.3360} {arXiv:0706.3360 [cond-mat.other]}
  \BibitemShut {NoStop}%
\bibitem [{\citenamefont {Grimm}()}]{GrimmPC}%
  \BibitemOpen
  \bibfield  {author} {\bibinfo {author} {\bibfnamefont {R.}~\bibnamefont
  {Grimm}},\ }\href@noop {} {\bibinfo  {journal} {private communication}\
  }\BibitemShut {NoStop}%
\bibitem [{\citenamefont {{Lu}}\ \emph {et~al.}(2012)\citenamefont {{Lu}},
  \citenamefont {{Burdick}},\ and\ \citenamefont
  {{Lev}}}]{2012PhRvL.108u5301L}%
  \BibitemOpen
\bibfield  {journal} {  }\bibfield  {author} {\bibinfo {author} {\bibfnamefont
  {M.}~\bibnamefont {{Lu}}}, \bibinfo {author} {\bibfnamefont {N.~Q.}\
  \bibnamefont {{Burdick}}}, \ and\ \bibinfo {author} {\bibfnamefont {B.~L.}\
  \bibnamefont {{Lev}}},\ }\href {\doibase 10.1103/PhysRevLett.108.215301}
  {\bibfield  {journal} {\bibinfo  {journal} {Phys. Rev. Lett.}\ }\textbf
  {\bibinfo {volume} {108}},\ \bibinfo {eid} {215301} (\bibinfo {year}
  {2012})},\ \Eprint {http://arxiv.org/abs/1202.4444} {arXiv:1202.4444
  [cond-mat.quant-gas]} \BibitemShut {NoStop}%
\bibitem [{\citenamefont {{Frisch}}\ \emph {et~al.}(2013)\citenamefont
  {{Frisch}}, \citenamefont {{Aikawa}}, \citenamefont {{Mark}}, \citenamefont
  {{Ferlaino}}, \citenamefont {{Berseneva}},\ and\ \citenamefont
  {{Kotochigova}}}]{2013PhRvA..88c2508F}%
  \BibitemOpen
  \bibfield  {author} {\bibinfo {author} {\bibfnamefont {A.}~\bibnamefont
  {{Frisch}}}, \bibinfo {author} {\bibfnamefont {K.}~\bibnamefont {{Aikawa}}},
  \bibinfo {author} {\bibfnamefont {M.}~\bibnamefont {{Mark}}}, \bibinfo
  {author} {\bibfnamefont {F.}~\bibnamefont {{Ferlaino}}}, \bibinfo {author}
  {\bibfnamefont {E.}~\bibnamefont {{Berseneva}}}, \ and\ \bibinfo {author}
  {\bibfnamefont {S.}~\bibnamefont {{Kotochigova}}},\ }\href {\doibase
  10.1103/PhysRevA.88.032508} {\bibfield  {journal} {\bibinfo  {journal}
  {\pra}\ }\textbf {\bibinfo {volume} {88}},\ \bibinfo {eid} {032508} (\bibinfo
  {year} {2013})},\ \Eprint {http://arxiv.org/abs/1304.3326} {arXiv:1304.3326
  [physics.atom-ph]} \BibitemShut {NoStop}%
\bibitem [{\citenamefont {Fulde}\ and\ \citenamefont
  {Ferrell}(1964)}]{FuldeFerrell64}%
  \BibitemOpen
  \bibfield  {author} {\bibinfo {author} {\bibfnamefont {P.}~\bibnamefont
  {Fulde}}\ and\ \bibinfo {author} {\bibfnamefont {R.~A.}\ \bibnamefont
  {Ferrell}},\ }\href {\doibase 10.1103/PhysRev.135.A550} {\bibfield  {journal}
  {\bibinfo  {journal} {Phys. Rev.}\ }\textbf {\bibinfo {volume} {135}},\
  \bibinfo {pages} {A550} (\bibinfo {year} {1964})}\BibitemShut {NoStop}%
\bibitem [{\citenamefont {Larkin}\ and\ \citenamefont
  {Ovchinnikov}(1964)}]{LarkinOvchinnikov64}%
  \BibitemOpen
  \bibfield  {author} {\bibinfo {author} {\bibfnamefont {A.}~\bibnamefont
  {Larkin}}\ and\ \bibinfo {author} {\bibfnamefont {Y.}~\bibnamefont
  {Ovchinnikov}},\ }\href@noop {} {\bibfield  {journal} {\bibinfo  {journal}
  {Zh.Eksp.Teor.Fiz.}\ }\textbf {\bibinfo {volume} {47}},\ \bibinfo {pages}
  {1136} (\bibinfo {year} {1964})}\BibitemShut {NoStop}%
\bibitem [{\citenamefont {Zwerger}(2012)}]{Zwerger-book}%
  \BibitemOpen
  \bibinfo {editor} {\bibfnamefont {W.}~\bibnamefont {Zwerger}},\ ed.,\
  \href@noop {} {\emph {\bibinfo {title} {{The BCS-BEC Crossover and the
  Unitary Fermi Gas}}}}\ (\bibinfo  {publisher} {Springer, Berlin},\ \bibinfo
  {year} {2012})\BibitemShut {NoStop}%
\bibitem [{\citenamefont {Bulgac}\ \emph {et~al.}(2006)\citenamefont {Bulgac},
  \citenamefont {Drut},\ and\ \citenamefont {Magierski}}]{Bulgac:2005pj}%
  \BibitemOpen
  \bibfield  {author} {\bibinfo {author} {\bibfnamefont {A.}~\bibnamefont
  {Bulgac}}, \bibinfo {author} {\bibfnamefont {J.~E.}\ \bibnamefont {Drut}}, \
  and\ \bibinfo {author} {\bibfnamefont {P.}~\bibnamefont {Magierski}},\ }\href
  {\doibase 10.1103/PhysRevLett.96.090404} {\bibfield  {journal} {\bibinfo
  {journal} {Phys. Rev. Lett.}\ }\textbf {\bibinfo {volume} {96}},\ \bibinfo
  {pages} {090404} (\bibinfo {year} {2006})}\BibitemShut {NoStop}%
\bibitem [{\citenamefont {Drut}\ \emph {et~al.}(2012)\citenamefont {Drut},
  \citenamefont {Lahde}, \citenamefont {Wlazlowski},\ and\ \citenamefont
  {Magierski}}]{Drut:2011tf}%
  \BibitemOpen
  \bibfield  {author} {\bibinfo {author} {\bibfnamefont {J.~E.}\ \bibnamefont
  {Drut}}, \bibinfo {author} {\bibfnamefont {T.~A.}\ \bibnamefont {Lahde}},
  \bibinfo {author} {\bibfnamefont {G.}~\bibnamefont {Wlazlowski}}, \ and\
  \bibinfo {author} {\bibfnamefont {P.}~\bibnamefont {Magierski}},\ }\href
  {\doibase 10.1103/PhysRevA.85.051601} {\bibfield  {journal} {\bibinfo
  {journal} {Phys. Rev.}\ }\textbf {\bibinfo {volume} {A85}},\ \bibinfo {pages}
  {051601} (\bibinfo {year} {2012})},\ \Eprint {http://arxiv.org/abs/1111.5079}
  {arXiv:1111.5079 [cond-mat.quant-gas]} \BibitemShut {NoStop}%
\bibitem [{\citenamefont {{Combescot}}\ \emph {et~al.}(2007)\citenamefont
  {{Combescot}}, \citenamefont {{Recati}}, \citenamefont {{Lobo}},\ and\
  \citenamefont {{Chevy}}}]{CRLC}%
  \BibitemOpen
  \bibfield  {author} {\bibinfo {author} {\bibfnamefont {R.}~\bibnamefont
  {{Combescot}}}, \bibinfo {author} {\bibfnamefont {A.}~\bibnamefont
  {{Recati}}}, \bibinfo {author} {\bibfnamefont {C.}~\bibnamefont {{Lobo}}}, \
  and\ \bibinfo {author} {\bibfnamefont {F.}~\bibnamefont {{Chevy}}},\ }\href
  {\doibase 10.1103/PhysRevLett.98.180402} {\bibfield  {journal} {\bibinfo
  {journal} {Phys. Rev. Lett.}\ }\textbf {\bibinfo {volume} {98}},\ \bibinfo
  {eid} {180402} (\bibinfo {year} {2007})},\ \Eprint
  {http://arxiv.org/abs/cond-mat/0702314} {cond-mat/0702314} \BibitemShut
  {NoStop}%
\bibitem [{\citenamefont {Parish}\ \emph {et~al.}(2007)\citenamefont {Parish},
  \citenamefont {Marchetti}, \citenamefont {Lamacraft},\ and\ \citenamefont
  {Simons}}]{PhysRevLett.98.160402}%
  \BibitemOpen
  \bibfield  {author} {\bibinfo {author} {\bibfnamefont {M.~M.}\ \bibnamefont
  {Parish}}, \bibinfo {author} {\bibfnamefont {F.~M.}\ \bibnamefont
  {Marchetti}}, \bibinfo {author} {\bibfnamefont {A.}~\bibnamefont
  {Lamacraft}}, \ and\ \bibinfo {author} {\bibfnamefont {B.~D.}\ \bibnamefont
  {Simons}},\ }\href {\doibase 10.1103/PhysRevLett.98.160402} {\bibfield
  {journal} {\bibinfo  {journal} {Phys. Rev. Lett.}\ }\textbf {\bibinfo
  {volume} {98}},\ \bibinfo {pages} {160402} (\bibinfo {year}
  {2007})}\BibitemShut {NoStop}%
\bibitem [{\citenamefont {Gezerlis}\ \emph {et~al.}(2009)\citenamefont
  {Gezerlis}, \citenamefont {Gandolfi}, \citenamefont {Schmidt},\ and\
  \citenamefont {Carlson}}]{Gezerlis:2009xp}%
  \BibitemOpen
  \bibfield  {author} {\bibinfo {author} {\bibfnamefont {A.}~\bibnamefont
  {Gezerlis}}, \bibinfo {author} {\bibfnamefont {S.}~\bibnamefont {Gandolfi}},
  \bibinfo {author} {\bibfnamefont {K.}~\bibnamefont {Schmidt}}, \ and\
  \bibinfo {author} {\bibfnamefont {J.}~\bibnamefont {Carlson}},\ }\href
  {\doibase 10.1103/PhysRevLett.103.060403} {\bibfield  {journal} {\bibinfo
  {journal} {Phys. Rev. Lett.}\ }\textbf {\bibinfo {volume} {103}},\ \bibinfo
  {pages} {060403} (\bibinfo {year} {2009})},\ \Eprint
  {http://arxiv.org/abs/0901.3148} {arXiv:0901.3148 [cond-mat.other]}
  \BibitemShut {NoStop}%
\bibitem [{\citenamefont {{Gandolfi}}\ and\ \citenamefont
  {{Carlson}}()}]{Gandolfi:2010}%
  \BibitemOpen
  \bibfield  {author} {\bibinfo {author} {\bibfnamefont {S.}~\bibnamefont
  {{Gandolfi}}}\ and\ \bibinfo {author} {\bibfnamefont {J.}~\bibnamefont
  {{Carlson}}},\ }\href@noop {} {\ }\Eprint {http://arxiv.org/abs/1006.5186}
  {arXiv:1006.5186 [cond-mat.quant-gas]} \BibitemShut {NoStop}%
\bibitem [{\citenamefont {{Baarsma}}\ \emph {et~al.}(2010)\citenamefont
  {{Baarsma}}, \citenamefont {{Gubbels}},\ and\ \citenamefont
  {{Stoof}}}]{2010PhRvA..82a3624B}%
  \BibitemOpen
  \bibfield  {author} {\bibinfo {author} {\bibfnamefont {J.~E.}\ \bibnamefont
  {{Baarsma}}}, \bibinfo {author} {\bibfnamefont {K.~B.}\ \bibnamefont
  {{Gubbels}}}, \ and\ \bibinfo {author} {\bibfnamefont {H.~T.~C.}\
  \bibnamefont {{Stoof}}},\ }\href {\doibase 10.1103/PhysRevA.82.013624}
  {\bibfield  {journal} {\bibinfo  {journal} {{Phys. Rev. A}}\ }\textbf
  {\bibinfo {volume} {82}},\ \bibinfo {eid} {013624} (\bibinfo {year}
  {2010})},\ \Eprint {http://arxiv.org/abs/0912.4205} {arXiv:0912.4205
  [cond-mat.quant-gas]} \BibitemShut {NoStop}%
\bibitem [{\citenamefont {Mathy}\ \emph {et~al.}(2011)\citenamefont {Mathy},
  \citenamefont {Parish},\ and\ \citenamefont {Huse}}]{PhysRevLett.106.166404}%
  \BibitemOpen
  \bibfield  {author} {\bibinfo {author} {\bibfnamefont {C.~J.~M.}\
  \bibnamefont {Mathy}}, \bibinfo {author} {\bibfnamefont {M.~M.}\ \bibnamefont
  {Parish}}, \ and\ \bibinfo {author} {\bibfnamefont {D.~A.}\ \bibnamefont
  {Huse}},\ }\href {\doibase 10.1103/PhysRevLett.106.166404} {\bibfield
  {journal} {\bibinfo  {journal} {Phys. Rev. Lett.}\ }\textbf {\bibinfo
  {volume} {106}},\ \bibinfo {pages} {166404} (\bibinfo {year}
  {2011})}\BibitemShut {NoStop}%
\bibitem [{\citenamefont {Baarsma}\ and\ \citenamefont
  {Stoof}(2013)}]{BaarsmaStoof12}%
  \BibitemOpen
  \bibfield  {author} {\bibinfo {author} {\bibfnamefont {J.~E.}\ \bibnamefont
  {Baarsma}}\ and\ \bibinfo {author} {\bibfnamefont {H.~T.~C.}\ \bibnamefont
  {Stoof}},\ }\href {\doibase 10.1103/PhysRevA.87.063612} {\bibfield  {journal}
  {\bibinfo  {journal} {Phys. Rev. A}\ }\textbf {\bibinfo {volume} {87}},\
  \bibinfo {pages} {063612} (\bibinfo {year} {2013})}\BibitemShut {NoStop}%
\bibitem [{\citenamefont {{Chevy}}(2006)}]{Chevy:2006}%
  \BibitemOpen
  \bibfield  {author} {\bibinfo {author} {\bibfnamefont {F.}~\bibnamefont
  {{Chevy}}},\ }\href {\doibase 10.1103/PhysRevA.74.063628} {\bibfield
  {journal} {\bibinfo  {journal} {{Phys. Rev. A}}\ }\textbf {\bibinfo {volume}
  {74}},\ \bibinfo {eid} {063628} (\bibinfo {year} {2006})},\ \Eprint
  {http://arxiv.org/abs/cond-mat/0605751} {cond-mat/0605751} \BibitemShut
  {NoStop}%
\bibitem [{\citenamefont {{Lobo}}\ \emph {et~al.}(2006)\citenamefont {{Lobo}},
  \citenamefont {{Recati}}, \citenamefont {{Giorgini}},\ and\ \citenamefont
  {{Stringari}}}]{Lobo:2006}%
  \BibitemOpen
  \bibfield  {author} {\bibinfo {author} {\bibfnamefont {C.}~\bibnamefont
  {{Lobo}}}, \bibinfo {author} {\bibfnamefont {A.}~\bibnamefont {{Recati}}},
  \bibinfo {author} {\bibfnamefont {S.}~\bibnamefont {{Giorgini}}}, \ and\
  \bibinfo {author} {\bibfnamefont {S.}~\bibnamefont {{Stringari}}},\ }\href
  {\doibase 10.1103/PhysRevLett.97.200403} {\bibfield  {journal} {\bibinfo
  {journal} {Phys. Rev. Lett.}\ }\textbf {\bibinfo {volume} {97}},\ \bibinfo
  {eid} {200403} (\bibinfo {year} {2006})},\ \Eprint
  {http://arxiv.org/abs/cond-mat/0607730} {cond-mat/0607730} \BibitemShut
  {NoStop}%
\bibitem [{\citenamefont {{Bulgac}}\ and\ \citenamefont
  {{Forbes}}(2007)}]{Bulgac:2007}%
  \BibitemOpen
  \bibfield  {author} {\bibinfo {author} {\bibfnamefont {A.}~\bibnamefont
  {{Bulgac}}}\ and\ \bibinfo {author} {\bibfnamefont {M.~M.}\ \bibnamefont
  {{Forbes}}},\ }\href {\doibase 10.1103/PhysRevA.75.031605} {\bibfield
  {journal} {\bibinfo  {journal} {Phys. Rev. A}\ }\textbf {\bibinfo {volume}
  {75}},\ \bibinfo {eid} {031605} (\bibinfo {year} {2007})},\ \Eprint
  {http://arxiv.org/abs/cond-mat/0606043} {cond-mat/0606043} \BibitemShut
  {NoStop}%
\bibitem [{\citenamefont {{Prokof'Ev}}\ and\ \citenamefont
  {{Svistunov}}(2008)}]{ProkSvist07}%
  \BibitemOpen
  \bibfield  {author} {\bibinfo {author} {\bibfnamefont {N.}~\bibnamefont
  {{Prokof'Ev}}}\ and\ \bibinfo {author} {\bibfnamefont {B.}~\bibnamefont
  {{Svistunov}}},\ }\href@noop {} {\bibfield  {journal} {\bibinfo  {journal}
  {{Phys. Rev. B}}\ }\textbf {\bibinfo {volume} {77}},\ \bibinfo {eid} {020408}
  (\bibinfo {year} {2008})},\ \Eprint {http://arxiv.org/abs/0707.4259}
  {arXiv:0707.4259 [cond-mat.stat-mech]} \BibitemShut {NoStop}%
\bibitem [{\citenamefont {{Chevy}}(2007)}]{Chevy}%
  \BibitemOpen
  \bibfield  {author} {\bibinfo {author} {\bibfnamefont {F.}~\bibnamefont
  {{Chevy}}},\ }\href@noop {} {\  (\bibinfo {year} {2007})},\ \Eprint
  {http://arxiv.org/abs/cond-mat/0701350} {cond-mat/0701350} \BibitemShut
  {NoStop}%
\bibitem [{\citenamefont {{Ku}}\ \emph {et~al.}(2009)\citenamefont {{Ku}},
  \citenamefont {{Braun}},\ and\ \citenamefont {{Schwenk}}}]{KBS}%
  \BibitemOpen
  \bibfield  {author} {\bibinfo {author} {\bibfnamefont {M.}~\bibnamefont
  {{Ku}}}, \bibinfo {author} {\bibfnamefont {J.}~\bibnamefont {{Braun}}}, \
  and\ \bibinfo {author} {\bibfnamefont {A.}~\bibnamefont {{Schwenk}}},\ }\href
  {\doibase 10.1103/PhysRevLett.102.255301} {\bibfield  {journal} {\bibinfo
  {journal} {Phys. Rev. Lett.}\ }\textbf {\bibinfo {volume} {102}},\ \bibinfo
  {eid} {255301} (\bibinfo {year} {2009})},\ \Eprint
  {http://arxiv.org/abs/0812.3430} {arXiv:0812.3430 [cond-mat.other]}
  \BibitemShut {NoStop}%
\bibitem [{\citenamefont {Schmidt}\ and\ \citenamefont
  {Enss}(2011)}]{Schmidt:2011zu}%
  \BibitemOpen
  \bibfield  {author} {\bibinfo {author} {\bibfnamefont {R.}~\bibnamefont
  {Schmidt}}\ and\ \bibinfo {author} {\bibfnamefont {T.}~\bibnamefont {Enss}},\
  }\href {\doibase 10.1103/PhysRevA.83.063620} {\bibfield  {journal} {\bibinfo
  {journal} {Phys. Rev. A}\ }\textbf {\bibinfo {volume} {83}},\ \bibinfo
  {pages} {063620} (\bibinfo {year} {2011})},\ \Eprint
  {http://arxiv.org/abs/1104.1379} {arXiv:1104.1379 [cond-mat.quant-gas]}
  \BibitemShut {NoStop}%
\bibitem [{\citenamefont {Boettcher}\ \emph {et~al.}(2015)\citenamefont
  {Boettcher}, \citenamefont {Braun}, \citenamefont {Herbst}, \citenamefont
  {Pawlowski}, \citenamefont {Roscher} \emph {et~al.}}]{hImb3DFRG}%
  \BibitemOpen
  \bibfield  {author} {\bibinfo {author} {\bibfnamefont {I.}~\bibnamefont
  {Boettcher}}, \bibinfo {author} {\bibfnamefont {J.}~\bibnamefont {Braun}},
  \bibinfo {author} {\bibfnamefont {T.}~\bibnamefont {Herbst}}, \bibinfo
  {author} {\bibfnamefont {J.}~\bibnamefont {Pawlowski}}, \bibinfo {author}
  {\bibfnamefont {D.}~\bibnamefont {Roscher}},  \emph {et~al.},\ }\href
  {\doibase 10.1103/PhysRevA.91.013610} {\bibfield  {journal} {\bibinfo
  {journal} {Phys. Rev.}\ }\textbf {\bibinfo {volume} {A91}},\ \bibinfo {pages}
  {013610} (\bibinfo {year} {2015})},\ \Eprint {http://arxiv.org/abs/1409.5070}
  {arXiv:1409.5070 [cond-mat.quant-gas]} \BibitemShut {NoStop}%
\bibitem [{\citenamefont {{Chevy}}\ and\ \citenamefont
  {{Mora}}(2010)}]{ChevyMora}%
  \BibitemOpen
  \bibfield  {author} {\bibinfo {author} {\bibfnamefont {F.}~\bibnamefont
  {{Chevy}}}\ and\ \bibinfo {author} {\bibfnamefont {C.}~\bibnamefont
  {{Mora}}},\ }\href {\doibase 10.1088/0034-4885/73/11/112401} {\bibfield
  {journal} {\bibinfo  {journal} {Reports on Progress in Physics}\ }\textbf
  {\bibinfo {volume} {73}},\ \bibinfo {eid} {112401} (\bibinfo {year}
  {2010})},\ \Eprint {http://arxiv.org/abs/1003.0801} {arXiv:1003.0801
  [cond-mat.quant-gas]} \BibitemShut {NoStop}%
\bibitem [{\citenamefont {{Gubbels}}\ and\ \citenamefont
  {{Stoof}}(2013)}]{StoofGubbels}%
  \BibitemOpen
  \bibfield  {author} {\bibinfo {author} {\bibfnamefont {K.~B.}\ \bibnamefont
  {{Gubbels}}}\ and\ \bibinfo {author} {\bibfnamefont {H.~T.~C.}\ \bibnamefont
  {{Stoof}}},\ }\href {\doibase 10.1016/j.physrep.2012.11.004} {\bibfield
  {journal} {\bibinfo  {journal} {{Phys. Rept.}}\ }\textbf {\bibinfo {volume}
  {525}},\ \bibinfo {pages} {255} (\bibinfo {year} {2013})},\ \Eprint
  {http://arxiv.org/abs/1205.0568} {arXiv:1205.0568 [cond-mat.quant-gas]}
  \BibitemShut {NoStop}%
\bibitem [{\citenamefont {Goulko}\ and\ \citenamefont
  {Wingate}(2010)}]{GoulkoWingate_hImbMC10}%
  \BibitemOpen
  \bibfield  {author} {\bibinfo {author} {\bibfnamefont {O.}~\bibnamefont
  {Goulko}}\ and\ \bibinfo {author} {\bibfnamefont {M.}~\bibnamefont
  {Wingate}},\ }\href {\doibase 10.1103/PhysRevA.82.053621} {\bibfield
  {journal} {\bibinfo  {journal} {Phys. Rev. A}\ }\textbf {\bibinfo {volume}
  {82}},\ \bibinfo {pages} {053621} (\bibinfo {year} {2010})}\BibitemShut
  {NoStop}%
\bibitem [{\citenamefont {Drut}\ and\ \citenamefont
  {Nicholson}(2013)}]{Drut:2012md}%
  \BibitemOpen
  \bibfield  {author} {\bibinfo {author} {\bibfnamefont {J.~E.}\ \bibnamefont
  {Drut}}\ and\ \bibinfo {author} {\bibfnamefont {A.~N.}\ \bibnamefont
  {Nicholson}},\ }\href {\doibase 10.1088/0954-3899/40/4/043101} {\bibfield
  {journal} {\bibinfo  {journal} {J.Phys.}\ }\textbf {\bibinfo {volume}
  {G40}},\ \bibinfo {pages} {043101} (\bibinfo {year} {2013})},\ \Eprint
  {http://arxiv.org/abs/1208.6556} {arXiv:1208.6556 [cond-mat.stat-mech]}
  \BibitemShut {NoStop}%
\bibitem [{\citenamefont {Braun}\ \emph {et~al.}(2013)\citenamefont {Braun},
  \citenamefont {Chen}, \citenamefont {Deng}, \citenamefont {Drut},
  \citenamefont {Friman}, \citenamefont {Ma},\ and\ \citenamefont
  {Tsai}}]{ImhMC13}%
  \BibitemOpen
  \bibfield  {author} {\bibinfo {author} {\bibfnamefont {J.}~\bibnamefont
  {Braun}}, \bibinfo {author} {\bibfnamefont {J.-W.}\ \bibnamefont {Chen}},
  \bibinfo {author} {\bibfnamefont {J.}~\bibnamefont {Deng}}, \bibinfo {author}
  {\bibfnamefont {J.~E.}\ \bibnamefont {Drut}}, \bibinfo {author}
  {\bibfnamefont {B.}~\bibnamefont {Friman}}, \bibinfo {author} {\bibfnamefont
  {C.-T.}\ \bibnamefont {Ma}}, \ and\ \bibinfo {author} {\bibfnamefont {Y.-D.}\
  \bibnamefont {Tsai}},\ }\href {\doibase 10.1103/PhysRevLett.110.130404}
  {\bibfield  {journal} {\bibinfo  {journal} {Phys. Rev. Lett.}\ }\textbf
  {\bibinfo {volume} {110}},\ \bibinfo {pages} {130404} (\bibinfo {year}
  {2013})}\BibitemShut {NoStop}%
\bibitem [{\citenamefont {Roscher}\ \emph
  {et~al.}(2014{\natexlab{a}})\citenamefont {Roscher}, \citenamefont {Braun},
  \citenamefont {Chen},\ and\ \citenamefont {Drut}}]{ImMbarMC14}%
  \BibitemOpen
  \bibfield  {author} {\bibinfo {author} {\bibfnamefont {D.}~\bibnamefont
  {Roscher}}, \bibinfo {author} {\bibfnamefont {J.}~\bibnamefont {Braun}},
  \bibinfo {author} {\bibfnamefont {J.-W.}\ \bibnamefont {Chen}}, \ and\
  \bibinfo {author} {\bibfnamefont {J.~E.}\ \bibnamefont {Drut}},\ }\href
  {\doibase 10.1088/0954-3899/41/5/055110} {\bibfield  {journal} {\bibinfo
  {journal} {J.Phys.}\ }\textbf {\bibinfo {volume} {G41}},\ \bibinfo {pages}
  {055110} (\bibinfo {year} {2014}{\natexlab{a}})},\ \Eprint
  {http://arxiv.org/abs/1306.0798} {arXiv:1306.0798 [cond-mat.stat-mech]}
  \BibitemShut {NoStop}%
\bibitem [{\citenamefont {Braun}\ \emph {et~al.}(2014)\citenamefont {Braun},
  \citenamefont {Drut},\ and\ \citenamefont {Roscher}}]{ImRidgeT0MC14}%
  \BibitemOpen
  \bibfield  {author} {\bibinfo {author} {\bibfnamefont {J.}~\bibnamefont
  {Braun}}, \bibinfo {author} {\bibfnamefont {J.~E.}\ \bibnamefont {Drut}}, \
  and\ \bibinfo {author} {\bibfnamefont {D.}~\bibnamefont {Roscher}},\
  }\href@noop {} {\  (\bibinfo {year} {2014})},\ \Eprint
  {http://arxiv.org/abs/1407.2924} {arXiv:1407.2924 [cond-mat.quant-gas]}
  \BibitemShut {NoStop}%
\bibitem [{\citenamefont {Sarma}(1963)}]{Sarma63}%
  \BibitemOpen
  \bibfield  {author} {\bibinfo {author} {\bibfnamefont {G.}~\bibnamefont
  {Sarma}},\ }\href {\doibase http://dx.doi.org/10.1016/0022-3697(63)90007-6}
  {\bibfield  {journal} {\bibinfo  {journal} {Journal of Physics and Chemistry
  of Solids}\ }\textbf {\bibinfo {volume} {24}},\ \bibinfo {pages} {1029 }
  (\bibinfo {year} {1963})}\BibitemShut {NoStop}%
\bibitem [{\citenamefont {Roscher}\ \emph
  {et~al.}(2014{\natexlab{b}})\citenamefont {Roscher}, \citenamefont {Braun},\
  and\ \citenamefont {Drut}}]{Inho1DMF}%
  \BibitemOpen
  \bibfield  {author} {\bibinfo {author} {\bibfnamefont {D.}~\bibnamefont
  {Roscher}}, \bibinfo {author} {\bibfnamefont {J.}~\bibnamefont {Braun}}, \
  and\ \bibinfo {author} {\bibfnamefont {J.~E.}\ \bibnamefont {Drut}},\ }\href
  {\doibase 10.1103/PhysRevA.89.063609} {\bibfield  {journal} {\bibinfo
  {journal} {Phys. Rev. A}\ }\textbf {\bibinfo {volume} {89}},\ \bibinfo
  {pages} {063609} (\bibinfo {year} {2014}{\natexlab{b}})}\BibitemShut
  {NoStop}%
\bibitem [{\citenamefont {Liao}\ \emph {et~al.}(2010)\citenamefont {Liao},
  \citenamefont {Rittner}, \citenamefont {Paprotta}, \citenamefont {Li},
  \citenamefont {Partridge}, \citenamefont {Hulet}, \citenamefont {Baur},\ and\
  \citenamefont {Mueller}}]{Liao_etal_1DFFLO10}%
  \BibitemOpen
  \bibfield  {author} {\bibinfo {author} {\bibfnamefont {Y.-A.}\ \bibnamefont
  {Liao}}, \bibinfo {author} {\bibfnamefont {A.~S.~C.}\ \bibnamefont
  {Rittner}}, \bibinfo {author} {\bibfnamefont {T.}~\bibnamefont {Paprotta}},
  \bibinfo {author} {\bibfnamefont {W.}~\bibnamefont {Li}}, \bibinfo {author}
  {\bibfnamefont {G.~B.}\ \bibnamefont {Partridge}}, \bibinfo {author}
  {\bibfnamefont {R.~G.}\ \bibnamefont {Hulet}}, \bibinfo {author}
  {\bibfnamefont {S.~K.}\ \bibnamefont {Baur}}, \ and\ \bibinfo {author}
  {\bibfnamefont {E.~J.}\ \bibnamefont {Mueller}},\ }\href {\doibase
  10.1038/nature09393} {\bibfield  {journal} {\bibinfo  {journal} {Nature}\
  }\textbf {\bibinfo {volume} {467}},\ \bibinfo {pages} {567} (\bibinfo {year}
  {2010})}\BibitemShut {NoStop}%
\bibitem [{\citenamefont {Sheehy}\ and\ \citenamefont
  {Radzihovsky}(2006)}]{SheehyRadzihovskyPRL06}%
  \BibitemOpen
  \bibfield  {author} {\bibinfo {author} {\bibfnamefont {D.~E.}\ \bibnamefont
  {Sheehy}}\ and\ \bibinfo {author} {\bibfnamefont {L.}~\bibnamefont
  {Radzihovsky}},\ }\href {\doibase 10.1103/PhysRevLett.96.060401} {\bibfield
  {journal} {\bibinfo  {journal} {Phys. Rev. Lett.}\ }\textbf {\bibinfo
  {volume} {96}},\ \bibinfo {pages} {060401} (\bibinfo {year}
  {2006})}\BibitemShut {NoStop}%
\bibitem [{\citenamefont {Hu}\ and\ \citenamefont {Liu}(2006)}]{HuLiuMFh06}%
  \BibitemOpen
  \bibfield  {author} {\bibinfo {author} {\bibfnamefont {H.}~\bibnamefont
  {Hu}}\ and\ \bibinfo {author} {\bibfnamefont {X.-J.}\ \bibnamefont {Liu}},\
  }\href {\doibase 10.1103/PhysRevA.73.051603} {\bibfield  {journal} {\bibinfo
  {journal} {Phys. Rev. A}\ }\textbf {\bibinfo {volume} {73}},\ \bibinfo
  {pages} {051603} (\bibinfo {year} {2006})}\BibitemShut {NoStop}%
\bibitem [{\citenamefont {Bulgac}\ and\ \citenamefont
  {Forbes}(2008)}]{Bulgac:2008tm}%
  \BibitemOpen
  \bibfield  {author} {\bibinfo {author} {\bibfnamefont {A.}~\bibnamefont
  {Bulgac}}\ and\ \bibinfo {author} {\bibfnamefont {M.~M.}\ \bibnamefont
  {Forbes}},\ }\href {\doibase 10.1103/PhysRevLett.101.215301} {\bibfield
  {journal} {\bibinfo  {journal} {Phys. Rev. Lett.}\ }\textbf {\bibinfo
  {volume} {101}},\ \bibinfo {pages} {215301} (\bibinfo {year} {2008})},\
  \Eprint {http://arxiv.org/abs/0804.3364} {arXiv:0804.3364
  [cond-mat.supr-con]} \BibitemShut {NoStop}%
\bibitem [{\citenamefont {{Wang}}\ \emph {et~al.}(2014)\citenamefont {{Wang}},
  \citenamefont {{Che}}, \citenamefont {{Zhang}},\ and\ \citenamefont
  {{Chen}}}]{Wang_etal_mIFFLO14}%
  \BibitemOpen
  \bibfield  {author} {\bibinfo {author} {\bibfnamefont {J.}~\bibnamefont
  {{Wang}}}, \bibinfo {author} {\bibfnamefont {Y.}~\bibnamefont {{Che}}},
  \bibinfo {author} {\bibfnamefont {L.}~\bibnamefont {{Zhang}}}, \ and\
  \bibinfo {author} {\bibfnamefont {Q.}~\bibnamefont {{Chen}}},\ }\href@noop {}
  {\bibfield  {journal} {\bibinfo  {journal} {ArXiv e-prints}\ } (\bibinfo
  {year} {2014})},\ \Eprint {http://arxiv.org/abs/1404.5696} {arXiv:1404.5696
  [cond-mat.quant-gas]} \BibitemShut {NoStop}%
\bibitem [{\citenamefont {Diehl}\ \emph
  {et~al.}(2007{\natexlab{a}})\citenamefont {Diehl}, \citenamefont {Gies},
  \citenamefont {Pawlowski},\ and\ \citenamefont {Wetterich}}]{Diehl:2007th}%
  \BibitemOpen
  \bibfield  {author} {\bibinfo {author} {\bibfnamefont {S.}~\bibnamefont
  {Diehl}}, \bibinfo {author} {\bibfnamefont {H.}~\bibnamefont {Gies}},
  \bibinfo {author} {\bibfnamefont {J.}~\bibnamefont {Pawlowski}}, \ and\
  \bibinfo {author} {\bibfnamefont {C.}~\bibnamefont {Wetterich}},\ }\href
  {\doibase 10.1103/PhysRevA.76.021602} {\bibfield  {journal} {\bibinfo
  {journal} {Phys. Rev.}\ }\textbf {\bibinfo {volume} {A76}},\ \bibinfo {pages}
  {021602} (\bibinfo {year} {2007}{\natexlab{a}})},\ \Eprint
  {http://arxiv.org/abs/cond-mat/0701198} {arXiv:cond-mat/0701198 [cond-mat]}
  \BibitemShut {NoStop}%
\bibitem [{\citenamefont {Diehl}\ \emph
  {et~al.}(2007{\natexlab{b}})\citenamefont {Diehl}, \citenamefont {Gies},
  \citenamefont {Pawlowski},\ and\ \citenamefont {Wetterich}}]{Diehl:2007ri}%
  \BibitemOpen
  \bibfield  {author} {\bibinfo {author} {\bibfnamefont {S.}~\bibnamefont
  {Diehl}}, \bibinfo {author} {\bibfnamefont {H.}~\bibnamefont {Gies}},
  \bibinfo {author} {\bibfnamefont {J.}~\bibnamefont {Pawlowski}}, \ and\
  \bibinfo {author} {\bibfnamefont {C.}~\bibnamefont {Wetterich}},\ }\href
  {\doibase 10.1103/PhysRevA.76.053627} {\bibfield  {journal} {\bibinfo
  {journal} {Phys. Rev.}\ }\textbf {\bibinfo {volume} {A76}},\ \bibinfo {pages}
  {053627} (\bibinfo {year} {2007}{\natexlab{b}})},\ \Eprint
  {http://arxiv.org/abs/cond-mat/0703366} {cond-mat/0703366} \BibitemShut
  {NoStop}%
\bibitem [{\citenamefont {Diehl}\ \emph {et~al.}(2010)\citenamefont {Diehl},
  \citenamefont {Floerchinger}, \citenamefont {Gies}, \citenamefont
  {Pawlowski},\ and\ \citenamefont {Wetterich}}]{Diehl:2009ma}%
  \BibitemOpen
  \bibfield  {author} {\bibinfo {author} {\bibfnamefont {S.}~\bibnamefont
  {Diehl}}, \bibinfo {author} {\bibfnamefont {S.}~\bibnamefont {Floerchinger}},
  \bibinfo {author} {\bibfnamefont {H.}~\bibnamefont {Gies}}, \bibinfo {author}
  {\bibfnamefont {J.}~\bibnamefont {Pawlowski}}, \ and\ \bibinfo {author}
  {\bibfnamefont {C.}~\bibnamefont {Wetterich}},\ }\href {\doibase
  10.1002/andp.201010458} {\bibfield  {journal} {\bibinfo  {journal} {Annalen
  Phys.}\ }\textbf {\bibinfo {volume} {522}},\ \bibinfo {pages} {615} (\bibinfo
  {year} {2010})},\ \Eprint {http://arxiv.org/abs/0907.2193} {arXiv:0907.2193
  [cond-mat.quant-gas]} \BibitemShut {NoStop}%
\bibitem [{\citenamefont {Scherer}\ \emph {et~al.}(2011)\citenamefont
  {Scherer}, \citenamefont {Floerchinger},\ and\ \citenamefont
  {Gies}}]{Scherer:2010sv}%
  \BibitemOpen
  \bibfield  {author} {\bibinfo {author} {\bibfnamefont {M.~M.}\ \bibnamefont
  {Scherer}}, \bibinfo {author} {\bibfnamefont {S.}~\bibnamefont
  {Floerchinger}}, \ and\ \bibinfo {author} {\bibfnamefont {H.}~\bibnamefont
  {Gies}},\ }\href@noop {} {\bibfield  {journal} {\bibinfo  {journal} {Phil.
  Trans. R. Soc. A}\ }\textbf {\bibinfo {volume} {368}},\ \bibinfo {pages}
  {2779} (\bibinfo {year} {2011})}\BibitemShut {NoStop}%
\bibitem [{\citenamefont {Boettcher}\ \emph {et~al.}(2012)\citenamefont
  {Boettcher}, \citenamefont {Pawlowski},\ and\ \citenamefont
  {Diehl}}]{Boettcher:2012cm}%
  \BibitemOpen
  \bibfield  {author} {\bibinfo {author} {\bibfnamefont {I.}~\bibnamefont
  {Boettcher}}, \bibinfo {author} {\bibfnamefont {J.~M.}\ \bibnamefont
  {Pawlowski}}, \ and\ \bibinfo {author} {\bibfnamefont {S.}~\bibnamefont
  {Diehl}},\ }\href {\doibase 10.1016/j.nuclphysbps.2012.06.004} {\bibfield
  {journal} {\bibinfo  {journal} {Nucl. Phys. Proc. Suppl.}\ }\textbf {\bibinfo
  {volume} {228}},\ \bibinfo {pages} {63} (\bibinfo {year} {2012})},\ \Eprint
  {http://arxiv.org/abs/1204.4394} {arXiv:1204.4394 [cond-mat.quant-gas]}
  \BibitemShut {NoStop}%
\bibitem [{\citenamefont {Cooper}(1956)}]{Cooper56}%
  \BibitemOpen
  \bibfield  {author} {\bibinfo {author} {\bibfnamefont {L.~N.}\ \bibnamefont
  {Cooper}},\ }\href {\doibase 10.1103/PhysRev.104.1189} {\bibfield  {journal}
  {\bibinfo  {journal} {Phys. Rev.}\ }\textbf {\bibinfo {volume} {104}},\
  \bibinfo {pages} {1189} (\bibinfo {year} {1956})}\BibitemShut {NoStop}%
\bibitem [{\citenamefont {Pitaevskii}(2008)}]{PitaevskiiBook}%
  \BibitemOpen
  \bibfield  {author} {\bibinfo {author} {\bibfnamefont {L.}~\bibnamefont
  {Pitaevskii}},\ }\href {\doibase 10.1007/978-3-540-73253-2_2} {\emph
  {\bibinfo {title} {Superconductivity}}},\ edited by\ \bibinfo {editor}
  {\bibfnamefont {K.}~\bibnamefont {Bennemann}}\ and\ \bibinfo {editor}
  {\bibfnamefont {J.}~\bibnamefont {Ketterson}}\ (\bibinfo  {publisher}
  {Springer Berlin Heidelberg},\ \bibinfo {year} {2008})\ pp.\ \bibinfo {pages}
  {27--71}\BibitemShut {NoStop}%
\bibitem [{\citenamefont {Hubbard}(1959)}]{Hubbard59}%
  \BibitemOpen
  \bibfield  {author} {\bibinfo {author} {\bibfnamefont {J.}~\bibnamefont
  {Hubbard}},\ }\href {\doibase 10.1103/PhysRevLett.3.77} {\bibfield  {journal}
  {\bibinfo  {journal} {Phys. Rev. Lett.}\ }\textbf {\bibinfo {volume} {3}},\
  \bibinfo {pages} {77} (\bibinfo {year} {1959})}\BibitemShut {NoStop}%
\bibitem [{\citenamefont {Stratonovich}(1957)}]{Stratonovich57}%
  \BibitemOpen
  \bibfield  {author} {\bibinfo {author} {\bibfnamefont {R.}~\bibnamefont
  {Stratonovich}},\ }\href@noop {} {\bibfield  {journal} {\bibinfo  {journal}
  {Dokl. Akad. Nauk.}\ }\textbf {\bibinfo {volume} {115}},\ \bibinfo {pages}
  {1097} (\bibinfo {year} {1957})}\BibitemShut {NoStop}%
\bibitem [{\citenamefont {Diehl}\ and\ \citenamefont
  {Wetterich}(2007)}]{Diehl:2005ae}%
  \BibitemOpen
  \bibfield  {author} {\bibinfo {author} {\bibfnamefont {S.}~\bibnamefont
  {Diehl}}\ and\ \bibinfo {author} {\bibfnamefont {C.}~\bibnamefont
  {Wetterich}},\ }\href {\doibase 10.1016/j.nuclphysb.2007.02.026} {\bibfield
  {journal} {\bibinfo  {journal} {Nucl.Phys.}\ }\textbf {\bibinfo {volume}
  {B770}},\ \bibinfo {pages} {206} (\bibinfo {year} {2007})}\BibitemShut
  {NoStop}%
\bibitem [{\citenamefont {Chen}\ and\ \citenamefont
  {Kaplan}(2004)}]{Chen:2003vy}%
  \BibitemOpen
  \bibfield  {author} {\bibinfo {author} {\bibfnamefont {J.-W.}\ \bibnamefont
  {Chen}}\ and\ \bibinfo {author} {\bibfnamefont {D.~B.}\ \bibnamefont
  {Kaplan}},\ }\href {\doibase 10.1103/PhysRevLett.92.257002} {\bibfield
  {journal} {\bibinfo  {journal} {Phys. Rev. Lett.}\ }\textbf {\bibinfo
  {volume} {92}},\ \bibinfo {pages} {257002} (\bibinfo {year} {2004})},\
  \Eprint {http://arxiv.org/abs/hep-lat/0308016} {arXiv:hep-lat/0308016
  [hep-lat]} \BibitemShut {NoStop}%
\bibitem [{\citenamefont {Thies}(2006)}]{ThiesInhoRev06}%
  \BibitemOpen
  \bibfield  {author} {\bibinfo {author} {\bibfnamefont {M.}~\bibnamefont
  {Thies}},\ }\href {\doibase 10.1088/0305-4470/39/41/S04} {\bibfield
  {journal} {\bibinfo  {journal} {J.Phys.}\ }\textbf {\bibinfo {volume}
  {A39}},\ \bibinfo {pages} {12707} (\bibinfo {year} {2006})},\ \Eprint
  {http://arxiv.org/abs/hep-th/0601049} {arXiv:hep-th/0601049 [hep-th]}
  \BibitemShut {NoStop}%
\bibitem [{\citenamefont {Ba\ifmmode~\mbox{\c{s}}\else \c{s}\fi{}ar}\ and\
  \citenamefont {Dunne}(2008)}]{BasarDunnePRD08}%
  \BibitemOpen
  \bibfield  {author} {\bibinfo {author} {\bibfnamefont {G.}~\bibnamefont
  {Ba\ifmmode~\mbox{\c{s}}\else \c{s}\fi{}ar}}\ and\ \bibinfo {author}
  {\bibfnamefont {G.~V.}\ \bibnamefont {Dunne}},\ }\href {\doibase
  10.1103/PhysRevD.78.065022} {\bibfield  {journal} {\bibinfo  {journal} {Phys.
  Rev. D}\ }\textbf {\bibinfo {volume} {78}},\ \bibinfo {pages} {065022}
  (\bibinfo {year} {2008})}\BibitemShut {NoStop}%
\bibitem [{\citenamefont {{Braun}}\ \emph {et~al.}(2014)\citenamefont
  {{Braun}}, \citenamefont {{Finkbeiner}}, \citenamefont {{Karbstein}},\ and\
  \citenamefont {{Roscher}}}]{GNDTrick}%
  \BibitemOpen
  \bibfield  {author} {\bibinfo {author} {\bibfnamefont {J.}~\bibnamefont
  {{Braun}}}, \bibinfo {author} {\bibfnamefont {S.}~\bibnamefont
  {{Finkbeiner}}}, \bibinfo {author} {\bibfnamefont {F.}~\bibnamefont
  {{Karbstein}}}, \ and\ \bibinfo {author} {\bibfnamefont {D.}~\bibnamefont
  {{Roscher}}},\ }\href@noop {} {\  (\bibinfo {year} {2014})},\ \Eprint
  {http://arxiv.org/abs/1410.8181} {arXiv:1410.8181 [hep-ph]} \BibitemShut
  {NoStop}%
\bibitem [{\citenamefont {Wu}\ \emph {et~al.}(2006)\citenamefont {Wu},
  \citenamefont {Pao},\ and\ \citenamefont {Yip}}]{WuPaoYip06}%
  \BibitemOpen
  \bibfield  {author} {\bibinfo {author} {\bibfnamefont {S.-T.}\ \bibnamefont
  {Wu}}, \bibinfo {author} {\bibfnamefont {C.-H.}\ \bibnamefont {Pao}}, \ and\
  \bibinfo {author} {\bibfnamefont {S.-K.}\ \bibnamefont {Yip}},\ }\href
  {\doibase 10.1103/PhysRevB.74.224504} {\bibfield  {journal} {\bibinfo
  {journal} {Phys. Rev. B}\ }\textbf {\bibinfo {volume} {74}},\ \bibinfo
  {pages} {224504} (\bibinfo {year} {2006})}\BibitemShut {NoStop}%
\bibitem [{\citenamefont {Braun}\ \emph {et~al.}(2014)\citenamefont {Braun},
  \citenamefont {Drut}, \citenamefont {Jahn}, \citenamefont {Pospiech},\ and\
  \citenamefont {Roscher}}]{Braun:2014ewa}%
  \BibitemOpen
  \bibfield  {author} {\bibinfo {author} {\bibfnamefont {J.}~\bibnamefont
  {Braun}}, \bibinfo {author} {\bibfnamefont {J.~E.}\ \bibnamefont {Drut}},
  \bibinfo {author} {\bibfnamefont {T.}~\bibnamefont {Jahn}}, \bibinfo {author}
  {\bibfnamefont {M.}~\bibnamefont {Pospiech}}, \ and\ \bibinfo {author}
  {\bibfnamefont {D.}~\bibnamefont {Roscher}},\ }\href@noop {} {\bibfield
  {journal} {\bibinfo  {journal} {Phys. Rev.}\ }\textbf {\bibinfo {volume}
  {A89}},\ \bibinfo {pages} {053613} (\bibinfo {year} {2014})}\BibitemShut
  {NoStop}%
\bibitem [{\citenamefont {Boettcher}\ \emph
  {et~al.}(2014{\natexlab{a}})\citenamefont {Boettcher}, \citenamefont
  {Herbst}, \citenamefont {Pawlowski}, \citenamefont {Strodthoff},
  \citenamefont {von Smekal} \emph {et~al.}}]{BoettcherSarma}%
  \BibitemOpen
  \bibfield  {author} {\bibinfo {author} {\bibfnamefont {I.}~\bibnamefont
  {Boettcher}}, \bibinfo {author} {\bibfnamefont {T.}~\bibnamefont {Herbst}},
  \bibinfo {author} {\bibfnamefont {J.}~\bibnamefont {Pawlowski}}, \bibinfo
  {author} {\bibfnamefont {N.}~\bibnamefont {Strodthoff}}, \bibinfo {author}
  {\bibfnamefont {L.}~\bibnamefont {von Smekal}},  \emph {et~al.},\ }\href@noop
  {} {\  (\bibinfo {year} {2014}{\natexlab{a}})},\ \Eprint
  {http://arxiv.org/abs/1409.5232} {arXiv:1409.5232 [cond-mat.quant-gas]}
  \BibitemShut {NoStop}%
\bibitem [{\citenamefont {Wetterich}(1993)}]{Wetterich93}%
  \BibitemOpen
  \bibfield  {author} {\bibinfo {author} {\bibfnamefont {C.}~\bibnamefont
  {Wetterich}},\ }\href {\doibase 10.1016/0370-2693(93)90726-X} {\bibfield
  {journal} {\bibinfo  {journal} {Phys. Lett. B}\ }\textbf {\bibinfo {volume}
  {301}},\ \bibinfo {pages} {90} (\bibinfo {year} {1993})}\BibitemShut
  {NoStop}%
\bibitem [{\citenamefont {Braun}(2012)}]{Braun:2011pp}%
  \BibitemOpen
  \bibfield  {author} {\bibinfo {author} {\bibfnamefont {J.}~\bibnamefont
  {Braun}},\ }\href {\doibase 10.1088/0954-3899/39/3/033001} {\bibfield
  {journal} {\bibinfo  {journal} {J. Phys. G}\ }\textbf {\bibinfo {volume}
  {39}},\ \bibinfo {pages} {033001} (\bibinfo {year} {2012})},\ \Eprint
  {http://arxiv.org/abs/1108.4449} {arXiv:1108.4449 [hep-ph]} \BibitemShut
  {NoStop}%
\bibitem [{\citenamefont {Bartosch}\ \emph {et~al.}(2009)\citenamefont
  {Bartosch}, \citenamefont {Kopietz},\ and\ \citenamefont
  {Ferraz}}]{Bartosch:2009zr}%
  \BibitemOpen
  \bibfield  {author} {\bibinfo {author} {\bibfnamefont {L.}~\bibnamefont
  {Bartosch}}, \bibinfo {author} {\bibfnamefont {P.}~\bibnamefont {Kopietz}}, \
  and\ \bibinfo {author} {\bibfnamefont {A.}~\bibnamefont {Ferraz}},\ }\href
  {\doibase 10.1103/PhysRevB.80.104514} {\bibfield  {journal} {\bibinfo
  {journal} {{Phys. Rev. B}}\ }\textbf {\bibinfo {volume} {80}},\ \bibinfo
  {pages} {104514} (\bibinfo {year} {2009})},\ \Eprint
  {http://arxiv.org/abs/0907.2687} {arXiv:0907.2687 [cond-mat.quant-gas]}
  \BibitemShut {NoStop}%
\bibitem [{\citenamefont {Boettcher}\ \emph
  {et~al.}(2014{\natexlab{b}})\citenamefont {Boettcher}, \citenamefont
  {Pawlowski},\ and\ \citenamefont {Wetterich}}]{PhysRevA.89.053630}%
  \BibitemOpen
  \bibfield  {author} {\bibinfo {author} {\bibfnamefont {I.}~\bibnamefont
  {Boettcher}}, \bibinfo {author} {\bibfnamefont {J.~M.}\ \bibnamefont
  {Pawlowski}}, \ and\ \bibinfo {author} {\bibfnamefont {C.}~\bibnamefont
  {Wetterich}},\ }\href {\doibase 10.1103/PhysRevA.89.053630} {\bibfield
  {journal} {\bibinfo  {journal} {Phys. Rev. A}\ }\textbf {\bibinfo {volume}
  {89}},\ \bibinfo {pages} {053630} (\bibinfo {year}
  {2014}{\natexlab{b}})}\BibitemShut {NoStop}%
\bibitem [{\citenamefont {Krippa}(2014)}]{Krippa:2014kra}%
  \BibitemOpen
  \bibfield  {author} {\bibinfo {author} {\bibfnamefont {B.}~\bibnamefont
  {Krippa}},\ }\href@noop {} {\  (\bibinfo {year} {2014})},\ \Eprint
  {http://arxiv.org/abs/1407.5438} {arXiv:1407.5438 [cond-mat.quant-gas]}
  \BibitemShut {NoStop}%
\bibitem [{\citenamefont {Krahl}\ \emph {et~al.}(2009)\citenamefont {Krahl},
  \citenamefont {Friederich},\ and\ \citenamefont
  {Wetterich}}]{KrahlFriedWett_Hubbard09}%
  \BibitemOpen
  \bibfield  {author} {\bibinfo {author} {\bibfnamefont {H.~C.}\ \bibnamefont
  {Krahl}}, \bibinfo {author} {\bibfnamefont {S.}~\bibnamefont {Friederich}}, \
  and\ \bibinfo {author} {\bibfnamefont {C.}~\bibnamefont {Wetterich}},\ }\href
  {\doibase 10.1103/PhysRevB.80.014436} {\bibfield  {journal} {\bibinfo
  {journal} {Phys. Rev. B}\ }\textbf {\bibinfo {volume} {80}},\ \bibinfo
  {pages} {014436} (\bibinfo {year} {2009})}\BibitemShut {NoStop}%
\bibitem [{\citenamefont {S\'a~de Melo}\ \emph {et~al.}(1993)\citenamefont
  {S\'a~de Melo}, \citenamefont {Randeria},\ and\ \citenamefont
  {Engelbrecht}}]{PhysRevLett.71.3202}%
  \BibitemOpen
  \bibfield  {author} {\bibinfo {author} {\bibfnamefont {C.~A.~R.}\
  \bibnamefont {S\'a~de Melo}}, \bibinfo {author} {\bibfnamefont
  {M.}~\bibnamefont {Randeria}}, \ and\ \bibinfo {author} {\bibfnamefont
  {J.~R.}\ \bibnamefont {Engelbrecht}},\ }\href {\doibase
  10.1103/PhysRevLett.71.3202} {\bibfield  {journal} {\bibinfo  {journal}
  {Phys. Rev. Lett.}\ }\textbf {\bibinfo {volume} {71}},\ \bibinfo {pages}
  {3202} (\bibinfo {year} {1993})}\BibitemShut {NoStop}%
\bibitem [{\citenamefont {Randeria}\ and\ \citenamefont
  {Trivedi}(1998)}]{Randeria19981754}%
  \BibitemOpen
  \bibfield  {author} {\bibinfo {author} {\bibfnamefont {M.}~\bibnamefont
  {Randeria}}\ and\ \bibinfo {author} {\bibfnamefont {N.}~\bibnamefont
  {Trivedi}},\ }\href {\doibase
  http://dx.doi.org/10.1016/S0022-3697(98)00099-7} {\bibfield  {journal}
  {\bibinfo  {journal} {Journal of Physics and Chemistry of Solids}\ }\textbf
  {\bibinfo {volume} {59}},\ \bibinfo {pages} {1754 } (\bibinfo {year}
  {1998})}\BibitemShut {NoStop}%
\bibitem [{\citenamefont {Ku}\ \emph {et~al.}(2012)\citenamefont {Ku},
  \citenamefont {Sommer}, \citenamefont {Cheuk},\ and\ \citenamefont
  {Zwierlein}}]{Ku_etal12}%
  \BibitemOpen
  \bibfield  {author} {\bibinfo {author} {\bibfnamefont {M.~J.~H.}\
  \bibnamefont {Ku}}, \bibinfo {author} {\bibfnamefont {A.~T.}\ \bibnamefont
  {Sommer}}, \bibinfo {author} {\bibfnamefont {L.~W.}\ \bibnamefont {Cheuk}}, \
  and\ \bibinfo {author} {\bibfnamefont {M.~W.}\ \bibnamefont {Zwierlein}},\
  }\href {\doibase 10.1126/science.1214987} {\bibfield  {journal} {\bibinfo
  {journal} {Science}\ }\textbf {\bibinfo {volume} {335}},\ \bibinfo {pages}
  {563} (\bibinfo {year} {2012})}\BibitemShut {NoStop}%
\bibitem [{\citenamefont {Efimov}(1972)}]{Efimov1}%
  \BibitemOpen
  \bibfield  {author} {\bibinfo {author} {\bibfnamefont {V.}~\bibnamefont
  {Efimov}},\ }\href@noop {} {\bibfield  {journal} {\bibinfo  {journal} {Sov.
  Phys. JETP Lett.}\ }\textbf {\bibinfo {volume} {16}},\ \bibinfo {pages} {34}
  (\bibinfo {year} {1972})}\BibitemShut {NoStop}%
\bibitem [{\citenamefont {Efimov}(1973)}]{Efimov2}%
  \BibitemOpen
  \bibfield  {author} {\bibinfo {author} {\bibfnamefont {V.}~\bibnamefont
  {Efimov}},\ }\href@noop {} {\bibfield  {journal} {\bibinfo  {journal} {Nucl.
  Phys. A}\ }\textbf {\bibinfo {volume} {210}},\ \bibinfo {pages} {157}
  (\bibinfo {year} {1973})}\BibitemShut {NoStop}%
\bibitem [{\citenamefont {Niemann}\ \emph {et~al.}()\citenamefont {Niemann},
  \citenamefont {Roscher}, \citenamefont {Braun},\ and\ \citenamefont
  {Hammer}}]{NRBH}%
  \BibitemOpen
  \bibfield  {author} {\bibinfo {author} {\bibfnamefont {P.}~\bibnamefont
  {Niemann}}, \bibinfo {author} {\bibfnamefont {D.}~\bibnamefont {Roscher}},
  \bibinfo {author} {\bibfnamefont {J.}~\bibnamefont {Braun}}, \ and\ \bibinfo
  {author} {\bibfnamefont {H.-W.}\ \bibnamefont {Hammer}},\ }\href@noop {} {\
  }\Eprint {http://arxiv.org/abs/in preparation (2015)} {in preparation (2015)}
  \BibitemShut {NoStop}%
\bibitem [{\citenamefont {Litim}(2000)}]{Litim:2000ci}%
  \BibitemOpen
  \bibfield  {author} {\bibinfo {author} {\bibfnamefont {D.~F.}\ \bibnamefont
  {Litim}},\ }\href {\doibase 10.1016/S0370-2693(00)00748-6} {\bibfield
  {journal} {\bibinfo  {journal} {Phys. Lett.}\ }\textbf {\bibinfo {volume}
  {B486}},\ \bibinfo {pages} {92} (\bibinfo {year} {2000})}\BibitemShut
  {NoStop}%
\bibitem [{\citenamefont {Litim}(2001{\natexlab{a}})}]{Litim:2001fd}%
  \BibitemOpen
  \bibfield  {author} {\bibinfo {author} {\bibfnamefont {D.~F.}\ \bibnamefont
  {Litim}},\ }\href {\doibase 10.1142/S0217751X01004748} {\bibfield  {journal}
  {\bibinfo  {journal} {Int.J.Mod.Phys.}\ }\textbf {\bibinfo {volume} {A16}},\
  \bibinfo {pages} {2081} (\bibinfo {year} {2001}{\natexlab{a}})}\BibitemShut
  {NoStop}%
\bibitem [{\citenamefont {Litim}(2001{\natexlab{b}})}]{Litim:2001up}%
  \BibitemOpen
  \bibfield  {author} {\bibinfo {author} {\bibfnamefont {D.~F.}\ \bibnamefont
  {Litim}},\ }\href {\doibase 10.1103/PhysRevD.64.105007} {\bibfield  {journal}
  {\bibinfo  {journal} {Phys. Rev.}\ }\textbf {\bibinfo {volume} {D64}},\
  \bibinfo {pages} {105007} (\bibinfo {year} {2001}{\natexlab{b}})}\BibitemShut
  {NoStop}%
\bibitem [{\citenamefont {Pawlowski}(2007)}]{Pawlowski20072831}%
  \BibitemOpen
  \bibfield  {author} {\bibinfo {author} {\bibfnamefont {J.~M.}\ \bibnamefont
  {Pawlowski}},\ }\href {\doibase 10.1016/j.aop.2007.01.007} {\bibfield
  {journal} {\bibinfo  {journal} {Annals of Physics}\ }\textbf {\bibinfo
  {volume} {322}},\ \bibinfo {pages} {2831 } (\bibinfo {year}
  {2007})}\BibitemShut {NoStop}%
\end{thebibliography}%

\end{document}